\documentclass[12pt,preprint]{aastex}

\newcommand{\logg} {\log g}
\newcommand{\Te} {T_{\rm eff}}

\newcommand{\msun} {$M_\odot$}
\newcommand{\lsun}{$L_{\odot}$}
\newcommand\gta{\lower 0.5ex\hbox{$\buildrel > \over \sim\ $}} 
\newcommand\lta{\lower 0.5ex\hbox{$\buildrel < \over \sim\ $}} 
\newcommand{\nh} {{\rm H}/{\rm He}}

\shortauthors{Bergeron et al.}
\shorttitle{Spectroscopic Study of DB White Dwarfs}

\begin{document}


\title{A Comprehensive Spectroscopic Analysis of DB White Dwarfs}

\author{P. Bergeron$^1$, F. Wesemael$^1$, Pierre Dufour$^1$, A. Beauchamp$^{1,2}$, C. Hunter$^{1,3}$, 
Rex~A. Saffer$^4$, A. Gianninas$^1$, M.~T. Ruiz$^5$, M.-M. Limoges$^1$, Patrick Dufour$^1$, G. Fontaine$^1$, 
James Liebert$^6$}

\affil{$^1$D\'epartement de Physique, Universit\'e de Montr\'eal, C.P.~6128, Succ.~Centre-Ville, Montr\'eal, QC H3C 3J7, Canada; bergeron@astro.umontreal.ca, wesemael@astro.umontreal.ca, gianninas@astro.umontreal.ca, limoges@astro.umontreal.ca, dufourpa@astro.umontreal.ca, fontaine@astro.umontreal.ca}

\affil{$^2$Forensic Technologies Wai Inc., 5757 Boulevard Cavendish, Montr\'eal, QC H4W 2W8, Canada;
alain.beauchamp@fti-ibis.com}

\affil{$^3$ITS Yale University, Box 27387, West Haven, CT 06516, USA; chris.hunter@yale.edu}

\affil{$^4$DataTime Consulting, 109 Forrest Avenue, Narberth, PA 19072, USA; rex.saffer@comcast.net}

\affil{$^5$Departamento de Astronom\'\i a, Universidad de Chile, Casilla 36-D, Santiago, Chile;
mtruiz@das.uchile.cl}

\affil{$^6$Steward Observatory, University of Arizona, Tucson, AZ 85721, USA}

\begin{abstract}

We present a detailed analysis of 108 helium-line (DB) white dwarfs
based on model atmosphere fits to high signal-to-noise optical spectroscopy. 
We derive a mean mass of 0.67 \msun\ for our sample,
with a dispersion of only 0.09 \msun. White dwarfs also showing
hydrogen lines, the DBA stars, comprise 44\% of our sample, and their
mass distribution appears similar to that of DB stars. As in our
previous investigation, we find no evidence for the existence of
low-mass ($M<0.5$ \msun) DB white dwarfs. We derive a luminosity
function based on a subset of DB white dwarfs identified in the
Palomar-Green survey. We show that 20\% of all white dwarfs in the
temperature range of interest are DB stars, although the fraction
drops to half this value above $\Te\sim20,000$~K. We also show that
the persistence of DB stars with no hydrogen features at low
temperatures is difficult to reconcile with a scenario involving
accretion from the interstellar medium, often invoked to account for
the observed hydrogen abundances in DBA stars. We present evidence for
the existence of two different evolutionary channels that produce DB
white dwarfs: the standard model where DA stars are transformed into
DB stars through the convective dilution of a thin hydrogen
layer, and a second channel where DB stars retain a
helium-atmosphere throughout their evolution. We finally demonstrate
that the instability strip of pulsating V777 Her white dwarfs contains
no nonvariables, if the hydrogen content of these stars is properly
accounted for.

\end{abstract}


\section{Introduction}

DB white dwarfs are characterized by the extreme purity of their
helium-dominated atmospheres with, occasionally, only minute traces of
hydrogen (DBA stars) and heavy elements (DBQ or DBZ stars) seen in
their optical spectra. It has been 15 years since we last reviewed the
work carried out in Montreal on the optical spectra of DB and DBA
white dwarfs \citep{beauchamp96}. Parts of this work have been
published over the years
\citep{beauchamp95,BWB97,beauchamp98,BW98,beauchamp99,hunter01,wesemael01},
and full and definitive results from this long-term project are
presented here. 

Progress on several fronts relevant to the study of DB stars has been
significant since our review
\citep{beauchamp96}: the discovery of stars in the DB gap through the Sloan
Digital Sky Survey (SDSS; \citealt{eisen06}), the discovery of a new
class of carbon-rich stars --- the so-called Hot DQ stars
\citep{dufour07b,dufour08}, the analysis of the DB stars in the ESO
Supernova Ia Progenitor Survey (SPY; \citealt{voss07}), the
far-ultraviolet observations of DB stars with FUSE
\citep{petitclerc05,desharnais08,dufour10a}, the likelihood of
atmospheric contamination of the photospheres of DB stars through
disk-fed accretion (e.g., \citealt{dufour10b}), etc. Yet our
knowledge of some of the fundamental properties of DB stars, embodied
in a set of questions posed by
\citet{beauchamp96}, remains sketchy. These questions referred to the
respective mass distributions for the DA and DB stars and to the
possible existence of a bimodal mass distribution for DB degenerates,
to the existence of statistical evidence for differences between the
DB and DBA samples, as well as to the existence of atmospheric
peculiarities in hot DB stars that might afford possible evidence for
a recent episode of convective mixing. 

One particular problem with the analysis of DB white dwarfs is related to
the effect of invisible traces of hydrogen that can affect significantly
the temperature scale of DB stars based on line profile fitting
techniques. For instance, \citet[][see their Table 1]{beauchamp99} showed
that small amounts of hydrogen, of the order of $\nh\sim10^{-5}-10^{-4}$,
can produce differences as large as 3000~K in $\Te$ for the hottest DB
stars. In the case of the Beauchamp et al.~analysis, upper limits on the
hydrogen abundance were based only on the absence of H$\beta$. To improve
on this aspect of the analysis, \citet{voss07} secured high-resolution
spectra in the region near H$\alpha$ for a sample of $\sim 60$ DB stars
in the SPY survey. A similar analysis had been carried out earlier by
\citet{hunter01}, but on a smaller sample of hot DB stars. With their
high-resolution data, Voss et al. were able to secure more stringent
limits on the hydrogen abundance. In their Table 1, they contrast
the effective temperatures and surface gravities of DB stars without
visible hydrogen obtained from two distinct grids of model atmospheres:
one with a pure helium composition, and one with a trace of hydrogen,
$\nh=10^{-5}$, the upper limit imposed by their data.
As discussed in Voss et al., the average temperature difference they
measure, around 230~K, is significantly smaller than that obtained by
\citet{beauchamp99} based on much higher upper limits on the hydrogen
abundance. However, this evaluation is somewhat misleading since the
upper limit imposed by the SPY data is strongly temperature dependent
according to the results shown in Figure 9 of Voss et al. In fact, at
$\Te\sim24,000$~K, their detection limit of 300 m\AA\ for the
H$\alpha$ equivalent width indicates an upper limit of
$\nh\sim10^{-4}$, or a factor of ten {\it larger} than the value used
in their Table 1. Hence we feel that the influence of invisible
amounts of hydrogen on the determination of the atmospheric parameters
of DB stars still needs to be properly addressed.

Another complication with the spectroscopic analysis of DB stars
is related to the problem of the convective efficiency in their atmospheres.
While the spectroscopic analysis of 
\citet{beauchamp99} relied on model atmospheres calculated
with the so-called ML2/$\alpha=1.25$ version of the mixing-length theory,
\citet{voss07} adopted in their study a much less efficient $\alpha=0.6$ version,
i.e.~the same convective efficiency as for DA white dwarfs
\citep{bergeron95}. This may represent an important source of
discrepancy when comparing the atmospheric parameters from various
analyses. While some authors argued that there is no reason to expect
the mixing length parameterization to be different in DA and DB white
dwarfs \citep[e.g.,][]{castan06}, we may as well argue that there is
no reason to expect it to be identical. Hence we feel the problem of
the convective efficiency in the atmospheres of DB stars needs to be
reexamined as well.

Of deeper astrophysical interest is the question of the possible
evolutionary channels that could produce DB white dwarfs. The
progenitors of DB white dwarfs are believed to be DA stars,
transformed into helium-atmosphere white dwarfs as a result of the
convective dilution of a thin radiative hydrogen layer ($M_{\rm
H}\sim 10^{-15}$ \msun) with the deeper convective helium envelope
\citep{fontaine87}. Evidence for this mechanism rests on the complete
absence of helium-atmosphere white dwarfs in the Palomar-Green (PG)
survey between $\Te\sim45,000$~K --- where the coolest DO stars are
found, and $\sim 30,000$~K --- where the hottest DB stars are
located. This so-called DB gap has partially been filled in by the
discovery of extremely hot DB stars in the SDSS
\citep{eisen06}, although the number of DB stars in the gap remains a factor
of 2.5 lower than what is expected based on the number of cooler DB
stars in the SDSS. We are thus dealing with a DB {\it
deficiency}, rather than with a real gap, in this particular range of
effective temperature. How do these DB stars in the gap fit in our understanding
of white dwarf evolution in general?

All these questions remain current, and need to be addressed for a
complete understanding of the physical properties and evolutionary
status of DB stars. With the hope of shedding some light on these
issues, we present in this paper a complete and comprehensive
spectroscopic analysis of over 100 DB stars. We first begin in Section
2 with a theoretical investigation of model atmospheres of DB stars to
explore the observational and theoretical issues at stake. The
observational data used in our analysis are then presented in Section
3. Our spectroscopic analysis follows in Section 4 and the results are
interpreted in Section 5.  We present our conclusions in
Section 6.

\section{Theoretical Considerations}

\subsection{Model Atmospheres}

Our model atmospheres and synthetic spectra are built from the model
atmosphere code described in \citet{TB09} and references therein, in
which we have incorporated the improved Stark profiles of neutral
helium of \citet{BWB97}.  These detailed profiles for over twenty
lines of neutral helium take into account the transition from
quadratic to linear Stark broadening, the transition from impact to
quasi-static regime for electrons, as well as forbidden components.
In the context of DB stars, these models are comparable to those
described by \citet{B95} and \citet{beauchamp96}, with the exception
that at low temperatures ($\Te<10,800$~K), we now use the free-free
absorption coefficient of the negative helium ion of
\citet{john94}. For the hydrogen lines in DBA stars, we rely on the
improved calculations for Stark broadening of \citet{TB09}. Our models
are in LTE and include convective energy transport treated within the
mixing-length theory.  The theoretical spectra are calculated using
the occupation formalism of
\citet{HM88} for both the hydrogen and helium populations and corresponding
bound-bound, bound-free, and pseudo-continuum opacities. 

An important issue related to the modeling of DB spectra is the
inclusion of van der Waals broadening at low effective temperatures
($\Te
\lesssim15,000$~K), where Stark broadening is no longer the dominant 
source of line broadening. \citet[][see their Figure 6]{beauchamp96}
show the effect of this broadening mechanism in a $\logg$ vs $\Te$ diagram
for DB stars with parameters determined from optical spectroscopy.
The neglect of van der Waals broadening results in spuriously high
$\logg$ values at low effective temperatures. In this series of synthetic
spectrum calculations, we include the treatment of this broadening
mechanism provided by \citet{deridder76}.

Our model grid covers a range of effective temperature between $\Te=
11,000$ K and 40,000~K by steps of 1000 K, while the $\logg$ ranges
from 7.0 to 9.0 by steps of 0.5 dex.  In addition to pure helium
models, we also calculated models with $\log \nh = -6.5$ to $-2.0$ by
steps of 0.5. In order to explore the effect of varying the convective
efficiency on the predicted emergent fluxes, models have been
calculated with the ML2 version of the mixing-length theory, which
corresponds to the prescription of \citet{ML2}, with various values of
$\alpha=\ell/H$, i.e.~the ratio of the mixing length to the pressure
scale height, ranging from 0.75 to 1.75 by steps of 0.25.

\subsection{Exploration of the Parameter Space}

The determination of the atmospheric parameters, $\Te$, $\logg$, and
H/He, of DB stars represents an inherent difficulty in any analysis of their properties.
While the atmospheric parameters for DA stars derived from
various studies agree generally well (see, e.g., Figure 9 of
\citealt{LBH05}), a similar comparison for DB stars is less than
satisfactory, as shown for instance in Figures 3 and 4 of
\citet{voss07}, which compare the values obtained from the SPY survey
with those obtained from independent studies
\citep{beauchamp99,friedrich00,castan06}. 

In addition to being rarer than their hydrogen-line DA counterparts,
the hotter DB stars are characterized by an optical spectrum where the
neutral helium transitions exhibit little sensitivity to effective
temperature. As an illustration, we show in the left panel of Figure
\ref{fg:f1} the model spectra\footnote{Unless
specified, the models displayed here assume a value of $\alpha=1.25$.}
for various values of $\Te$ (shown in blue), compared to a model
template at $\Te=24,000$~K (shown in red). There is a wide
range of temperature, between $\Te\sim20,000$~K and 30,000~K, where
the optical spectra are virtually identical. This underscores the need to secure
well calibrated, high signal-to-noise spectroscopic
observations to study these stars. Equivalently, we show in the left
panel of Figure
\ref{fg:f2} the variation of the equivalent width of
He~\textsc{i} $\lambda$4471 as a function of effective temperature for
various hydrogen abundances. In all cases, the variation of the
equivalent width shows a wide plateau, about 10,000 K wide, that
inhibits the determination of accurate effective temperatures from the
optical spectrum of DB stars in this particular range of
temperature. Also shown in this figure are the {\it measured}
equivalent widths in our sample of DB stars, arbitrarily located at
$\Te=22,000$~K. Interestingly enough, the DB stars in our sample with
the strongest lines have equivalent widths that are larger than
predicted by any of our models shown in this panel.

Luckily, we fare somewhat better with gravity indicators, as
illustrated in the middle panel of Figure \ref{fg:f1}. The 
large gravity-sensitivity of several segments containing
neutral helium transitions (notably the $2^3P-6^3D$ $\lambda$3819
line, the 4100-4200 \AA\ region that contains the $2^3P-5^3S$
$\lambda$4121, $2^1P-6^1D$ $\lambda$4144, and $2^1P-6^1S$
$\lambda$4169 lines, as well as the strong $2^1P-5^1D$ $\lambda$4388
transition; \citealt{W79}) can be used successfully through the entire
temperature range. We disagree in that respect with the statement of
\citet[][see Section 3.1]{voss07} that the dependence of line
profiles on $\logg$ is smaller for the cooler DB stars. Interestingly,
the variation of the He~\textsc{i} $\lambda$4471
equivalent width with $\Te$ is also strongly $\logg$ dependent,
as can be observed in the middle panel of Figure
\ref{fg:f2}. For low-gravity DB white dwarfs, the maximum
is now strongly peaked near 22,000~K, while for high-gravity DB stars,
the plateau becomes even wider, and the maximum equivalent
width is reached at temperatures near 33,000~K. Again, variations
of $\logg$ help very little in matching the maximum
{\it observed} equivalent widths of the DB stars in our sample.

Less well documented is the sensitivity of the optical spectrum of DB
stars to the convective efficiency. Helium atmospheres of white dwarfs
become convective below $\Te\sim 60,000$~K, but it is not until the
temperature drops below $\sim30,000$~K that the helium convection zone
becomes sufficiently important, depending on the assumed convective
efficiency (see Figure 10 of \citealt{tassoul90}).  For completeness
and future reference, we show in Figure
\ref{fg:f3} the extent of the helium convection zone in a 0.6
\msun\ DB white dwarf. Convective energy transport is generally
treated in model atmosphere calculations within the mixing-length
theory, and unless convection becomes adiabatic, the emergent model fluxes
will depend on the assumed convective efficiency, parameterized in this
case by the free parameter $\alpha$. This additional parameterization
was first introduced in the context of DB stars by \citet{B95} on the basis of work carried out in
the hydrogen-line DA stars \citep{bergeron95}. 

To illustrate this sensitivity, we compare in the
right panel of Figure \ref{fg:f1} model spectra
calculated with $\alpha=0.75$ and 1.75. The effect is 
subtle, and only noticeable between $\Te\sim 18,000$~K and
28,000~K. At higher effective temperatures, the atmospheres of DB
stars become almost completely radiative (see Figure
\ref{fg:f3}), while at lower temperatures convection becomes
adiabatic, and the convective efficiency no longer depends on the
specific value of $\alpha$. Despite the apparent lack of sensitivity to
variations of $\alpha$, the effects on the normalized line profiles
are more pronounced. This is illustrated in the right panel of Figure
\ref{fg:f2}, where the equivalent width of He \textsc{i} $\lambda$4471
as a function of effective temperature is shown for model atmospheres
calculated with various convective efficiencies. 
Once again, the line strength is sensitive to $\alpha$ for
effective temperatures between 18,000 K and 28,000 K and, in contrast
to the results shown in the previous panels, the largest
predicted equivalent widths (achieved with $\alpha=0.75$) are larger than the
largest measured values. This means that, were we to overestimate
$\alpha$ in our models, all the stars would be shifted toward the
maximum of the curve (i.e., near 30,000 K for $\alpha=1.75$) and form
a clump. In contrast, were we to underestimate $\alpha$, the stars we fit would
be lumped on respective sides of the maximum of the curve (i.e., near
18,000 K and 30,000 K for $\alpha=0.75$), and gaps would develop in the
distribution of stars with respect to effective temperature. This
behavior can be used to constrain $\alpha$, as we show below.

We finally note that the results presented here are consistent with those
of \citet{voss07} who mention that the helium lines in their models,
calculated with ML2/$\alpha=0.6$ (i.e. a value even smaller than the
smallest value used in Figure \ref{fg:f2}), reach a maximum
strength around $\Te=20,000$~K.

\section{Spectroscopic Observations}

Our sample of DB stars was selected from the online version of the
Villanova White Dwarf
Catalog\footnote{http://www.astronomy.villanova.edu/WDCatalog/index.html}. One
of the first goals of this project was to secure spectroscopic
observations of all DB stars identified in the PG Survey \citep{PG} in
order to compare the luminosity function of DB stars with that
obtained by
\citet{LBH05} for DA stars. Otherwise, DB white dwarfs were routinely
observed over the years and included in our sample.  Due to
observational constraints, we focused mainly on the brightest white
dwarfs in the catalog, and as such, there is little overlap between
our sample and the SDSS sample. The distribution of $V$
magnitudes\footnote{These can be photographic magnitudes in some
cases, or even $B$ magnitudes.} for the 118 objects observed
spectroscopically is presented in Figure \ref{fg:f4} (we exclude
here the 10 misclassified stars discussed further below). Several
objects in this original sample are excluded from our analysis. Three
are DA + DB double degenerate systems and have been analyzed
elsewhere: PG 1115+166
\citep{BL02}, KUV 02196+2816
\citep{limoges09}, and KUV 14197+2514 \citep{limoges10}. Two are
magnetic DB stars: PG 0853+164 \citep{wesemael01} and Feige 7
\citep{achilleos92}. Finally, five white dwarfs do not show enough
helium lines in their optical spectra to be analyzed quantitatively: LP 891-12
(0443$-$275.1), G102-39 (0551+123), GD 84 (0714+458), G227-5
(1727+560), and LDS 678A (1917$-$077).

We are thus left with a total of 108 DB white dwarfs, including several
DBA and DBZ stars (the analysis of \citet{beauchamp96} included close
to 80 objects).  The complete list of objects is provided in Table 1. For
comparison, a recent analysis based on the SPY sample \citep{voss07}, the
largest completed up to now, includes 69 white dwarfs. Most of our objects
were observed at high signal-to-noise ratio (S/N) in the blue region of
the optical spectrum (3700--5200 \AA) with the Steward Observatory 2.3 m
Bok Telescope equipped with the Boller \& Chivens spectrograph. The 4$\farcs$5
slit together with the 600 line mm$^{-1}$ grating blazed at 3568~\AA\
in first order provides a spectral coverage from about 3500 to 5250
\AA\ at a resolution of $\sim6$ \AA\ FWHM. Additional blue spectra in
the southern hemisphere were secured at Carnegie Observatories' 2.5 m
Ir\'en\'ee du Pont Telescope at Las Campanas (Chile) with the Boller \&
Chivens spectrograph. The 1$\farcs$5 slit with the 600 line mm$^{-1}$ grating
blazed at 5000 \AA\ provided a spectral coverage from about 3500 to
6600 \AA\ at a slightly better resolution of $\sim$ 3 \AA\ FWHM. Finally,
the spectrum of GD 554 (2250+746) was obtained at the Mt.~Hopkins 6.5 m
MMT telescope using the Blue Channel of the MMT Spectrograph. The 500
lines mm$^{-1}$ grating and a 1$\farcs$0 slit provided a spectral coverage
from about 3400 to 6300~\AA\ with a resolution of $\sim$ 4 \AA\ FWHM.

Our blue optical spectra are displayed in Figure
\ref{fg:f5} in order of decreasing
effective temperature.  In addition to the numerous neutral helium
lines that dominate the spectra of these stars, hydrogen lines
(H$\beta$) and the Ca~\textsc{ii} H \& K doublet are also visible in
some objects. The spectrum of L7-44 (1708$-$871) suffers from residual
flux calibration problem, while that of HE 0429$-$1651 (0429$-$168) is
significantly contaminated by the presence of an M dwarf
companion. Note that the spectra of the DB stars recently identified
by \citet{lepine11} represent new observations at higher S/N. One of
the most obvious qualities of our sample is its homogeneity, both in
terms of wavelength coverage and signal-to-noise. While its size remains
small compared to that of the SDSS sample, there is no comparison
in terms of the quality of the spectra, as can be seen from Figure
\ref{fg:f6}, where we compare the distribution of S/N of our observations
with that of the SDSS spectra from the Data Release 4. The
majority of our spectra have S/N well above 50, while the SDSS spectra
are strongly peaked between $\rm{S/N}\sim5$ and 20.  Since the
exposure time of a given SDSS spectrum is set by that of the entire
plate, the corresponding S/N is necessarily a function of the
magnitude of the star, and SDSS white dwarfs are much fainter than
those studied here.

As part of our survey, we observed several DB white dwarf candidates
that turned out to be lower surface gravity objects, mostly sdOB stars
(with the glaring exception of one DA star). For completeness, we
show these misclassified objects in Figure
\ref{fg:f7}. Note that several of these have already been
reclassified properly in the online version of the Villanova White
Dwarf Catalog.

Because coverage of the H$\alpha$ line is essential for the
determination of accurate atmospheric parameters for DB stars, we
complemented our blue data with red spectra obtained from several
distinct sources: the KPNO 4 m data used by \citet{hunter01}, spectra
secured at the du Pont 2.5 m telescope, spectra extracted from the
Data Release 4 of the SDSS, and SPY spectra from \citet[][only those
with H$\alpha$ detected]{voss07}, available from the ESO archive web
site. We have also recently secured additional spectra at the Steward
Observatory 2.3 m Bok Telescope and at the KPNO 4 m telescope.  Our
own spectroscopic observations (36 objects) as well as SDSS spectra
(17 objects) are displayed in Figure \ref{fg:f8} in order of
right ascension; these have typically 3 to 5 \AA\ resolution
(FWHM). SPY/UVES spectra for 13 additional objects can be visualized
as online material in
\citet{voss07}. To guide the eye, we also show in this figure the location
of the H$\alpha$ absorption feature. Hydrogen is detected in 50\% of
the objects shown here, a value in good agreement with the 55\%
fraction of DBA stars estimated by \citet{voss07}. This fraction
drops to $\sim 44$\% if we consider all stars in our sample, since
42 objects remain without H$\alpha$ data. Note how the strength of H$\alpha$ 
varies considerably from object to object, and how it is
particularly strong in LP 497-114 (1311+129). 

Finally, in Section 4.4, we also make use of {\it IUE} low-dispersion
spectra issued from the reduction procedure of \citet{holberg03},
which follows the prescription of \citet{massa} for the correction of
residual temporal and thermal effects and for the absolute flux
calibration. A total of 34 DB stars in our sample have UV spectra suitable for
analysis.

\section{Atmospheric Parameter Determination}

\subsection{Fitting Technique}

Our fitting technique is similar to that outlined in \citet[][and
references therein]{LBH05} for the analysis of DA stars. The first
step is to normalize the flux from individual predefined wavelength
segments, in both observed and model spectra, to a continuum set to
unity.  The comparison with model spectra, which are convolved with
the appropriate Gaussian instrumental profile, is then carried out in
terms of this normalized spectrum only. The most sensitive aspect of
this approach is to properly define the continuum of the observed
spectra. Here we rely on the procedure developed by \citet{LBH05},
where the fluxed spectrum is first fitted with model spectra,
multiplied by a high order polynomial (up to $\lambda^5$), in order to
account for any residual error in the flux calibration.  The nonlinear
least-squares minimization method of Levenberg-Marquardt
\citep{press86} is used throughout. Note that the values of $\Te$ and
$\logg$ derived in this first step are meaningless since too many
fitting parameters are considered, and the model fit just serves as a
smooth fitting function used to define the continuum of the observed
spectrum. Normal points are then fixed at several wavelengths defined
by this smooth function, and used to normalize the spectrum. This
method turns out to be quite accurate when a glitch is present in the
spectrum at the location where the continuum is set. It also provides
a precise value of the line center, which can be corrected to the
laboratory wavelength.  An example of continuum fitting using this
procedure is illustrated in the top panel of Figure
\ref{fg:f9}.

Once the spectrum is normalized to a continuum set to unity, as shown
in the bottom panel of Figure \ref{fg:f9}, we use our
grid of model spectra to determine $\Te$, $\logg$, and $\nh$ in terms
of the normalized spectrum only. Our three-dimensional minimization
technique again relies on the nonlinear least-squares method of
Levenberg-Marquardt, which is based on a steepest descent method.

\subsection{Determination of $\Te$, $\logg$, and H/He}

As shown in Figure \ref{fg:f2}, two solutions exist for a
given DB spectrum, one on each side of the maximum strength of the
neutral helium lines. In most cases, it is easy to distinguish these
cool and hot solutions from an examination of our best fits, as
displayed in the left panel of Figure
\ref{fg:f10} for the cool DB star LP 475-242 (0437+138). In addition, we also
look at our spectroscopic solution in terms of {\it absolute} fluxes
(not shown here) by normalizing both the observed and predicted model spectra
at 4250 \AA. In general, the slopes agree well, except in some
cases where the object has been observed at high airmasses, in
which case the slope of the observed spectrum might be distorted.

However, when the star is close to the temperature where the helium
lines reach their maximum strength --- a temperature range that
strongly depends on $\logg$ and on the assumed convective efficiency
according to Figure \ref{fg:f2} --- the cool and hot
solutions cannot be distinguished, as shown in the right panel of
Figure \ref{fg:f10} for the hot DB star PG 1115+158. The
comparison of the slopes of the energy distributions does not help
either in this case. Fortunately, we discovered that this ambiguity
could be resolved with the help of spectroscopic observations at
H$\alpha$, which add an additional constraint in our fitting
procedure. Since our H$\alpha$ data are independent of our blue
observations, we use an iterative procedure where we fit the blue
spectrum with an assumed hydrogen abundance to obtain a first estimate
of $\Te$ and $\logg$. The H$\alpha$ spectra are then used to determine
the hydrogen abundance, or upper limits on H/He, at these particular
values of $\Te$ and $\logg$.  The procedure is then repeated until the
value of H/He has converged. An example of our solution for PG
1115+158 is displayed in Figure \ref{fg:f11}. The weak H$\alpha$
absorption feature shown in the insert (from a SDSS spectrum, in this case) serves as
an important constraint to determine the hydrogen abundance, which
otherwise could only be inferred from the weak depression observed
near H$\beta$. Interestingly, our final solution for this object lies
between the cool and hot solutions displayed in Figure
\ref{fg:f10}.

The procedure described above works well when any hydrogen absorption
feature (H$\alpha$ or H$\beta$) is visible in the optical
spectrum. When no such feature is present, only upper limits on the
hydrogen abundance can be determined. These limits depend on the
signal-to-noise ratio of the observations, and on whether they are
based on our blue spectra for H$\beta$, or our red spectra for
H$\alpha$. We show, in Figure \ref{fg:f12}, the limits on the
hydrogen abundance imposed by the absence of H$\alpha$ and H$\beta$
features in our spectra (we adopt a detection limit of 200 m\AA\ and
300 m\AA\ for the equivalent widths of H$\alpha$ and H$\beta$,
respectively); the results for H$\alpha$ are qualitatively similar to
those shown in Figure 9 of \citet{voss07}. For high S/N spectra with
no detectable hydrogen feature\footnote{We use in particular the fact
that no detectable H$\alpha$ absorption feature has been reported by
\citet[][see their Table 2]{voss07} in 0308$-$565, 0429$-$168,
0845$-$188, 0900+142, 1046$-$017, 1144$-$084, 1336+123, 2129+000, and
2144$-$079.}, we set the hydrogen abundance in our fitting procedure
at the value given by these upper limits at the appropriate
temperature. In cases where the spectrum is noisier, our fitting
procedure may find an upper limit on the hydrogen abundance that is
larger than those of Figure
\ref{fg:f12}.

\subsection{DBZ White Dwarfs}

Four DBZ stars in our sample have sufficiently strong calcium lines
that the inclusion of the Ca~\textsc{ii} H \& K doublet is required to
fit them properly. These are GD 40 (0300$-$013), CBS 78 (0838+375),
KUV 15519+1730 (1551+175), and G241-6 (2222+683). We take a simple
approach here of including only calcium in our equation of state and
opacity calculations. A more detailed analysis of these DBZ stars,
including additional heavy elements, will be presented elsewhere. A
smaller grid of models between $\Te=12,000$~K and 17,000~K has been
calculated for this purpose, with calcium abundances of log Ca/He
$=-7.5$ to $-$6.0 by steps of 0.5, and the same $\logg$ and H/He
values as above. For each star we adopt the calcium abundance that
best reproduces the calcium doublet. Our best fits for these 4 DBZ
stars are presented in Figure \ref{fg:f13}.

\subsection{Convective Efficiency in DB White Dwarfs}

The problem of the convective efficiency in the atmosphere of DA stars has been tackled
by \citet{bergeron95} who used optical spectroscopic observations
combined with UV energy distributions to show that the so-called
ML2/$\alpha=0.6$ parametrization of the mixing-length theory provides
the best internal consistency between optical and UV effective
temperatures, trigonometric parallaxes, $V$ magnitudes, and
gravitational redshifts\footnote{The more recent analysis by
\citet{TB09}, based on improved Stark profiles, suggests a more
efficient value of $\alpha=0.8$.}.  Similarly, \citet[][see their
Figure 5]{beauchamp96} showed a comparison of optical and UV
temperatures of DB white dwarfs using model atmospheres calculated
with the ML1, ML2, and ML3 versions of the mixing-length theory; the
authors only concluded then that ML2 and ML3 provided a more
satisfactory correlation than ML1. The spectroscopic analysis of
\citet{beauchamp99}, however, relied on model atmospheres calculated
with a convective efficiency of ML2/$\alpha=1.25$, although details of
this particular choice were not provided in that study. The
spectroscopic analysis of the DB stars in the SPY survey by
\citet{voss07}, on the other hand, relied on models calculated with
ML2/$\alpha=0.6$, i.e.~the same convective efficiency as for DA
models. As shown in Figure \ref{fg:f2}, the strength of the
helium lines in DB stars are significantly affected by the assumed
parameterization of the mixing-length theory used in the model
atmosphere calculations, in particular between $\Te\sim18,000$~K and
30,000~K. It is thus of utmost importance to calibrate the convective
efficiency in DB stars as accurately as possible.

The $\logg$ versus $\Te$ distribution of all DB stars in our sample is
displayed in Figure \ref{fg:f14} for three versions of the
mixing-length theory, namely ML2/$\alpha=0.75$, 1.25, and 1.75. As
expected, only objects with 18,000~K $\lesssim\Te\lesssim30,000$~K are
affected by this parameterization. As discussed at the end of Section
2.2, if the convective efficiency is underestimated, the helium lines
are predicted too strong near the maximum equivalent widths, and DB
stars will have a tendency to be lumped on either side of this
maximum.  This is precisely what is observed in Figure
\ref{fg:f14} for $\alpha=0.75$ where the maximum occurs near
23,000~K according to Figure \ref{fg:f2} (see also Figure 3
of \citealt{bergeron95} for DA stars near the ZZ Ceti instability
strip).  On the other hand, if the convective efficiency is
overestimated, helium lines are predicted too weak near the maximum,
and DB stars in this case will tend to accumulate at this location.
Again, this is what is observed in Figure \ref{fg:f14} for
$\alpha=1.75$ where DB stars form a clump at $\Te\sim25,000$~K, even
leaving a void near 20,000~K (see a similar result for DA stars in
Bergeron et al.~1995).  The results with $\alpha=1.25$, however,
provide a smoother distribution and a uniform increase of the number
of DB stars as a function of effective temperature, which is exactly
what is expected in terms of the white dwarf luminosity function.
Based on these results alone, a value around $\alpha=1.25$ seems
entirely reasonable, even though we cannot easily rule out values
between $\alpha\sim1.0$ and 1.5 at this stage.

\citet{bergeron95} went a step further to calibrate the
convective efficiency in DA stars, by comparing optical temperatures
determined from spectroscopy with those obtained from UV energy
distributions. Such a comparison is displayed in Figure
\ref{fg:f15} for two versions of the convective efficiency, namely
ML2/$\alpha=1.25$ and 1.75. Also shown in the bottom panel is a
comparison of our UV temperatures, derived from fits to {\it IUE}
spectra with ML2/$\alpha=0.6$ models, with those obtained by
\citet{castan06} based on similar assumptions; note the almost perfect agreement here. In all UV fits, we simply assume
$\logg=8$ and pure helium compositions for all objects.  We first
begin by examining the results for $\alpha=1.25$. Although there is a
good overall agreement between optical and UV temperatures, there are
also several discrepant results at the hot end of the sequence.  The
two objects with $T_{\rm opt}\sim 30,000$~K correspond to PG 0112+104,
recently analyzed in great detail by \citet{dufour10a}, and PG
1654+160. Although the {\it IUE} spectrum of the first object is of
good quality, the spectrum of the second object suffers from a
significant discontinuity between the SWP and LWP images. It is also  possible
that the UV energy distributions may become increasingly less sensitive to
temperature at the hot of the DB sequence, while interstellar extinction
may have to be taken into account for the intrinsically brighter, and
more distant, DB stars. Indeed, for PG 0112+104, \citet{dufour10a} found
the energy distribution extending into the FUV region ($\sim 925$\AA)
globally consistent with the $\sim 31,000$ K effective temperature
derived from the optical spectroscopy provided that a small amount
of reddening were included in the fits ($E(B-V) = 0.015$).  The most
important outlier in Figure \ref{fg:f15} is LP 497-114 (1311+129)
with a $\sim 8000$~K temperature difference; this peculiar white dwarf
is further discussed in Section 5.4.

More worrisome in the top panel of Figure \ref{fg:f15} is the
systematic offset between both temperature scales (with $T_{\rm
UV}>T_{\rm opt}$) for 15,000~K $<T_{\rm UV}< 22,000$~K.  Note that a
more efficient prescription of the mixing-length theory, shown in the
middle panel of Figure \ref{fg:f15}, improves the agreement
between both temperature scales near $T_{\rm UV}\sim22,000$~K, but the
agreement gets worse for hot DB stars, and the systematic shift still
remains for 15,000~K $<T_{\rm UV}< 20,000$~K. Hence, it is not clear
that increasing the convective efficiency would improve the overall agreement.

Since UV temperatures may depend on surface gravity,
hydrogen abundance, and convective efficiency, we explore in Figure
\ref{fg:f16} the effects of variations of these parameters on our
temperature determinations. All in all, we see that UV temperatures
are fairly independent of our initial assumption of $\logg=8$ and pure
helium compositions in our fits\footnote{In light of these results, we
are surprised that \citet{castan06} succeeded in
constraining $\logg$ by using only {\it IUE} data (see their Table 1); note
that their $\logg$ determinations were omitted in the comparison
displayed in Figure 4 of \citet{voss07}.}. The same conclusion applies
to our particular choice of $\alpha$. We thus conclude that the
systematic shifts observed in Figure \ref{fg:f15} with our
ML2/$\alpha=1.25$ models cannot be explained in terms of inadequate
assumptions in our fits to UV data. What the overall results suggest,
instead, is that theoretical improvements probably need to be made at
the level of the model atmospheres, perhaps even with the treatment of
convective energy transport (see below).

\subsection{Error Estimation}

The {\it internal} uncertainties of the atmospheric parameters ---
$\Te$, $\logg$, and H/He --- can be obtained directly from the
covariance matrix of our fitting method. These depend mostly on the
signal-to-noise ratio of the observations and on the sensitivity of
the models to each parameter.  For DA white dwarfs, these internal
errors can become negligibly small with sufficiently high
signal-to-noise spectroscopic observations (S/N $\gtrsim50$;
\citealt{BSL92}, \citealt{LBH05}).  Since our DB spectra are rather
homogeneous in terms of S/N (see Figures \ref{fg:f5} and
\ref{fg:f6}) and of sufficiently high quality (S/N $>$ 50), the
internal errors in our analysis will be mostly dominated by the
sensitivity of the models to each fitted parameter. The theoretical
spectra displayed in Figure \ref{fg:f1} indicate that
the internal errors will be relatively small for $\logg$ (similar
conclusions apply to H/He), but can be relatively large for $\Te$,
particularly in the temperature range where the strength of the helium
lines reaches its maximum.

The true error budget, however, must also take into account {\it
external} uncertainties originating mostly from flux calibration,
which can be estimated from multiple observations of the same star, as
performed by \citet[][Figure 8]{LBH05} for DA stars, or by
\citet[][Figures 1 and 2]{voss07} for DB stars.  Even though we have
not reobserved DB stars for that specific purpose (except in a few
cases), we still managed to secure repeated observations of 28 DB
stars in our sample, either because of flux calibration problems
(e.g., objects observed at high airmass), or insufficient S/N. Hence
the external errors determined below will tend to be overestimated, if
anything.

We compare in Figure \ref{fg:f17} the $\Te$ and $\logg$
measurements for the 28 DB stars in our sample with mutliple
spectra. As expected, the external error on $\Te$ are particularly
large between $\sim 20,000$~K and 25,000~K, where the helium lines
reach their maximum strength. Otherwise, outside this temperature
range, the temperature estimates are in excellent agreement. The
spread in $\logg$ measurements is more significant, however, but some
of the most extreme outliers are white dwarfs with a first
spectroscopic observation at very low S/N; these include the DB stars
reported by \citet{lepine11} for instance. If we take the average
uncertainties of both parameters, we obtain
$\langle\Delta\Te/\Te\rangle=4.6$\% and
$\langle\Delta\logg\rangle=0.069$, but if we remove the obvious
outliers, these numbers drop to $\langle\Delta\Te/\Te\rangle=2.3$\%
and $\langle\Delta\logg\rangle=0.052$. Given that these are probably
overestimated, as discussed in the previous paragraph, we adopt these
values as conservative estimates of the external uncertainties of our
atmospheric parameters. Note that these results are comparable to
those obtained by \citet{voss07}, $\langle\Delta\Te/\Te\rangle=2.03$\%
and $\langle\Delta\logg\rangle=0.058$, based on multiple observations
of 24 objects.

\subsection{Adopted Atmospheric Parameters}

The atmospheric parameters for the 108 DB and DBA stars in our sample
are reported in Table~1; the values in parentheses for $\Te$ and
$\logg$ represent the combined (in quadrature) internal and external
errors of the fitting technique, while only the internal error is
available for H/He. For DB stars without detectable hydrogen, upper
limits on the hydrogen abundance are given in Table 1.  The stellar
mass ($M$) and white dwarf cooling age ($\log\tau$) of each star are
obtained from evolutionary models similar to those described in
\citet{fon01} but with C/O cores, $q({\rm He})\equiv \log M_{\rm
He}/M_{\star}=10^{-2}$ and $q({\rm H})=10^{-10}$, which are
representative of helium-atmosphere white dwarfs. The absolute visual
magnitude ($M_V$) and luminosity ($L$) are determined with the
improved calibration from \citet{holberg06}, defined with the {\it
Hubble Space Telescope} absolute flux scale of Vega. The $V$
magnitudes\footnote{Note that these sometimes represent photographic
magnitudes, Str\"omgren $y$ magnitudes, or even $B$ magnitudes.} are
taken from the Villanova White Dwarf Catalog, which combined with
$M_V$ yield the distance $D$. Finally, for all PG stars in the
complete sample, we also provide the $1/V_{\rm max}$ weighting
(pc$^{-3}$) used in the calculation of the luminosity function, where
$V_{\rm max}$ represents the volume defined by the maximum distance at
which a given object would still appear in the sample given the
magnitude limit of the PG survey (see \citealt{LBH05} for details).

Sample fits for both DB and DBA stars covering the full temperature
range of our sample are displayed in Figure \ref{fg:f18}. The
left panels show the blue portion of our spectroscopic fits, while the
right panels show the corresponding region near H$\alpha$. The
importance of the red coverage is clearly illustrated, with H$\alpha$
being barely detected in several DBA stars, while H$\beta$ remains
spectroscopically invisible. In other objects, only an upper limit on H/He
could be set, based on the absence of H$\alpha$. We also show in this
figure the DBA star with the largest hydrogen abundance measured in
our sample (LP 497-114; 1311+129). A trend for the
coolest objects to have larger than average surface gravities is already apparent.

\section{Selected Results}

\subsection{Comparison with Results from SPY}

As mentioned in the Introduction, the spectroscopic analysis of $\sim
70$ DB and DBA stars in the SPY survey \citep{voss07} represents the
largest set of data against which our atmospheric parameter
determinations can be compared.  We have 44 white dwarfs in common
with the SPY sample, 22 of which are DBA stars. The comparison of effective
temperatures and surface gravities is displayed in Figure
\ref{fg:f19}. 

\citet{voss07} also included van der Waals broadening in their
models, in a simplified form. Even though they found a much better
agreement with the observed line profiles and strengths, their fits
were not completely satisfactory according to the authors; they also
mention that a fit in which $\logg$ is actually allowed to vary
converges to very high values. Hence a value of $\logg=8$ was simply
assumed for the coolest DB stars in their sample (shown as open
circles in Figure \ref{fg:f19}). The comparison of $\logg$
values is thus meaningless for these stars, although we note that the
corresponding temperatures are in excellent agreement, except for the
two coolest white dwarfs for which our values of $\logg\sim9$ are
likely overestimated. Worth mentioning in this context are the
results shown in Figure 5 of \citet{kepler07} for the DB stars
identified in the Data Release 4 of the SDSS, where the masses
gradually increase to very large values ($M>1.0$ \msun) at low
effective temperatures. Whether this increase is due to an
inappropriate treatment of van der Waals broadening, or to the neglect
of this broadening mechanism altogether, is not discussed in their
analysis.

For $\Te\lesssim19,000$~K, the $\Te$ determinations displayed in
Figure \ref{fg:f19} are in good agreement, although our $\logg$
values in this temperature range appear systematically larger than
those of Voss et al.~by 0.15 dex, on average. The $\logg$ values agree
much better at higher temperatures, however, with the exception of PG
1115+158 (labeled in the figure) for which we get a surface gravity
0.4 dex higher than the value derived by Voss et
al.~($\logg=7.52$). This white dwarf is of particular interest because
of its low inferred mass of only 0.385 \msun, and also because it is a
member of the pulsating V777 Her class.
\citet{beauchamp96} found no low-mass DB stars in their sample
and argued that evolutionary scenarios that could produce such
low-mass degenerates could simply not form DB white dwarfs.  Both
spectroscopic solutions for PG 1115+158 are compared in Figures
\ref{fg:f11} and \ref{fg:f20}. While the effective temperatures 
agree well within the uncertainties, the low surface gravity obtained
by Voss et al.~predicts helium lines that are much shallower than
observed. Hence we believe that our solution at $\logg=7.91$ (or
$M=0.56$ \msun) is more appropriate for this star.

The differences in temperature for $\Te>19,000$~K are significantly
more important, partly due to the use of different versions of the
mixing-length theory (ML2/$\alpha=0.6$ versus $\alpha=1.25$), but also
because of the intrinsic difficulty of measuring $\Te$ precisely in
this particular temperature range. The most extreme case here is for
KUV 03493+0131 (0349+015; labeled in Figure \ref{fg:f19}) for
which we obtain an effective temperature $\sim 6000$~K higher than
Voss et al.  Both spectroscopic solutions for this object are
displayed in Figures \ref{fg:f18} and
\ref{fg:f20}. Note that if we rely only on the blue spectrum for this
object, we find a cooler solution at $\Te=22,760$~K, still
significantly hotter than Voss et al. We find that our solution
at $\Te=24,860$~K reproduces the widths and overall strengths of the
neutral helium lines, as well of the observed slope of the spectral
energy distribution (not shown here), much better than a solution at
18,740~K.

\subsection{Mass Distribution}

\citet{beauchamp96} showed that the mass
distribution of DB stars was relatively narrow, with the mass of 73\%
of their 41 DB stars above $\Te=15,000$~K (excluding DBA stars) lying
between 0.5 and 0.6 \msun, for an average mass of $\langle M_{\rm
DB}\rangle=0.567$ \msun.  The 13 DBA stars were found at slightly
larger masses, with an average mass of $\langle M_{\rm
DBA}\rangle=0.642$ \msun, suggesting that more massive white dwarfs
may have a tendency to show hydrogen, perhaps through an increased
hydrogen accretion from the interstellar medium, or through a
decreased dilution of the accreted hydrogen throughout a thinner
helium convection zone \citep{beauchamp96}. One significant
distinction between the mass distributions of DB and DA stars, noted
by Beauchamp et al., was the almost complete absence of the low- and
high-mass tails in the mass distribution of DB stars, suggesting that
the formation of double degenerates and mergers, which are often
invoked to explain respectively these low- and high-mass tails for DA
stars, are not producing DB stars. The picture drawn by \citet{voss07}
using $\sim 50$ DB stars from the SPY survey differs significantly
from the one described above. While the results of \citet{beauchamp96}
indicate that 25\% of all DB white dwarfs are DBA stars, the
high-resolution spectroscopic observations at H$\alpha$ of the SPY
survey revealed a much larger fraction of 55\% of DBA stars in their
sample, as discussed above. The resulting mean mass of each subsample,
$\langle M_{\rm DB}\rangle=0.584$ \msun\ and $\langle M_{\rm
DBA}\rangle=0.607$ \msun, are now in much closer agreement, suggesting
that both populations have a common origin. Furthermore, Voss et
al.~found a significantly larger mass dispersion, with several
low-mass ($M\lesssim 0.5$ \msun) and high-mass white dwarfs in their
sample, in line with the results for DA stars.

The distribution of mass as a function of effective temperature for
all 108 DB and DBA stars in our sample is displayed in Figure
\ref{fg:f21}. Our final sample comprises 47 DBA stars, or 44\% of
the DB white dwarf population. Based on our discussion above, it is
clear that this fraction represents only a lower limit since we are
lacking the red spectral coverage for many of the DB stars in our
sample, the cool ones in particular. In contrast to the mass
distribution of DB stars in the SDSS obtained by \citet[][see their
Figure 5]{kepler07}, our mass distribution does not show this
overwhelming increase in mass at low effective temperature, even
though we do find several stars with masses in excess of 1
\msun. These are (below $\Te=15,000$~K) 0249+346, 1419+351,
1542$-$275, 2058+342, 2147+280, and 2316$-$273, all of which are shown
in the last panel of Figure \ref{fg:f5}. Note that there are
additional white dwarfs in the same range of temperature that have
more normal masses ($M<0.8$ \msun), for instance 0517+771, 1056+345,
1129+373, 1542+244, also shown in this last panel, with the exception
that these have well defined helium absorption features --- the
$\logg$-sensitive He \textsc{i} $\lambda$3819 line in particular ---
as opposed to the very massive ones. Instead of invoking a problem
with the model spectra, our results suggest that we have probably
reached the limit of the spectroscopic technique for these
objects. They are probably cooler, more normal mass DB stars, with
extremely weak absorption features (see further evidence below).

Even if we ignore the most massive objects in Figure
\ref{fg:f21}, we still note a trend for DB stars in the range
$13,000\ {\rm K}\lesssim\Te\lesssim18,000$~K to have larger masses, up
to $\sim 0.9$ \msun\ in some cases, a problem usually attributed to
the treatment of van der Waals broadening
\citep{beauchamp96}. What is more intriguing, however, is that white dwarfs 
with normal masses, $M\sim 0.6$ \msun, are also found {\it in the same
range of temperature as these massive stars}, which implies that the
large dispersion in mass that occurs at these temperatures might be
real after all. This point has already been made by \citet{limoges10}
who show in their Figure 7 the spectroscopic fits for two DB stars with
nearly identical temperatures ($\Te\sim15,000$~K), but with $\logg$
values that differ by 0.33 dex ($\logg=8.03$ and 8.36).  The
comparison reveals that the He~\textsc{i} $\lambda3819$ and
$\lambda4388$ lines, which are the most gravity sensitive in this
range of temperature, differ markedly in both stars, indicating that
the high $\logg$ value measured in one of these stars is probably
real, and not a simple artifact produced by the models.

The problem thus rests on our ability to confirm {\it independently}
the atmospheric parameters obtained here within the current
theoretical framework. To do so we rely on independent distance
estimates obtained from trigonometric parallax measurements taken
mostly from the Yale parallax catalog \citep[][hereafter YPC]{ypc} and
from the new {\it Hipparcos}-based parallaxes from \citet{gould04}. We
found good trigonometric parallaxes for 11 DB white dwarfs in our
sample. The parallax and corresponding distance for each object are
reported in Table 2 together with our atmospheric parameter solution,
including the spectroscopic distance. The comparison of spectroscopic
distances and those obtained from trigonometric parallaxes is
displayed in Figure \ref{fg:f22}. The agreement is surprisingly
good, especially given the fact that white dwarfs with the closest
match are found in the temperature range where the mass dispersion is
large in Figure \ref{fg:f21} ($\Te\sim15,000$~K; see Table 2). We
are thus fairly confident that our surface gravity and mass values
measured spectroscopically are fairly accurate.

We also observe some discrepant distance estimates, labeled in Figure
\ref{fg:f22}. For Feige 4 (0017+136; label 1), we obtain a spectroscopic distance
twice as large as that inferred from the trigonometric parallax. Our
fit to this DB star is excellent, and we find no easy way to reconcile
the two distance estimates. GD 358 (1645+325; label 2) is a pulsating
white dwarf, the only hot DB star in Table 2. Again we do not have an
easy explanation for the discrepancy. G188-27 (2147+280; label 3) has
a spectroscopic distance significantly smaller than that inferred from
the parallax. In this case, however, the spectroscopic $\logg$ value
of 8.85 (or $M=1.12$ \msun) is certainly overestimated, and this
object corresponds to one of the (almost) featureless DB stars
discussed earlier. A value of $\logg=8.2$ would actually reconcile
both distance estimates perfectly. This corresponds to a mass of $\sim
0.7$ \msun, i.e.~the average mass of the bulk of DB white dwarfs near
15,000~K.

The mass distribution of all DB and DBA stars in our sample,
regardless of their temperature, is displayed in Figure
\ref{fg:f23}. If we exclude from this distribution the most massive objects
near $\sim1.2$ \msun, which all correspond to the almost featureless
cool DB stars discussed above, we find a mean mass for our sample of
$\langle M\rangle=0.671$ \msun\ with a standard deviation of only
$\sigma_M=0.085$ \msun. In contrast to our preliminary conclusion
presented in \citet{beauchamp96}, we now find that there is no
significant mass difference between the DB and the DBA samples ---
$\langle M_{\rm DB}\rangle=0.657$ \msun\ and $\langle M_{\rm
DBA}\rangle=0.688$ \msun\ --- in agreement with the conclusions of
\citet{voss07}. This result was already apparent from a quick
examination of Figure \ref{fg:f21}. Note that the hottest white
dwarfs in this figure are all of the DB type, and that their masses
are slightly below 0.6 \msun, thus contributing to an average mass for
the DB white dwarfs that is only slightly lower than the mean for DBA
stars.

We also compare, in Figure \ref{fg:f23}, the mass distribution of DB
stars with that of DA stars from the PG survey, reanalyzed with our
improved models for hydrogen-atmosphere white dwarfs (see
\citealt{TB11} and references therein), and for which we obtain $\langle
M_{\rm DA}\rangle=0.631$ \msun\ and $\sigma_M=0.133$ \msun. While the
peak of the mass distribution of DB stars agrees with that of DA
stars, the former distribution is shifted towards higher masses,
resulting in an average mass $\sim 0.03$ \msun\ higher. The peak value
for the DB stars, between 0.60 and 0.65 \msun, also agrees with the
peak value determined by \citet[][see their Figure 8]{voss07},
although their mean mass of $\langle M\rangle=0.596$ \msun\ is much
smaller than ours. This result is consistent with our $\logg$
determinations being larger than their values by about 0.15 dex in the
temperature range where the bulk of DB white dwarfs is found (see
Figure \ref{fg:f19}).

Also of interest is the absence of low-mass DB stars in our sample, in
contrast with low-mass DA stars, which are present in large numbers in
all surveys, including the PG survey shown in Figure
\ref{fg:f23}. \citet[][see their Figure 8]{voss07} identified three
objects\footnote{Note that we could only find two white dwarfs with
$M<0.5$ \msun\ in their Tables 1 and 2.} in the SPY sample with
$M\lesssim0.5$ \msun, the lowest mass DB white dwarf being PG
1115+158, with a mass of only 0.385 \msun\ (or $\logg=7.52$). This
object has already been discussed in Section 5.1, and we find instead
a significantly larger value of $M=0.56$ \msun\ for the same
star. Similarly, Voss et al.~report a mass of $M=0.481$ \msun\
($\Te=16,904$~K) for HE 0423$-$1434, while we find 0.64
\msun\ ($\Te=16,900$~K). Hence we reaffirm the conclusion
of \citet{beauchamp96} that low-mass DB stars are rare, or absent,
a conclusion also reached by \citet{BLR01}. These authors found, in their
photometric analysis of 152 white dwarfs with trigonometric parallax
measurements, that all low-mass degenerates probably possess
hydrogen-rich atmospheres.  As discussed in that study, since
common envelope evolution is required to produce white dwarfs with
$M\lesssim 0.5$ \msun --- the Galaxy being too young to have produced
them from single star evolution --- we must conclude that this
particular evolutionary channel does not produce helium-rich
atmosphere white dwarfs. This could be the case because the objects which go
through this close-binary phase end up with hydrogen layers too
massive to allow the DA to DB conversion near $\Te\sim30,000$~K, or below.

\subsection{Luminosity Function}

\citet{LBH05} presented spectrophotometric observations 
of a complete sample of 348 DA white dwarfs identified in the Palomar
Green Survey and obtained robust values of $\Te$, $\logg$, masses,
radii, and cooling ages for all stars in their sample using the
spectroscopic technique. They also calculated the luminosity function
of the sample, weighted by $1/V_{\rm max}$, where $V_{\rm max}$
represents the volume defined by the maximum distance at which a given
object would still appear in the sample given the magnitude limit of
the PG survey.  An overall formation rate of white dwarfs in the local
Galactic disk of $1\pm0.25\times10^{-12}\ {\rm pc}^{-3}\ {\rm
yr}^{-1}$ was also reported.

Since we observed all DB stars in the PG survey, we can calculate
their luminosity function in the same fashion as for DA stars using
the $1/V_{\rm max}$ values given in Table 1\footnote{We ignore here
the magnetic DB star PG 0853+163 as well as the DB component of the
double degenerate binary PG 1115+166.}.  Our results are reported in
Table 3 and shown in Figure \ref{fg:f24}. Note that we rely here on
bolometric magnitudes rather than absolute visual magnitudes --- as
\citet{LBH05} did (see their Figure 10) --- since DA and DB stars have
different $M_V$ values at a given effective temperature and surface
gravity. Also shown in Figure \ref{fg:f24} is the contribution from the
DA stars in the PG survey, again reevaluated using our updated grid of
DA model spectra. In the bolometric magnitude range over which DB
stars are detected (roughly $M_{\rm bol}=7-11$), the DB/(DA+DB) ratio
is of the order of 0.2. We note, however, a large and significant
increase in the value of log $\phi$ in the range from $M_{\rm bol}= 9$
to 10, which corresponds to an effective temperature near 20,000
K. Thus, the DB/DA ratio is lower than the nominal value of 1 out of 4
white dwarfs at high effective temperatures, but increases sharply for
stars below 20,000 K. This, it turns out, is also the temperature at
which the bottom of the helium convection zone sinks sharply into the
envelope along the cooling sequence (see Figure
\ref{fg:f3}). It is also the temperature below which DBA stars
start to appear in large numbers in Figure \ref{fg:f21}, and also
where the average mass of all DB stars increases by $\sim 0.1$ \msun.

The total space density of DB stars in the PG survey can be obtained
by summing the values in Table 3 over all magnitudes bins. We obtain a
space density of $5.15\times10^{-5}$ DB star per pc$^{3}$. Hence we
expect in a volume defined within 20 pc from the Sun around two
genuine DB stars, yet none have been identified in our sample (see
also \citealt{holberg08}).

\subsection{Hydrogen Abundance and Mass Fraction in DB White Dwarfs}

About 44\% of all DB stars in our sample show traces of hydrogen, or
50\% if we consider only the objects for which we have spectroscopic
data covering the H$\alpha$ region. \citet{voss07} obtained an even
higher fraction of 55\% based on the SPY data. Hence the DBA
phenomenon is quite common among DB stars. The hydrogen abundances as
a function of effective temperature for all DBA stars in our sample
are displayed in Figure \ref{fg:f25}. Also shown are the {\it
upper limits} on the hydrogen abundance for DB stars, as determined
from the absence of H$\alpha$ or H$\beta$.  In general, these DB stars
are aligned on the observational limits reproduced here from Figure
\ref{fg:f12}, but some objects have noisier data and these limits
are simply not reached. The first striking result is that almost all
white dwarfs above $\Te\sim 20,000$~K are DB stars, with the exception
of 3 objects (KUV 05134+2605, PG 1115+158, and PG 1456+103).

There are two possible origins for hydrogen in these
helium-dominated atmospheres. First, hydrogen could be residual,
originating from the thin hydrogen atmosphere of the DA progenitor,
convectively diluted into the more massive helium envelope during the
DA to DB transition that occurred at higher effective
temperatures. \citet{MV91} showed that the exact temperature at which
this process occurs depends on the thickness of the hydrogen layer
(see their Table 1). For their S1 models, which correspond to the
Schwarzschild criterion with a value of $\alpha=1$, a DA star with
hydrogen layer masses of $M_{\rm H}=10^{-15}$, $10^{-14}$, and
$10^{-13}$ \msun\ would respectively mix at temperatures of
$\Te=27,400$~K, 17,900~K, and 11,700~K.  The main effect of adding
more hydrogen on top of the helium convection zone is to delay the
time it takes for helium to develop a sufficiently deep convection
zone, thus reducing the temperature at which mixing occurs. Upon
mixing, it is then assumed that the thin hydrogen atmosphere will be
thoroughly mixed with the more massive helium convection zone, and
that the resulting photospheric hydrogen abundance will simply
correspond to the ratio of the total hydrogen mass $M_{\rm H}$ to that
of the helium convection zone, $M_{\rm He-conv}$. One problem
immediately arises when trying to account for the hot DB stars above
$\Te\sim20,000$~K in Figure \ref{fg:f25}. Indeed, for convective
mixing to occur at these temperatures, the required hydrogen layer
mass has to be of the order of $10^{-15}$ \msun\ according to
MacDonald \& Vennes. This mass, it turns out, is comparable to the
mass of the helium convection zone in this temperature range (see
Figure \ref{fg:f3}). The complete dilution of hydrogen should
thus lead to a H/He ratio $\sim 1$ instead of the observed limits of
H/He $\sim 10^{-4}-10^{-5}$.

We can explore this problem more quantitatively by converting the
hydrogen abundances from Figure \ref{fg:f25} into total hydrogen
masses by assuming that hydrogen is indeed homogeneously mixed in the
helium convection zone (see also \citealt{dufour07a} and
\citealt{voss07}). For simplicity, we assume evolutionary models at
0.6 \msun\ (as in Figure \ref{fg:f3}) and we also ignore the
feedback effect of the presence of hydrogen on the depth of the mixed
H/He convection zone\footnote{The values of $M_{\rm H}$ inferred here
differ somewhat from the values obtained by \citet[][see Figure
13]{dufour07a} since in their calculations, the depth of the helium
convection zone was taken from Table 1 of \citet{dupuis93}, while we
rely here on the more recent evolutionary models of
\citet{fon01}.}. The total hydrogen mass as a function of effective
temperature for all DBA stars in our sample is displayed in Figure
\ref{fg:f26}. Again, the values for DB stars represent only
upper limits. Above $\Te=22,000$~K, these upper limits imply $M_{\rm
H}\lesssim 10^{-17}$ \msun.  Such small hydrogen layer masses are
simply unable to maintain a hydrogen-rich atmosphere at higher
temperatures. The only way to account for the existence of these hot
DB stars in our sample is thus to conclude that {\it they must have
maintained a helium-dominated atmosphere through their entire life
history}. This interpretation is consistent with the
existence of hot DB stars in the DB gap, i.e.~30,000~K
$\lesssim\Te\lesssim 45,000$~K \citep{eisen06}.  Even the three DBA
stars near 24,000~K have hydrogen layer masses ($M_{\rm H}\sim
10^{-16}$ \msun) too small to have been DA stars in the past according
to the results shown Figure 1 of \citet{MV91}. However, our estimates
are sufficiently approximate that larger masses of the order of
$M_{\rm H}\sim 10^{-15}$ \msun\ can probably be accommodated in these
objects, in which case they would have undergone the DA to DB
transition near $\sim 27,000$~K according to Table 1 of \citet{MV91}.

We note that a similar evolutionary channel is required to explain the
existence of the Hot DQ stars, whose atmospheres are dominated by
carbon, with only traces of helium and no hydrogen
\citep{dufour08}. Since the hottest DQ stars known have temperatures
around 24,000~K, their immediate progenitors must necessarily be DB
stars with no hydrogen. Hence it is tantalizing to suggest that some
of the hot DB stars observed here, as well as the hot DB stars in the
gap identified in the SDSS, are the immediate progenitors of the Hot
DQ stars. If this interpretation is correct, what fraction of DB stars
between $\Te\sim30,000$~K and 24,000~K are progenitors of Hot DQ stars
remains to be determined.

The situation is even more complicated below $\Te\sim20,000$~K where the
inferred hydrogen layer masses in the DBA stars are in the range
$10^{-12}\lesssim M_{\rm H}/M_\odot\lesssim10^{-10}$. According to
\citet{MV91}, DA stars with such large amounts of hydrogen
would only mix at temperatures below 12,000~K. In other words, we
observe much more hydrogen in those stars than would be expected on
the basis of the {\it complete mixing} of the hydrogen layer in the
underlying helium convection zone. Hence, the hydrogen abundances in
these DBA stars are too high to have a residual origin, and external
sources of hydrogen must be invoked, either from the interstellar
medium or from other bodies such as comets, disrupted asteroids, small
planets, etc., a conclusion also reached by \citet[][see also
\citealt{MV91}]{voss07}. It is important to recall that hydrogen can
only accumulate in the photospheric regions, as a result of accretion,
while heavier elements will slowly diffuse at the bottom of the helium
convection zone on time scales of the order of $10^5$ years or less
\citep[see Table 1 of][]{dupuis93}. Since the bottom of the helium
convection zone grows deeper with time (see Figure
\ref{fg:f3}), hydrogen becomes diluted in a larger volume. The
measured hydrogen abundances in cool DB stars, and eventually DC white
dwarfs, are thus governed by this balance between accretion and
dilution within the helium convection zone. It is thus important to
connect the constraints imposed by DBA stars with independent
determinations of the hydrogen abundances in cooler white dwarfs, such
as DZA stars \citep{dufour07a}.

We show in Figure \ref{fg:f27} the same results as in Figure
\ref{fg:f25}, but we added the H/He abundance ratios predicted from
continuous accretion from the interstellar medium at various rates,
from $10^{-21}$ to $10^{-17}$ \msun\ yr$^{-1}$ (see also Figure 10 of
\citealt{voss07} for a similar calculation). Also shown are the
hydrogen abundances for the DZA stars determined by \citet{dufour07a}.
As can be seen, with the exception of the three hottest DBA stars,
this range of accretion rates can easily account for the amount of
hydrogen observed in DBA and DZA stars. However, some of the upper
limits on the hydrogen abundance below $\sim 20,000$~K are quite
stringent, H/He $< 10^{-6}$. These DB stars are observed in the same
range as some DBA stars, characterized by hydrogen abundances of the
order of H/He $\sim 10^{-4}$, or a factor 100 higher. It is thus
difficult to understand why accretion of hydrogen could be effective
for some DBA stars, and not for other white dwarfs at the same
temperature.  The existence of such cool, hydrogen-deficient DB stars
in our sample, and probably a significant fraction of cool DC and DZ
stars as well, can only be interpreted as having evolved from hotter
DB stars that contain negligible amounts of hydrogen in their
atmospheres, without invoking any accretion mechanism whatsoever.

If this interpretation is correct, the presence of hydrogen in DBA
stars must be residual. Perhaps then, it is time to question the
assumption of {\it complete mixing} of the hydrogen layer: in that
case, it remains conceivable that the amount of hydrogen present is
indeed of the order of $\sim 10^{-14}$ \msun, i.e.~the amount expected
for the convective dilution of the hydrogen atmosphere to occur below
$\Te\sim20,000$~K according to \citet{MV91}, but that it somehow
floats on top of the photosphere rather than being forcefully mixed by
the helium convection zone.

One particular object in our sample is LP 497-114 (1311+129), the DBA
star with the largest hydrogen abundance, H/He $\sim10^{-3}$ (at
$\Te\sim19,000$~K in Figure \ref{fg:f25}). We have three
independent spectroscopic observations for this object, one of which
is a SDSS spectrum (which serves also as our H$\alpha$ spectrum). Our
best fit to these spectra are displayed in Figure
\ref{fg:f28}. The two bottom fits are clearly at odds with what 
we can usually achieve in this temperature range, especially given the
high signal-to-noise ratio of these observations. In particular, the
region around He \textsc{i} $\lambda$4471, as well as the core of most
neutral helium lines, are poorly reproduced. If we consider only the blue
spectra, we can achieve much better fits at $\Te\sim23,000$~K, but
then the line cores of H$\alpha$ and He \textsc{i} $\lambda$6678 are
predicted much too shallow. Note that the fit to the SDSS spectrum
(top) is less problematic, although the line cores are also predicted
too shallow. Incidentally, LP 497-114 shows the worst agreement
between optical and UV temperatures ($T_{\rm UV}\sim 27,000$~K in
Figure \ref{fg:f15}). Perhaps the problem with the modeling of this
star is the assumption of a homogeneously mixed hydrogen and helium
atmospheric composition. This star might be in the process of being convectively
mixed, with hydrogen constantly trying to float back to the
surface. This could even explain the small spectroscopic variations
from spectrum to spectrum observed in Figure \ref{fg:f28}.  If
this is the case, the chemically homogeneous models used here no longer
apply. This might also be the case, but to a lesser extent, with other
DBA stars in our sample.

\subsection{The Instability Strip of the Pulsating V777 Her Stars}

Our first assessment of the instability strip of the pulsating DB
(V777 Her) stars was presented in the spectroscopic study of
\citet{beauchamp99}. The picture at that time was complicated by the fact 
that only blue spectroscopic observations were available,
and the insufficient constraints on the hydrogen content is these
stars did not allow us to choose the most appropriate
solution. \citet{hunter01} presented an update of the instability
strip of the V777 Her stars using new spectroscopic observations at
H$\alpha$, with the specific aim of constraining the hydrogen
abundance in these stars. The results, shown in Figure 2 of Hunter et
al., reveal an instability strip that contains at least one
non-variable white dwarf, as well as a significant number of
unknowns. Since this last report, one object in this sample, PG
2246+121, has been identified as a new variable DB white dwarf by
\citet{handler01}.

We show in Figure \ref{fg:f29} the results from our revised
spectroscopic analysis. The V777 Her stars are identified in Table
1. We distinguish in this plot variable and non-variable white dwarfs,
but also DB and DBA stars. Above $\Te\sim 20,000$~K, there are only 3
DBA stars in our sample with hydrogen abundances around H/He $\sim
10^{-3.5}$, i.e.~about one dex larger than the upper limits derived
for DB stars in the same range of temperature, {\it and all three are
pulsating white dwarfs}. Given our results presented so far, DB and
DBA stars in this range of effective temperature seem to form two
distinct classes of objects, with probably different progenitors: hot
DB stars that once resided in the gap in one case, DA stars with thin
hydrogen layers ($M_{\rm H}\sim 10^{-15}$ \msun) in the other
case. Our results displayed in Figure
\ref{fg:f29} are consistent with this idea, in the sense that we can
easily identify two distinct {\it pure} instability strips, within the
uncertainties, one for the pure helium-atmosphere DB stars, and one
for the DBA stars.

Also displayed in Figure \ref{fg:f29} are the theoretical blue
edges for pure helium envelope models reproduced from Figure 3 of
\citet{beauchamp99}, based on the nonadiabatic calculations of
\citet{fontaine97} and \citet{brassard97}, for various versions of the
mixing-length theory. For DB stars, the observed blue edge near
$\Te\sim30,000$~K suggests a convective efficiency somewhere between
the ML2 ($\alpha=1.0$) and ML3 ($\equiv$ ML2/$\alpha=2.0$) versions of
the mixing-length theory, a parameterization entirely consistent with
that adopted in our model atmosphere calculations
(ML2/$\alpha=1.25$). The location of the red edge of the DB instability strip 
is located near $\Te\sim25,000$~K, for a corresponding
strip about 5000~K wide. Two objects, shown as black
open circles, are located at the red edge --- 
KUV 03493+0131 (0349+015) and PG 2354+159 --- but only the latter has
been confirmed as a non-variable, to our knowledge \citep{rw83}. 

The boundaries of the instability strip for the DBA stars are not as
well defined as for the DB white dwarfs since we lack {\it
non-variable} DBA stars both hotter and cooler than the DBA pulsators
studied here, which would allow us to define empirically the edges of
the instability strip (for instance, there is no DBA star in our
sample between $\Te\sim23,000$~K and 20,000~K). Nevertheless, the
existence of DBA pulsators {\it cooler} than the coolest variable DB
stars suggests that the presence of small traces of hydrogen (H/He
$\sim 10^{-3}$) in the partial ionization zone would lower the
temperature at which helium-atmosphere white dwarfs can
pulsate. Interestingly, the effect produced by the presence of small
amounts of hydrogen on the theoretical blue edge of the instability
strip has been presented in Figure 10 of \citet{fontaine08}. Indeed,
the blue edge becomes cooler, but the effect is rather small ($\sim
500$~K), although the authors considered a trace of only H/He =
$10^{-4}$ in their calculations, and it is expected that the effect
would be considerably more important with the higher hydrogen
abundances inferred in our analysis.

What finally comes out of these results is that DB white dwarfs cannot
be considered as a homogeneous class of objects when studying the
location of the instability strip because of this additional
parameter, the hydrogen abundance.

\subsection{The Case of HE 2149$-$0516}

HE 2149$-$0516 is a DAB star discussed briefly by \citet{voss07}, who
describe its spectrum as exhibiting strong Balmer lines and weaker but
strong neutral helium lines. The authors had difficulty fitting this
object with either pure hydrogen or pure helium models, and only a
pure hydrogen fit to H$\alpha$ suggested a temperature near
$\Te\sim30,000$~K, with an extremely low surface gravity of
$\logg\sim7$. Our normalized spectrum for this peculiar white dwarf is
displayed in Figure \ref{fg:f30}. The {\it absolute} energy
distribution (not shown here) is almost flat in the wavelength range
displayed here, so this cannot be a hot white dwarf. Our best solution
with our grid of DB models indeed yields $\Te=11,370$~K, $\logg=7.6$,
and log H/He $=-4.5$, but our spectroscopic fit completely fails to
reproduce both the helium and hydrogen lines adequately. Instead, we
could achieve a much better fit by assuming that HE 2149$-$0516 is an
unresolved DA + DB (or perhaps DBA) composite system. In this case,
each component of the system dilutes the absorption features of the
other component. Given that there are simply too many free parameters
to fit and that the observed features are relatively weak, we kept
both values of $\logg$ fixed in our fitting procedure, as well as the
value of H/He for the DB star, and experimented with various values of
these parameters. Our best fit is displayed in Figure
\ref{fg:f30}.  This is admittedly not a perfect fit, especially
in the blue region of the spectrum, but given that both the DA and DB
components fall in a temperature range which is problematic in terms
of the modeling (i.e., the high-$\logg$ problem), we are rather
satisfied with the quality of our fit.  Note that the combined model
fluxes also reproduce the slope of the observed energy distribution
perfectly (not shown here).

\section{Conclusion}

We presented a comprehensive analysis of 108 DB white dwarfs using
model atmosphere fits to blue and red spectroscopic observations. As
in previous investigations, we showed that high signal-to-noise
spectroscopic data near H$\alpha$ are crucial to determine, or
constrain, the hydrogen abundance in these white dwarfs.  For the
hottest stars in particular, such data allowed us to pinpoint more
accurately the effective temperatures and surface gravities. While we
demonstrated that these parameters depend on the assumed convective
efficiency used in our model atmospheres, we also showed that a
detailed parameterization of the mixing-length theory is more
difficult to achieve than in DA stars, partly because of the lack of
sensitivity of the UV energy distributions to this parameter.

The mean mass of DB white dwarfs appears to be significantly larger
than that of DA stars, especially at the cool end of the DB
sequence. Since this corresponds to a temperature regime where the
physics of line broadening becomes more questionable, van der Waals
broadening in particular, we used trigonometric parallax measurements
to demonstrate that the spectroscopic distances were in excellent
agreement with those obtained from parallaxes, giving us confidence in
our ability to model the DB stars with sufficient accuracy. We found
no significant differences between the mass of DB and DBA stars,
however, suggesting that both populations may have a common
origin. DBA stars, on the other hand, are usually found at much lower
temperatures than DB stars. While the luminosity functions of DA and
DB stars identified in the PG survey indicate that 20\% of all white
dwarfs are DB stars, this appears to be true only below
$\Te\sim17,000$~K (i.e.~$M_{\rm bol}>9.5$ in Figure \ref{fg:f24}). At
higher temperatures, only $\sim9$\% of all white dwarfs are DB
stars. This clearly indicates that the sudden increase in the ratio of
DB to DA white dwarfs is the result of the transformation of DA stars
into DB stars, but at a much lower temperature than the canonical
value defined by the red edge of the DB gap --- or DB deficiency ---
near 30,000~K, a conclusion also reached by \citet{eisen06}.

The global picture that ultimately emerges from our study is that
there are at least two evolutionary channels feeding the DB white
dwarf population. A small fraction of white dwarf stars are born as
hydrogen-deficient stars, with negligible amounts of hydrogen in their
envelope. These will remain hydrogen-deficient throughout their
subsequent evolution, and will never become DA stars. The existence of
extremely hot DB stars in the DB gap certainly supports this
conclusion. We suggest that the hottest DB stars in our sample with
$\Te>20,000$~K descend directly from this evolutionary channel. In the
bin centered on $M_{\rm bol}=7.0$ in Figure \ref{fg:f24}, we actually
find only 3 DB stars, which represent 5.6\% of the entire white dwarf
population in this particular bin. If we apply the same fraction to
the brighter bin centered on $M_{\rm bol}=6.0$ ($\Te\sim35,000$~K),
which contains 24 DA stars, we should only expect a single DB white
dwarf in this particular bin. It is then perhaps not surprising that
none were found in the PG survey. Note that a hot DB star in the gap
was actually identified in the KUV survey, but this star was hidden in
an unresolved DA + DB double degenerate system \citep{limoges09}.

The hydrogen-deficient evolutionary channel is also supported by
observations at low effective temperatures ($\Te\sim15,000$~K) where
we see a significant number of DB stars with extremely low hydrogen
abundances (H/He $< 10^{-6}$).  The mere existence of these almost
pure helium atmospheres rules out the accretion from the interstellar
medium scenario as the most likely source of hydrogen in the DBA stars
found in the same temperature range. The only alternative, which is
the second evolutionary channel, is that hydrogen must have a
primordial origin, resulting from the transformation of a thin
hydrogen-atmosphere DA white dwarf into a DB degenerate.  The observed
hydrogen abundances in DBA stars are too large, however, to be
explained by a simple model where hydrogen is thoroughly mixed within
the helium convection zone. Instead, we proposed a model where
hydrogen tends to float on top of the photosphere rather than being
completely mixed in the helium convective envelope. Since DBA stars
appear in large numbers only below $\Te\sim20,000$~K, this suggests
hydrogen layer masses of $M_{\rm H}\sim10^{-14}$ \msun\ for most DA
progenitors, although the existence of a few hotter DBA stars may
indicate even thinner hydrogen layers in some cases.

In principle, Hot DQ stars must also be affecting the luminosity
function of DB stars, either when DB stars are transformed into Hot DQ
stars near $\Te\sim24,000$~K (i.e., the hottest DQ stars currently
known), most likely through convective mixing, or when Hot DQ stars
apparently return to being DB stars near $\Te\sim18,000$~K (i.e. the
coolest DQ stars currently known), through a process currently
unknown. But the number of known Hot DQ stars in the SDSS sample is so
small, only 14 known to date, that the contribution of these stars to
the luminosity function of DB stars is most certainly negligible.

We may speculate as to the origin of a larger mean mass for DB stars
compared to DA stars. First, this may be an artifact associated with the model
atmospheres. Indeed, the increase in mass becomes particularly
important when the atmospheres of DB stars become strongly convective
(see Figures \ref{fg:f3} and \ref{fg:f21}). Perhaps this
increase in mass is analog to that observed in cool
($\Te\lesssim12,000$~K) DA stars, where the most likely explanation
for this phenomenon has been attributed to the limitations of the
mixing-length theory used in the model atmosphere calculations
\citep{koester09,TB10}. It is possible that such limitations also
apply to DB atmospheres. If the mean mass of DB stars is indeed
significantly larger than that of DA stars, this implies that their
progenitors on the main sequence must have been more massive as well.
According to \citet{WH06}, the mechanism by which hydrogen is
destroyed --- during the post-asymptotic giant branch (AGB) phase --- to the
level required to account for the almost pure helium-atmosphere DB
stars discussed here, or even the DBA stars in our sample, is probably
caused by a very late helium-shell flash, or an AGB final thermal
pulse. It is not inconceivable that such a mechanism is more efficient
in more massive stars on the main sequence.

Our next task will be to study the numerous DB white dwarfs discovered
in the Sloan Digital Sky Survey, even though the average
signal-to-noise ratio of the spectroscopic observations is admittedly
much lower than that of our sample (see Figure
\ref{fg:f6}). Future work should also include a proper treatment
of evolutionary models with mixed hydrogen and helium abundances to
evaluate better the total hydrogen content of DBA stars.

\acknowledgements The work reported here was supported in part by the NSERC
Canada and by the Fund FQRNT (Qu\'ebec). MTR received support from
FONDAP Center for Astrophysics and PFB06 (CATA). PB is a Cottrell
Scholar of the Research Corporation for Science Advancement, while PaD
is a CRAQ postdoctoral fellow. We are grateful to the Steward
Observatory, to the Kitt Peak National Observatory, and to the
Carnegie Observatories for providing observing time for this project.

\clearpage
\clearpage
\begin{deluxetable}{lllrrlccccccc}
\tabletypesize{\scriptsize}
\tablecolumns{13}
\tablewidth{0pt}
\rotate
\tablecaption{Atmospheric Parameters of DB and DBA Stars}
\tablehead{
\colhead{WD} &
\colhead{Name} &
\colhead{$\Te$ (K)} &
\colhead{log $g$} &
\colhead{log H/He} &
\colhead{$M/$\msun} &
\colhead{$M_V$} &
\colhead{log $L/$\lsun} &
\colhead{$V$} &
\colhead{$D ({\rm pc})$} &
\colhead{$1/V_{\rm max}$} &
\colhead{log $\tau$} &
\colhead{Notes}}
\startdata
0000$-$170  &G266-32             &13,880 (362)        &8.63 (0.13)    &$ -$5.64 (0.48)     &0.99 (0.08)    &12.49          &$-$2.67        & 14.69      &   27&               &8.84           &                    \\
0002$+$729  &GD 408              &14,410 (353)        &8.26 (0.10)    &$ -$5.97 (0.82)     &0.75 (0.07)    &11.79          &$-$2.36        & 14.33      &   32&               &8.54           &                    \\
0017$+$136  &Feige 4             &18,130 (440)        &8.09 (0.07)    &$ -$4.61 (0.20)     &0.65 (0.04)    &10.97          &$-$1.85        & 15.37      &   75&7.68($-$7)     &8.12           &                    \\
0031$-$186  &KUV 00312$-$1837    &15,020 (397)        &8.43 (0.12)    &$ -$5.36 (0.34)     &0.86 (0.08)    &11.96          &$-$2.39        & 16.66      &   87&               &8.61           &                    \\
0100$-$068  &G270-124            &19,800 (532)        &8.07 (0.07)    &$ -$5.08 (0.93)     &0.64 (0.04)    &10.78          &$-$1.68        & 13.95      &   43&               &7.96           &                    \\
\\
0112$+$104  &PG 0112+104         &31,040 (1058)       &7.83 (0.06)    &$<-$3.84 (0.84)     &0.53 (0.03)    & 9.89          &$-$0.74        & 15.36      &  124&1.80($-$7)     &7.17           &                    \\
0125$-$236  &G274-39             &16,610 (457)        &8.26 (0.09)    &$ -$5.31 (0.38)     &0.75 (0.06)    &11.44          &$-$2.11        & 15.38      &   61&               &8.36           &                    \\
0129$+$246  &PG 0129+247         &16,440 (463)        &8.27 (0.10)    &$ -$5.26 (0.40)     &0.76 (0.07)    &11.48          &$-$2.13        & 16.09      &   83&1.49($-$6)     &8.38           &                    \\
0211$+$646  &Lan 150             &20,060 (596)        &8.01 (0.07)    &$<-$3.91 (0.18)     &0.60 (0.04)    &10.66          &$-$1.62        & 17.43      &  226&               &7.88           &                    \\
0214$+$699  &Lan 158             &29,130 (1331)       &7.88 (0.07)    &$<-$4.01 (0.97)     &0.55 (0.04)    &10.06          &$-$0.89        & 16.60      &  203&               &7.06           &                    \\
\\
0224$+$683  &Lan 142             &17,830 (474)        &8.19 (0.10)    &$<-$4.88 (0.49)     &0.71 (0.06)    &11.18          &$-$1.95        & 17.78      &  209&               &8.21           &                    \\
0249$+$346  &KUV 02499+3442      &13,320 (454)        &8.96 (0.22)    &$<-$5.04 (0.41)     &1.17 (0.10)    &13.24          &$-$3.00        & 16.40      &   42&               &9.13           &                    \\
0249$-$052  &KUV 02498$-$0515    &17,700 (548)        &8.16 (0.09)    &$ -$5.47 (0.59)     &0.69 (0.05)    &11.13          &$-$1.94        & 16.60      &  123&               &8.20           &                    \\
0258$+$683  &Lan 143             &14,650 (370)        &8.33 (0.11)    &$ -$3.94 (0.04)     &0.80 (0.07)    &11.85          &$-$2.37        & 16.80      &   97&               &8.57           &                    \\
0300$-$013  &GD 40               &14,780 (362)        &8.13 (0.09)    &$ -$6.11 (1.00)     &0.67 (0.06)    &11.52          &$-$2.23        & 15.56      &   64&               &8.43           &1                   \\
\\
0308$-$565  &L175-34             &23,000 (2336)       &8.04 (0.07)    &$<-$4.80 (3.25)     &0.63 (0.04)    &10.55          &$-$1.40        & 14.07      &   50&               &7.64           &2                   \\
0336$+$625  &Lan 174             &21,280 (880)        &8.12 (0.07)    &$<-$4.05 (0.38)     &0.67 (0.04)    &10.74          &$-$1.58        & 17.15      &  191&               &7.86           &                    \\
0349$+$015  &KUV 03493+0131      &24,860 (1939)       &7.95 (0.07)    &$<-$4.59 (1.79)     &0.58 (0.04)    &10.38          &$-$1.21        & 17.20      &  231&               &7.39           &                    \\
0414$-$045  &HE 0414$-$0434      &13,470 (335)        &8.14 (0.11)    &$ -$5.60 (0.31)     &0.67 (0.07)    &11.76          &$-$2.40        & 15.70      &   61&               &8.55           &                    \\
0418$-$539  &BPM 17731           &19,050 (464)        &8.10 (0.06)    &$<-$4.58 (0.20)     &0.66 (0.04)    &10.90          &$-$1.77        & 15.32      &   76&               &8.05           &                    \\
\\
0423$-$145  &HE 0423$-$1434      &16,900 (402)        &8.08 (0.09)    &$<-$5.98 (1.22)     &0.64 (0.05)    &11.11          &$-$1.97        & 16.21      &  104&               &8.21           &                    \\
0429$-$168  &HE 0429$-$1651      &15,540 (416)        &7.99 (0.16)    &$<-$6.35 (3.11)     &0.59 (0.09)    &11.19          &$-$2.06        & 15.82      &   84&               &8.27           &2                   \\
0435$+$410  &GD 61               &16,810 (410)        &8.19 (0.09)    &$ -$4.20 (0.06)     &0.70 (0.06)    &11.28          &$-$2.04        & 14.86      &   52&               &8.29           &                    \\
0437$+$138  &LP 475-242          &15,120 (362)        &8.25 (0.09)    &$ -$4.68 (0.06)     &0.74 (0.06)    &11.64          &$-$2.27        & 14.92      &   45&               &8.47           &                    \\
0503$+$147  &KUV 05034+1445      &15,610 (382)        &8.08 (0.08)    &$ -$5.38 (0.22)     &0.64 (0.05)    &11.31          &$-$2.11        & 13.80      &   31&               &8.33           &                    \\
\\
0513$+$260  &KUV 05134+2605      &24,680 (1322)       &8.21 (0.06)    &$ -$3.78 (0.34)     &0.73 (0.04)    &10.76          &$-$1.39        & 16.70      &  154&               &7.66           &3                   \\
0517$+$771  &GD 435              &13,150 (338)        &8.13 (0.13)    &$ -$6.00 (0.80)     &0.67 (0.08)    &11.80          &$-$2.44        & 16.01      &   69&               &8.57           &                    \\
0615$-$591  &L182-61             &15,750 (374)        &8.04 (0.07)    &$<-$6.32 (1.09)     &0.61 (0.04)    &11.22          &$-$2.07        & 14.09      &   37&               &8.29           &                    \\
0716$+$404  &GD 85               &17,150 (429)        &8.08 (0.07)    &$<-$6.07 (1.38)     &0.64 (0.04)    &11.08          &$-$1.94        & 14.94      &   59&               &8.19           &                    \\
0825$+$367  &CBS 73              &15,960 (438)        &8.12 (0.11)    &$<-$5.30 (0.38)     &0.67 (0.07)    &11.32          &$-$2.09        & 17.00      &  136&               &8.32           &                    \\
0835$+$340  &CSO 197             &22,290 (1382)       &8.25 (0.07)    &$<-$4.67 (2.02)     &0.75 (0.05)    &10.89          &$-$1.59        & 16.00      &  105&               &7.91           &                    \\
0838$+$375  &CBS 78              &14,280 (426)        &8.61 (0.17)    &$<-$6.49 (2.01)     &0.97 (0.11)    &12.39          &$-$2.60        & 17.71      &  115&               &8.79           &1                   \\
0840$+$262  &TON 10              &17,770 (421)        &8.30 (0.07)    &$ -$3.98 (0.05)     &0.78 (0.04)    &11.32          &$-$2.01        & 14.78      &   49&1.21($-$6)     &8.30           &                    \\
0840$+$364  &CBS 82              &21,260 (864)        &8.15 (0.07)    &$<-$5.05 (2.60)     &0.69 (0.04)    &10.80          &$-$1.61        & 17.03      &  176&               &7.90           &                    \\
0845$-$188  &L748-70             &17,470 (420)        &8.15 (0.08)    &$<-$6.00 (1.55)     &0.68 (0.05)    &11.15          &$-$1.95        & 15.55      &   75&               &8.21           &2                   \\
\\
0900$+$142  &PG 0900+142         &14,860 (352)        &8.07 (0.10)    &$<-$6.44 (1.30)     &0.63 (0.06)    &11.42          &$-$2.19        & 16.48      &  102&1.40($-$6)     &8.39           &2                   \\
0902$+$293  &CBS 3               &18,590 (588)        &8.02 (0.08)    &$<-$5.47 (1.68)     &0.61 (0.05)    &10.82          &$-$1.76        & 17.00      &  172&               &8.02           &                    \\
0906$+$341  &CBS 94              &17,310 (535)        &8.06 (0.11)    &$<-$5.02 (0.52)     &0.63 (0.07)    &11.04          &$-$1.92        & 17.00      &  155&               &8.17           &                    \\
0921$+$091  &PG 0921+092         &19,430 (522)        &8.01 (0.07)    &$ -$4.63 (0.37)     &0.60 (0.04)    &10.73          &$-$1.68        & 16.19      &  123&5.60($-$7)     &7.94           &                    \\
0948$+$013  &PG 0948+013         &16,810 (432)        &8.09 (0.07)    &$ -$5.38 (0.29)     &0.65 (0.04)    &11.15          &$-$1.99        & 15.59      &   77&9.75($-$7)     &8.23           &                    \\
\\
0954$+$342  &CBS 114             &26,050 (1823)       &7.98 (0.08)    &$<-$3.99 (0.45)     &0.60 (0.04)    &10.37          &$-$1.15        & 17.20      &  232&               &7.31           &3                   \\
1006$+$413  &KUV 10064+4120      &15,100 (537)        &8.82 (0.20)    &$<-$5.44 (0.92)     &1.10 (0.11)    &12.68          &$-$2.67        & 17.83      &  107&               &8.93           &                    \\
1009$+$416  &KUV 10098+4138      &16,480 (424)        &8.65 (0.08)    &$<-$5.20 (0.30)     &1.00 (0.05)    &12.13          &$-$2.39        & 16.33      &   69&               &8.65           &                    \\
1011$+$570  &GD 303              &17,350 (424)        &8.13 (0.07)    &$<-$5.01 (0.21)     &0.67 (0.04)    &11.13          &$-$1.95        & 14.57      &   48&               &8.21           &                    \\
1026$-$056  &PG 1026$-$057       &17,650 (427)        &8.08 (0.06)    &$<-$4.93 (0.22)     &0.64 (0.04)    &11.02          &$-$1.89        & 16.94      &  153&8.18($-$7)     &8.15           &                    \\
\\
1038$+$290  &Ton 40              &16,630 (392)        &8.10 (0.09)    &$ -$5.89 (0.88)     &0.65 (0.05)    &11.19          &$-$2.01        & 16.94      &  141&1.02($-$6)     &8.25           &                    \\
1046$-$017  &GD 124              &14,620 (354)        &8.14 (0.13)    &$<-$6.46 (1.63)     &0.68 (0.08)    &11.57          &$-$2.26        & 15.81      &   70&1.71($-$6)     &8.45           &2                   \\
1056$+$345  &G119-47             &12,440 (337)        &8.23 (0.15)    &$ -$5.30 (0.22)     &0.73 (0.10)    &12.09          &$-$2.60        & 15.58      &   49&3.62($-$6)     &8.70           &                    \\
1107$+$265  &GD 128              &15,060 (361)        &8.09 (0.08)    &$ -$5.36 (0.18)     &0.65 (0.05)    &11.42          &$-$2.18        & 15.89      &   78&1.39($-$6)     &8.38           &                    \\
1115$+$158  &PG 1115+158         &23,770 (1624)       &7.91 (0.07)    &$ -$3.84 (0.41)     &0.56 (0.04)    &10.35          &$-$1.26        & 16.12      &  142&3.39($-$7)     &7.44           &3                   \\
\\
1129$+$373  &PG 1129+373         &13,030 (358)        &8.16 (0.16)    &$ -$6.06 (1.17)     &0.68 (0.10)    &11.86          &$-$2.47        & 16.23      &   74&2.63($-$6)     &8.60           &                    \\
1144$-$084  &PG 1144$-$085       &15,730 (379)        &8.06 (0.08)    &$<-$6.32 (1.37)     &0.63 (0.04)    &11.27          &$-$2.08        & 15.95      &   86&1.15($-$6)     &8.31           &2                   \\
1148$+$408  &KUV 11489+4052      &17,130 (443)        &8.30 (0.11)    &$<-$4.66 (0.24)     &0.78 (0.07)    &11.43          &$-$2.08        & 17.33      &  151&               &8.35           &                    \\
1149$-$133  &PG 1149$-$133       &20,370 (578)        &8.30 (0.06)    &$ -$3.82 (0.14)     &0.78 (0.04)    &11.07          &$-$1.77        & 16.29      &  110&8.59($-$7)     &8.11           &                    \\
1252$-$289  &EC 12522-2855       &21,880 (757)        &8.03 (0.06)    &$<-$4.85 (1.24)     &0.62 (0.03)    &10.59          &$-$1.49        & 15.85      &  112&               &7.74           &                    \\
\\
1311$+$129  &LP 497-114          &19,100 (479)        &7.96 (0.07)    &$ -$2.90 (0.06)     &0.58 (0.04)    &10.64          &$-$1.68        & 16.26      &  132&5.09($-$7)     &7.92           &                    \\
1326$-$037  &PG 1326$-$037       &19,890 (531)        &8.03 (0.06)    &$<-$4.66 (0.39)     &0.62 (0.04)    &10.72          &$-$1.65        & 15.60      &   94&5.51($-$7)     &7.92           &                    \\
1332$+$162  &PB 3990             &16,790 (424)        &8.17 (0.08)    &$ -$5.09 (0.26)     &0.70 (0.05)    &11.28          &$-$2.04        & 15.98      &   87&1.15($-$6)     &8.28           &                    \\
1333$+$487  &GD 325              &15,320 (377)        &8.03 (0.09)    &$<-$5.40 (0.26)     &0.61 (0.05)    &11.28          &$-$2.11        & 14.02      &   35&1.17($-$6)     &8.32           &                    \\
1336$+$123  &LP 498-26           &15,950 (407)        &8.01 (0.09)    &$<-$6.29 (1.90)     &0.60 (0.05)    &11.16          &$-$2.03        & 14.72      &   51&9.92($-$7)     &8.25           &2                   \\
1351$+$489  &PG 1351+489         &26,010 (1536)       &7.91 (0.07)    &$<-$4.37 (0.82)     &0.56 (0.04)    &10.27          &$-$1.11        & 16.38      &  166&               &7.27           &3, 4                \\
1352$+$004  &PG 1352+004         &13,980 (341)        &8.05 (0.10)    &$ -$5.30 (0.16)     &0.62 (0.06)    &11.53          &$-$2.28        & 15.72      &   69&1.64($-$6)     &8.45           &                    \\
1403$-$010  &G64-43              &15,420 (375)        &8.10 (0.08)    &$ -$6.10 (0.94)     &0.65 (0.05)    &11.37          &$-$2.14        & 15.90      &   80&1.30($-$6)     &8.35           &                    \\
1411$+$218  &PG 1411+219         &14,910 (361)        &8.04 (0.09)    &$<-$5.46 (0.25)     &0.62 (0.05)    &11.36          &$-$2.17        & 14.30      &   38&1.30($-$6)     &8.36           &                    \\
1415$+$234  &PG 1415+234         &17,380 (469)        &8.19 (0.08)    &$ -$5.06 (0.34)     &0.71 (0.05)    &11.23          &$-$1.99        & 16.80      &  130&1.07($-$6)     &8.25           &                    \\
\\
1416$+$229  &KUV 14161+2255      &17,410 (437)        &8.22 (0.10)    &$<-$4.67 (0.23)     &0.73 (0.06)    &11.26          &$-$2.00        & 16.60      &  116&               &8.27           &                    \\
1419$+$351  &GD 335              &12,830 (585)        &8.88 (0.36)    &$<-$5.58 (1.46)     &1.13 (0.18)    &13.15          &$-$3.00        & 16.89      &   55&               &9.14           &                    \\
1421$-$011  &PG 1421$-$011       &16,910 (413)        &8.20 (0.09)    &$ -$4.26 (0.07)     &0.71 (0.06)    &11.28          &$-$2.04        & 15.97      &   86&1.15($-$6)     &8.29           &                    \\
1425$+$540  &G200-39             &14,490 (345)        &7.95 (0.08)    &$ -$4.20 (0.03)     &0.56 (0.05)    &11.29          &$-$2.16        & 15.04      &   56&1.20($-$6)     &8.33           &                    \\
1444$-$096  &PG 1444$-$096       &17,040 (412)        &8.26 (0.11)    &$ -$5.82 (1.21)     &0.75 (0.07)    &11.38          &$-$2.06        & 14.98      &   52&1.30($-$6)     &8.32           &                    \\
\\
1445$+$152  &PG 1445+153         &20,960 (840)        &8.05 (0.08)    &$<-$5.12 (2.62)     &0.63 (0.04)    &10.66          &$-$1.57        & 15.55      &   95&5.08($-$7)     &7.84           &                    \\
1454$-$630.1&L151-81A            &14,050 (336)        &7.96 (0.09)    &$ -$4.79 (0.06)     &0.57 (0.05)    &11.39          &$-$2.23        & 16.60      &  110&               &8.38           &                    \\
1456$+$103  &PG 1456+103         &24,080 (1199)       &7.91 (0.08)    &$ -$3.24 (0.14)     &0.56 (0.04)    &10.32          &$-$1.24        & 15.89      &  129&3.27($-$7)     &7.41           &3                   \\
1459$+$821  &G256-18             &15,850 (395)        &8.09 (0.08)    &$<-$5.32 (0.24)     &0.65 (0.05)    &11.29          &$-$2.09        & 14.78      &   49&               &8.31           &                    \\
1540$+$680  &PG 1540+681         &22,140 (1246)       &7.96 (0.07)    &$<-$4.25 (0.62)     &0.58 (0.04)    &10.47          &$-$1.42        & 16.19      &  139&3.99($-$7)     &7.64           &                    \\
\\
1542$+$182  &GD 190              &22,630 (984)        &8.04 (0.06)    &$<-$4.84 (1.41)     &0.63 (0.03)    &10.57          &$-$1.43        & 14.72      &   67&4.48($-$7)     &7.67           &                    \\
1542$-$275  &LP 916-27           &12,700 (385)        &9.13 (0.15)    &$ -$4.95 (0.58)     &1.24 (0.06)    &13.72          &$-$3.23        & 15.49      &   22&               &9.19           &                    \\
1545$+$244  &Ton 249             &12,840 (332)        &8.18 (0.13)    &$ -$5.02 (0.11)     &0.70 (0.08)    &11.93          &$-$2.51        & 15.78      &   58&2.88($-$6)     &8.63           &                    \\
1551$+$175  &KUV 15519+1730      &15,550 (378)        &7.96 (0.11)    &$ -$4.38 (0.07)     &0.57 (0.06)    &11.13          &$-$2.04        & 17.50      &  188&               &8.24           &5                   \\
1557$+$192  &KUV 15571+1913      &19,570 (558)        &8.15 (0.07)    &$ -$4.37 (0.30)     &0.68 (0.04)    &10.92          &$-$1.75        & 15.40      &   78&               &8.04           &                    \\
\\
1610$+$239  &PG 1610+239         &13,360 (334)        &8.16 (0.12)    &$<-$5.58 (0.30)     &0.69 (0.07)    &11.81          &$-$2.43        & 15.34      &   50&2.42($-$6)     &8.57           &                    \\
1612$-$111  &GD 198              &23,420 (1781)       &7.96 (0.06)    &$<-$4.75 (2.11)     &0.59 (0.04)    &10.44          &$-$1.32        & 15.53      &  104&               &7.53           &                    \\
1644$+$198  &PG 1644+199         &15,190 (383)        &8.12 (0.10)    &$<-$5.42 (0.31)     &0.66 (0.06)    &11.43          &$-$2.18        & 15.20      &   56&1.41($-$6)     &8.38           &                    \\
1645$+$325  &GD 358              &24,940 (1115)       &7.92 (0.06)    &$<-$4.58 (0.88)     &0.57 (0.03)    &10.33          &$-$1.19        & 13.65      &   46&3.31($-$7)     &7.36           &3                   \\
1654$+$160  &PG 1654+160         &29,410 (1613)       &7.97 (0.08)    &$<-$3.98 (1.19)     &0.60 (0.04)    &10.18          &$-$0.92        & 16.55      &  187&2.62($-$7)     &7.07           &3                   \\
\\
1703$+$319  &PG 1703+319         &14,430 (362)        &8.45 (0.11)    &$ -$5.51 (0.34)     &0.88 (0.07)    &12.09          &$-$2.48        & 16.25      &   67&3.38($-$6)     &8.67           &                    \\
1708$-$871  &L7-44               &23,980 (1686)       &8.05 (0.06)    &$<-$4.69 (1.93)     &0.64 (0.03)    &10.54          &$-$1.33        & 14.38      &   58&               &7.55           &                    \\
1709$+$230  &GD 205              &19,610 (507)        &8.09 (0.06)    &$ -$4.00 (0.13)     &0.65 (0.04)    &10.81          &$-$1.71        & 14.90      &   65&               &7.99           &                    \\
1726$-$578  &L204-118            &14,320 (341)        &8.20 (0.08)    &$ -$5.46 (0.19)     &0.71 (0.05)    &11.70          &$-$2.33        & 15.27      &   51&               &8.51           &                    \\
1822$+$410  &GD 378              &16,230 (385)        &8.01 (0.08)    &$ -$4.45 (0.06)     &0.60 (0.05)    &11.09          &$-$2.00        & 14.39      &   45&               &8.22           &                    \\
1940$+$374  &L1573-31            &16,630 (431)        &8.07 (0.08)    &$<-$5.17 (0.28)     &0.64 (0.05)    &11.14          &$-$1.99        & 14.51      &   47&               &8.23           &                    \\
2034$-$532  &L279-25             &17,160 (406)        &8.48 (0.07)    &$<-$5.74 (0.57)     &0.90 (0.05)    &11.73          &$-$2.19        & 14.46      &   35&               &8.49           &                    \\
2058$+$342  &GD 392A             &12,220 (423)        &9.09 (0.20)    &$<-$5.54 (2.56)     &1.23 (0.08)    &13.72          &$-$3.26        & 15.68      &   24&               &9.22           &                    \\
2129$+$000  &G26-10              &14,380 (351)        &8.26 (0.14)    &$<-$6.48 (1.66)     &0.75 (0.09)    &11.79          &$-$2.36        & 15.27      &   49&               &8.55           &2                   \\
2130$-$047  &GD 233              &18,110 (427)        &8.11 (0.08)    &$ -$5.76 (1.32)     &0.66 (0.05)    &11.01          &$-$1.86        & 14.52      &   50&               &8.13           &                    \\
\\
2144$-$079  &G26-31              &16,340 (410)        &8.18 (0.07)    &$<-$6.22 (1.45)     &0.70 (0.04)    &11.36          &$-$2.09        & 14.82      &   49&               &8.33           &2                   \\
2147$+$280  &G188-27             &12,940 (400)        &8.85 (0.20)    &$<-$5.58 (0.72)     &1.12 (0.10)    &13.09          &$-$2.96        & 14.68      &   20&               &9.12           &                    \\
2222$+$683  &G241-6              &15,230 (380)        &8.20 (0.11)    &$<-$5.42 (0.43)     &0.71 (0.07)    &11.56          &$-$2.23        & 15.65      &   65&               &8.43           &1                   \\
2229$+$139  &PG 2229+139         &14,940 (357)        &8.18 (0.09)    &$ -$4.73 (0.07)     &0.70 (0.05)    &11.57          &$-$2.25        & 15.99      &   76&1.70($-$6)     &8.44           &                    \\
2234$+$064  &PG 2234+064         &23,770 (1770)       &8.07 (0.06)    &$<-$4.72 (2.10)     &0.65 (0.04)    &10.58          &$-$1.36        & 16.03      &  123&4.51($-$7)     &7.59           &                    \\
\\
2236$+$541  &KPD 2236+5410       &15,470 (386)        &8.30 (0.09)    &$<-$5.38 (0.27)     &0.78 (0.06)    &11.68          &$-$2.26        & 16.19      &   79&               &8.48           &                    \\
2246$+$120  &PG 2246+121         &27,070 (1521)       &7.92 (0.07)    &$<-$4.27 (0.67)     &0.57 (0.04)    &10.23          &$-$1.04        & 16.73      &  199&2.86($-$7)     &7.20           &3                   \\
2250$+$746  &GD 554              &16,370 (386)        &8.16 (0.06)    &$<-$5.22 (0.12)     &0.69 (0.04)    &11.32          &$-$2.07        & 16.69      &  118&               &8.31           &                    \\
2253$-$062  &GD 243              &17,190 (437)        &8.07 (0.10)    &$ -$4.33 (0.12)     &0.64 (0.06)    &11.05          &$-$1.93        & 15.06      &   63&               &8.18           &                    \\
2310$+$175  &KUV 23103+1736      &15,170 (373)        &8.37 (0.09)    &$<-$5.43 (0.26)     &0.82 (0.06)    &11.84          &$-$2.34        & 15.88      &   64&2.40($-$6)     &8.56           &                    \\
\\
2316$-$173  &G273-13             &12,610 (424)        &9.11 (0.19)    &$ -$4.85 (0.56)     &1.23 (0.07)    &13.70          &$-$3.23        & 14.08      &   11&               &9.19           &                    \\
2328$+$510  &GD 406              &14,460 (362)        &8.06 (0.12)    &$<-$5.51 (0.35)     &0.63 (0.07)    &11.47          &$-$2.23        & 15.09      &   53&               &8.42           &                    \\
2354$+$159  &PG 2354+159         &24,830 (1671)       &8.15 (0.06)    &$<-$4.59 (1.78)     &0.70 (0.04)    &10.67          &$-$1.34        & 15.78      &  105&5.00($-$7)     &7.57           &                    \\
\enddata
\tablecomments{
(1) Solution with log Ca/He $=-7.0$; 
(2) Limits on the hydrogen abundance based on the absence of H$\alpha$ from Table 1 of \citealt{voss07}; 
(3) Variable white dwarf of the V777 Her class; 
(4) PG star not in the complete sample; 
(5) Solution with log Ca/He $=-6.5$. }
\end{deluxetable}

\clearpage
\clearpage
\begin{deluxetable}{llcccccccc}
\tabletypesize{\scriptsize}
\tablecolumns{10}
\tablewidth{0pt}
\tablecaption{Parameters of DB Stars with Trigonometric Parallaxes}
\tablehead{
\colhead{} &
\colhead{} &
\colhead{$\pi$} &
\colhead{$\sigma_{\pi}$} &
\colhead{$D_{\pi}$} &
\colhead{$D_{\rm spec}$} &
\colhead{$\Te$} &
\colhead{} &
\colhead{} &
\colhead{}\\
\colhead{WD} &
\colhead{Name} &
\colhead{(mas)} &
\colhead{(mas)} &
\colhead{(pc)} &
\colhead{(pc)} &
\colhead{(K)} &
\colhead{log $g$} &
\colhead{$M/$\msun} &
\colhead{Notes}}
\startdata
0002$+$729  &GD 408              &28.80&(4.7)&34.7&32.2&14,410    &8.26&0.75&1   \\
0017$+$136  &Feige 4             &30.00&(9.4)&33.3&75.9&18,130    &8.09&0.65&1   \\
0615$-$591  &L182-61             &27.50&(1.0)&36.4&37.5&15,750    &8.04&0.61&2   \\
1333$+$487  &GD 325              &28.60&(3.2)&35.0&35.3&15,320    &8.03&0.61&1   \\
1425$+$540  &G200-39             &17.30&(3.9)&57.8&56.2&14,490    &7.95&0.56&1   \\
1645$+$325  &GD 358              &27.30&(3.3)&36.6&46.0&24,940    &7.92&0.57&1   \\
1940$+$374  &L1573-31            &20.30&(2.8)&49.3&47.2&16,630    &8.07&0.64&1   \\
2129$+$000  &G26-10              &20.26&(2.0)&49.4&49.6&14,380    &8.26&0.75&2   \\
2144$-$079  &G26-31              &14.40&(5.8)&69.4&49.2&16,340    &8.18&0.70&1   \\
2147$+$280  &G188-27             &28.30&(3.0)&35.3&20.8&12,940    &8.85&1.12&1   \\
2222$+$683  &G241-6              &14.70&(4.4)&68.0&65.8&15,230    &8.20&0.71&3   \\
\enddata
\tablecomments{
(1) YPC, (2) \citealt{gould04}, (3) \citealt{dahn88}. }
\end{deluxetable}

\clearpage
\clearpage
\begin{deluxetable}{cccc}
\tablecolumns{4}
\tablewidth{0pt}
\tablecaption{White Dwarf Space Density$^{\rm a}$ in the PG Survey}
\tablehead{
\colhead{$M_{\rm bol}$} &
\colhead{DB} &
\colhead{DA} &
\colhead{Total}}
\startdata
 6  &  ---     & $2.43\times10^{-6}$ & $2.43\times10^{-6}$ \\
 7  & $7.28\times10^{-7}$ & $1.23\times10^{-5}$ & $1.30\times10^{-5}$ \\
 8  & $3.10\times10^{-6}$ & $2.48\times10^{-5}$ & $2.79\times10^{-5}$ \\
 9  & $4.57\times10^{-6}$ & $5.15\times10^{-5}$ & $5.61\times10^{-5}$ \\
 10 & $2.57\times10^{-5}$ & $6.00\times10^{-5}$ & $8.58\times10^{-5}$ \\
 11 & $1.73\times10^{-5}$ & $6.11\times10^{-5}$ & $7.85\times10^{-5}$ \\
\enddata
\tablenotetext{a}{All units in pc$^{-3}$ mag$^{-1}$.}
\end{deluxetable}

\clearpage 

\figcaption[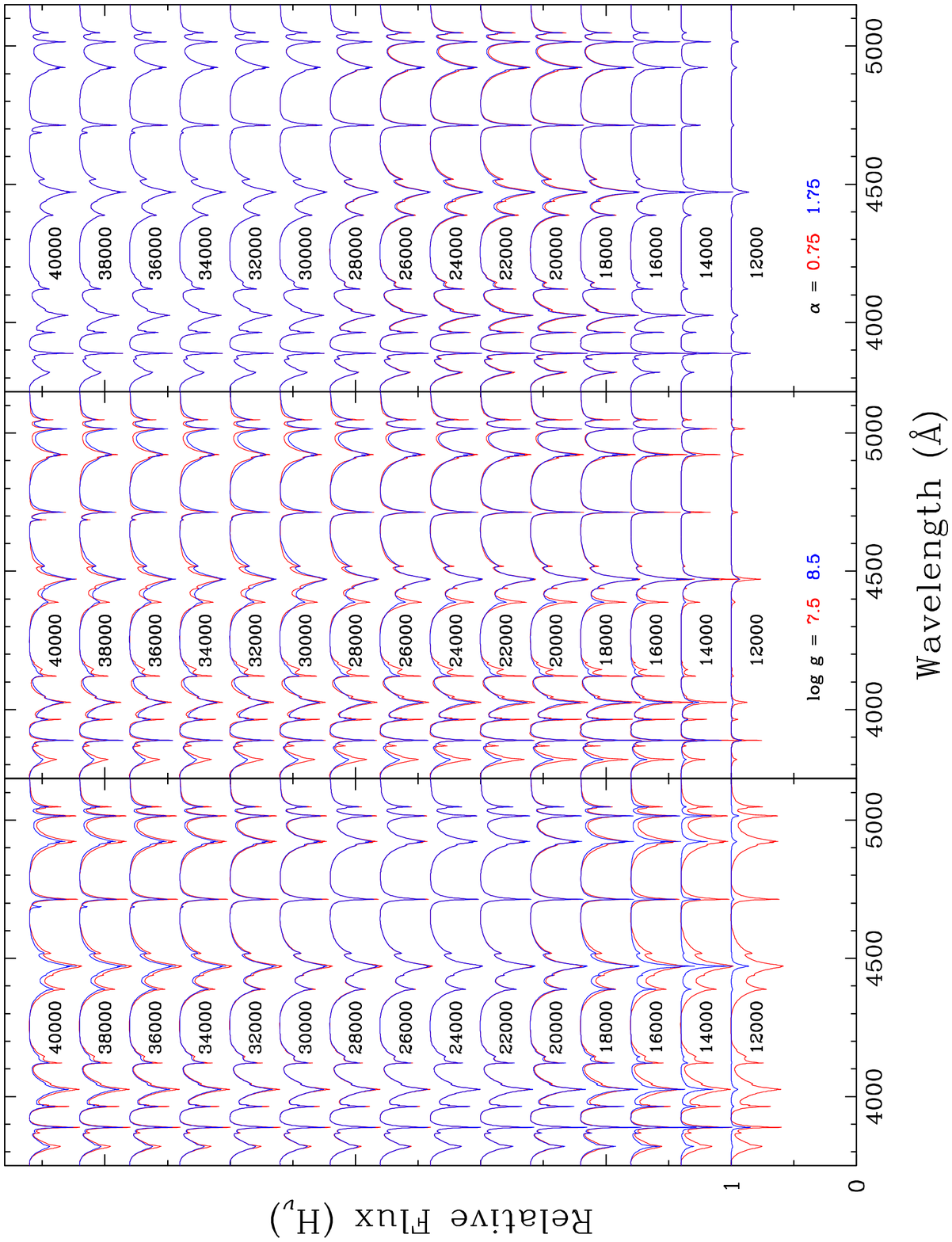] 
{Left panel: synthetic spectra of pure helium DB models ($\logg=8$ and
ML2/$\alpha=1.25$) at various effective temperatures (blue)
normalized to a continuum set to unity and offset from each other by a
factor of 0.4 for clarity; each model spectrum is compared to a
spectrum at $\Te=24,000$~K (red). Middle panel: same as left
panel but for synthetic spectra at $\logg=7.5$ (red) and
$\logg=8.5$ (blue). Right panel: same as left panels but for
synthetic spectra calculated with ML2/$\alpha=0.75$ (red) and
$\alpha=1.75$ (blue).\label{fg:f1}}

\figcaption[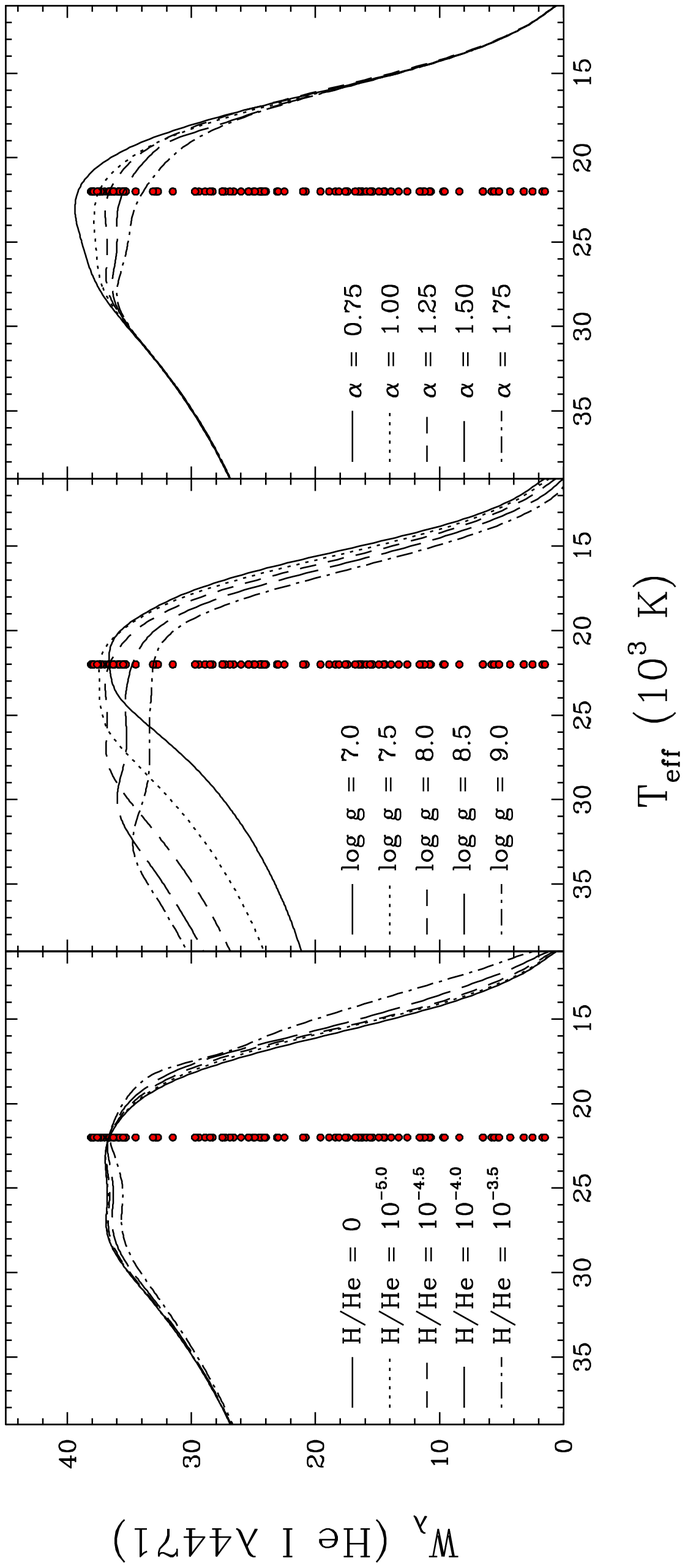] 
{Variation of the equivalent width of He~\textsc{i} $\lambda$4471
(measured from 4220 to 4625 \AA) as a function of effective
temperature for our grid of model spectra with various hydrogen
abundances (left panel), surface gravities (middle panel), and
convective efficiencies (right panel). Also shown as red circles are
the measured equivalent widths in our sample of DB stars, arbitrarily
located at $\Te=22,000$~K.\label{fg:f2}}

\figcaption[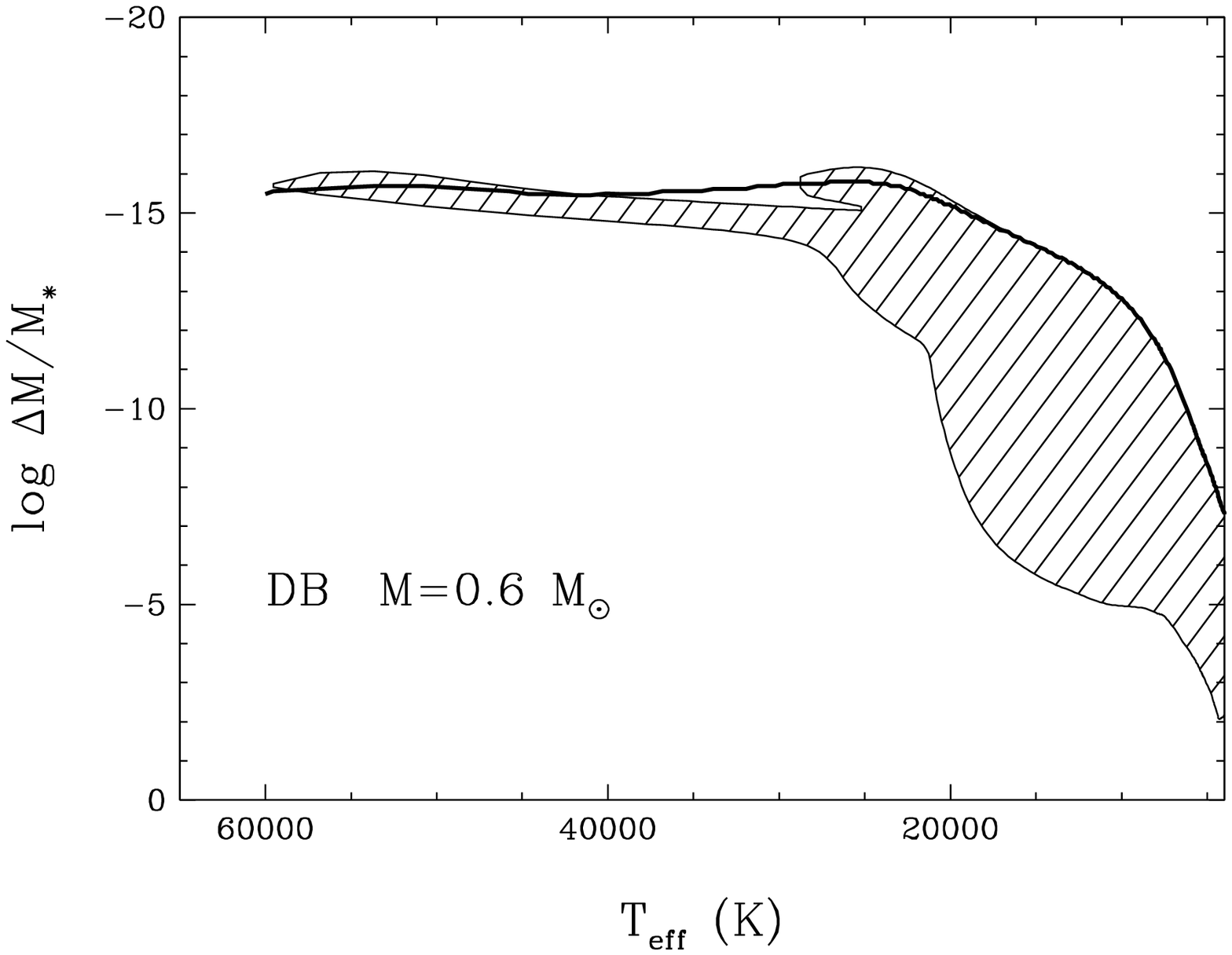] 
{Location of the helium convection zone (hatched region) as a
function of effective temperature in the pure helium envelope of a 0.6
\msun\ DB white dwarf calculated with the ML2/$\alpha=0.6$ version
of the mixing-length theory (from G.~Fontaine \&
P.~Brassard 2006, private communication). The depth is expressed as
the fractional mass above the point of interest with respect to the
total mass of the star. The thick solid line corresponds to the
photosphere ($\tau_R\sim1$).\label{fg:f3}}

\figcaption[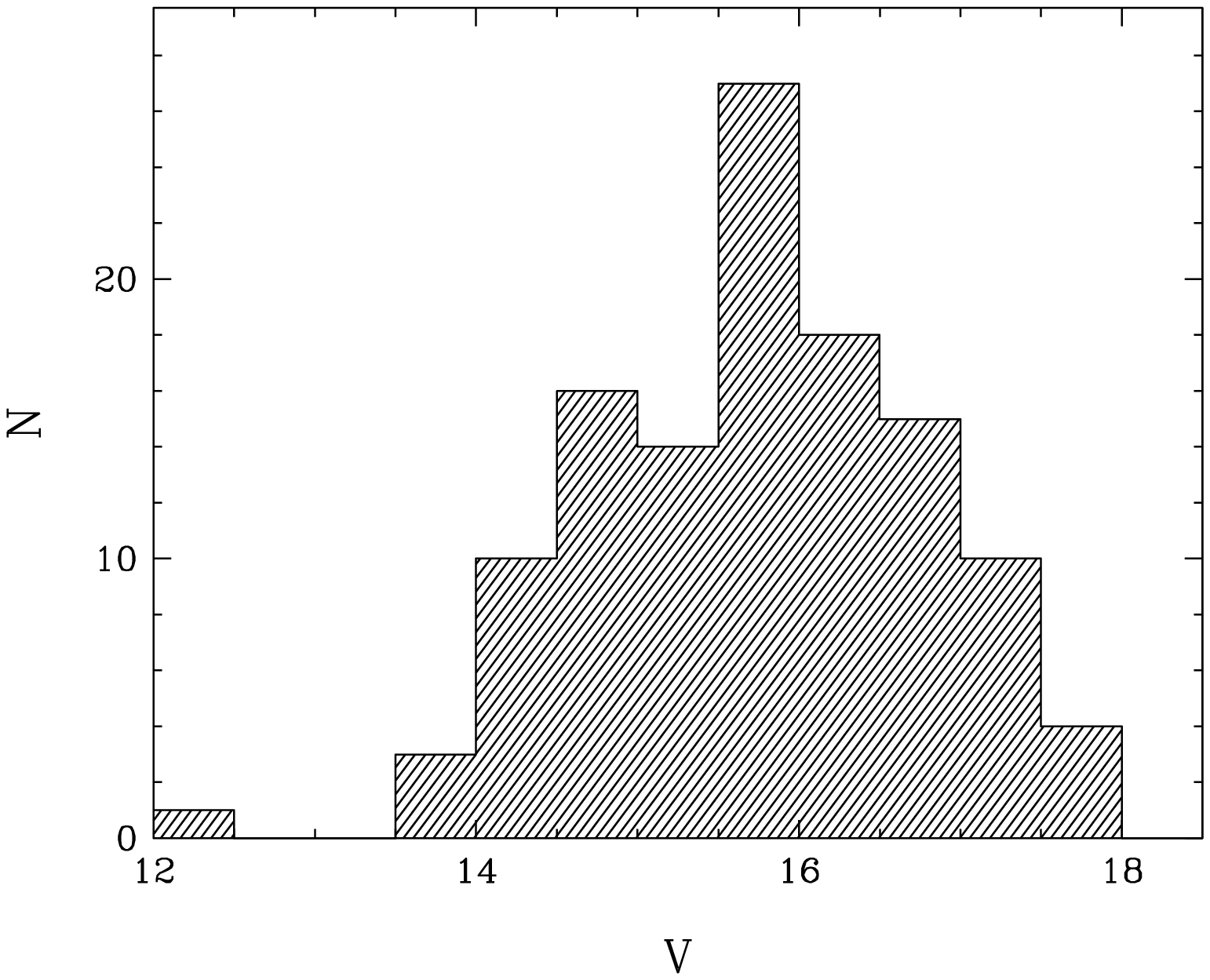] 
{Distribution of DB stars in our sample as a function of the $V$
magnitude.\label{fg:f4}}

\figcaption[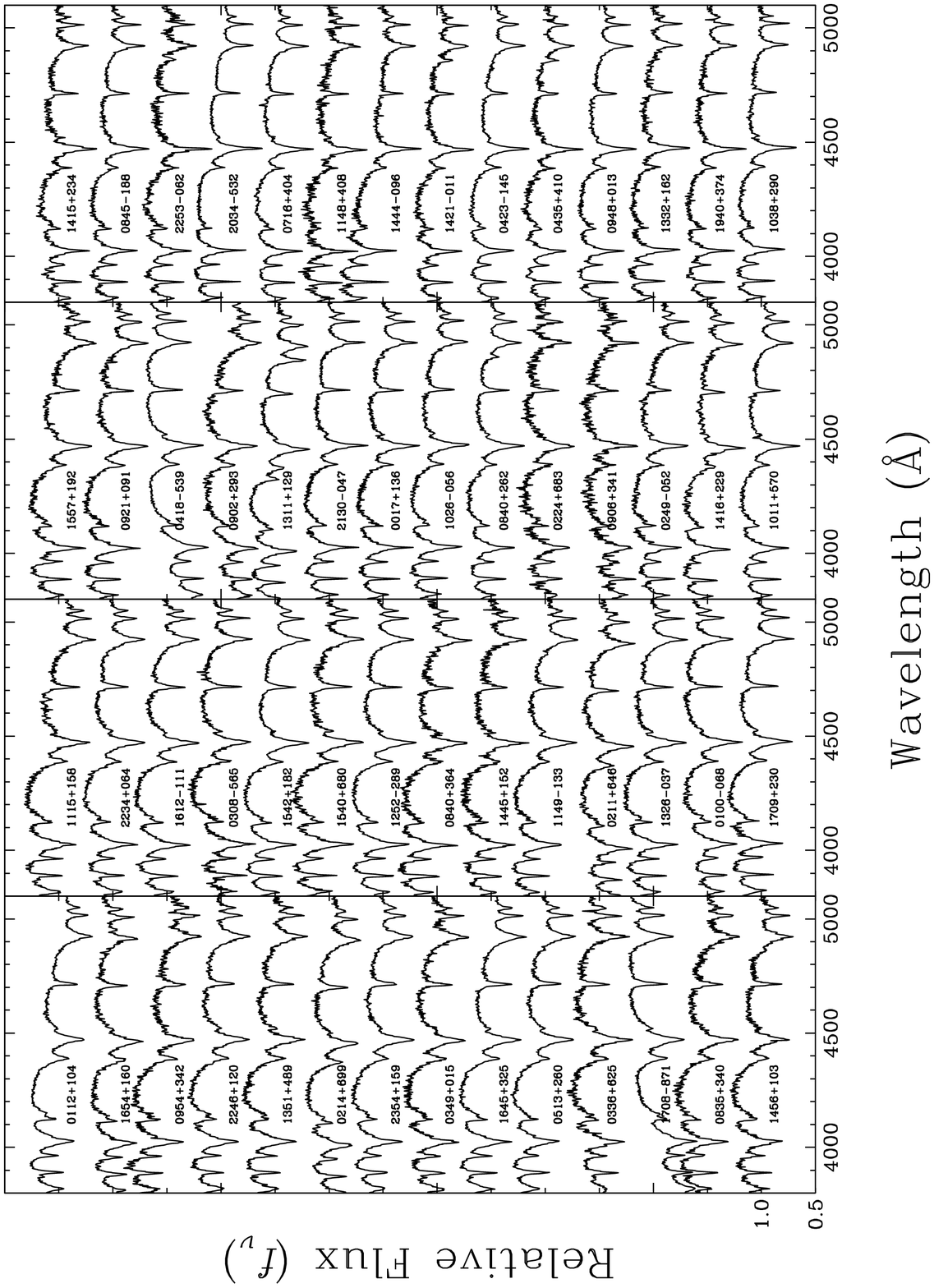] 
{Optical (blue) spectra for our complete sample of DB stars. The
spectra are normalized at 4500 \AA\ and shifted vertically from each
other by a factor of 0.5 for clarity. The effective temperature
decreases from upper left to bottom right.\label{fg:f5}}

\figcaption[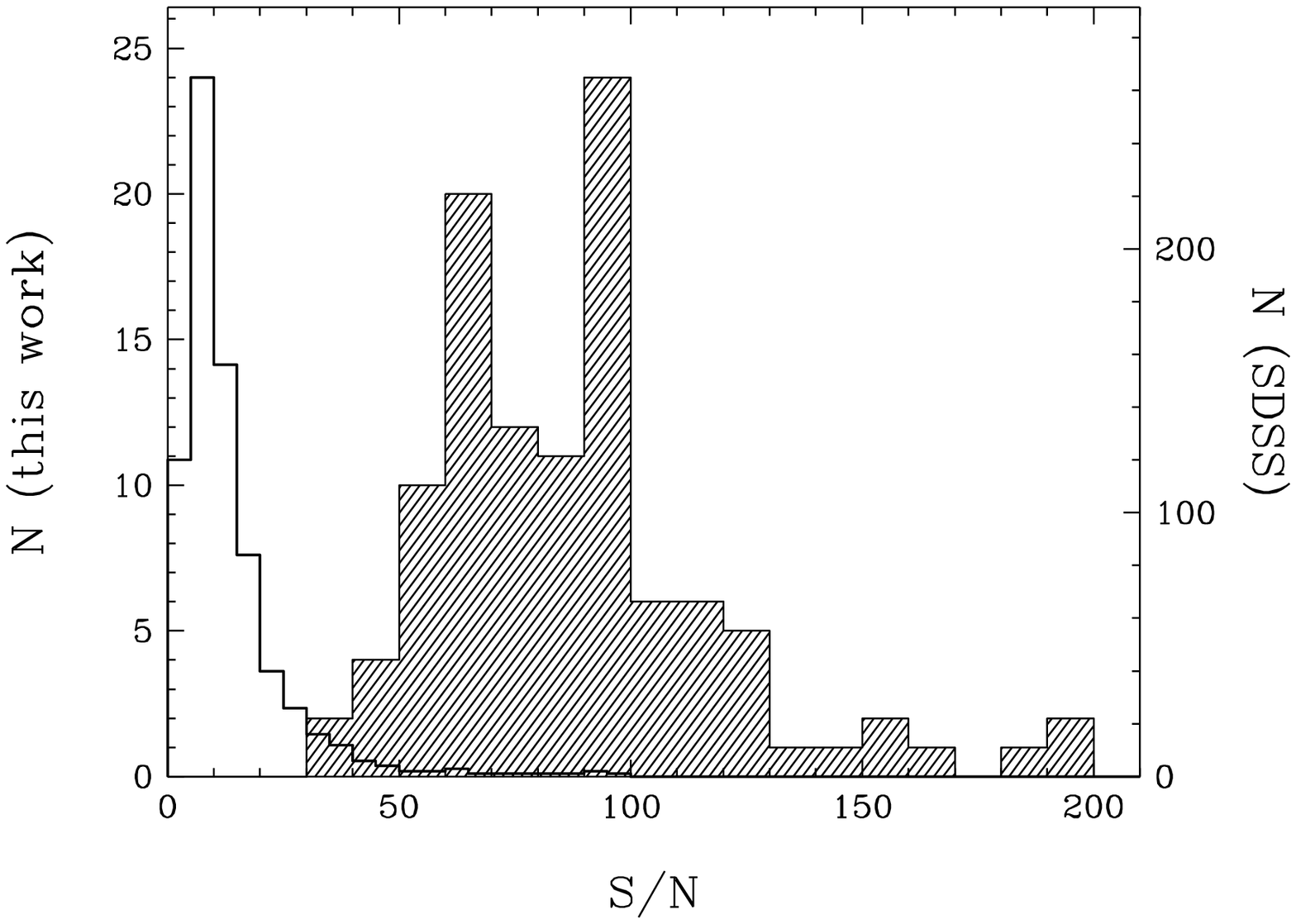] 
{Distribution of signal-to-noise ratios for the 108 optical spectra
secured for the analysis of the DB stars in our sample (hatched histogram). Also shown are
the corresponding values for 744 DB stars identified in the Data
Release 4 of the SDSS (thick solid line).\label{fg:f6}}

\figcaption[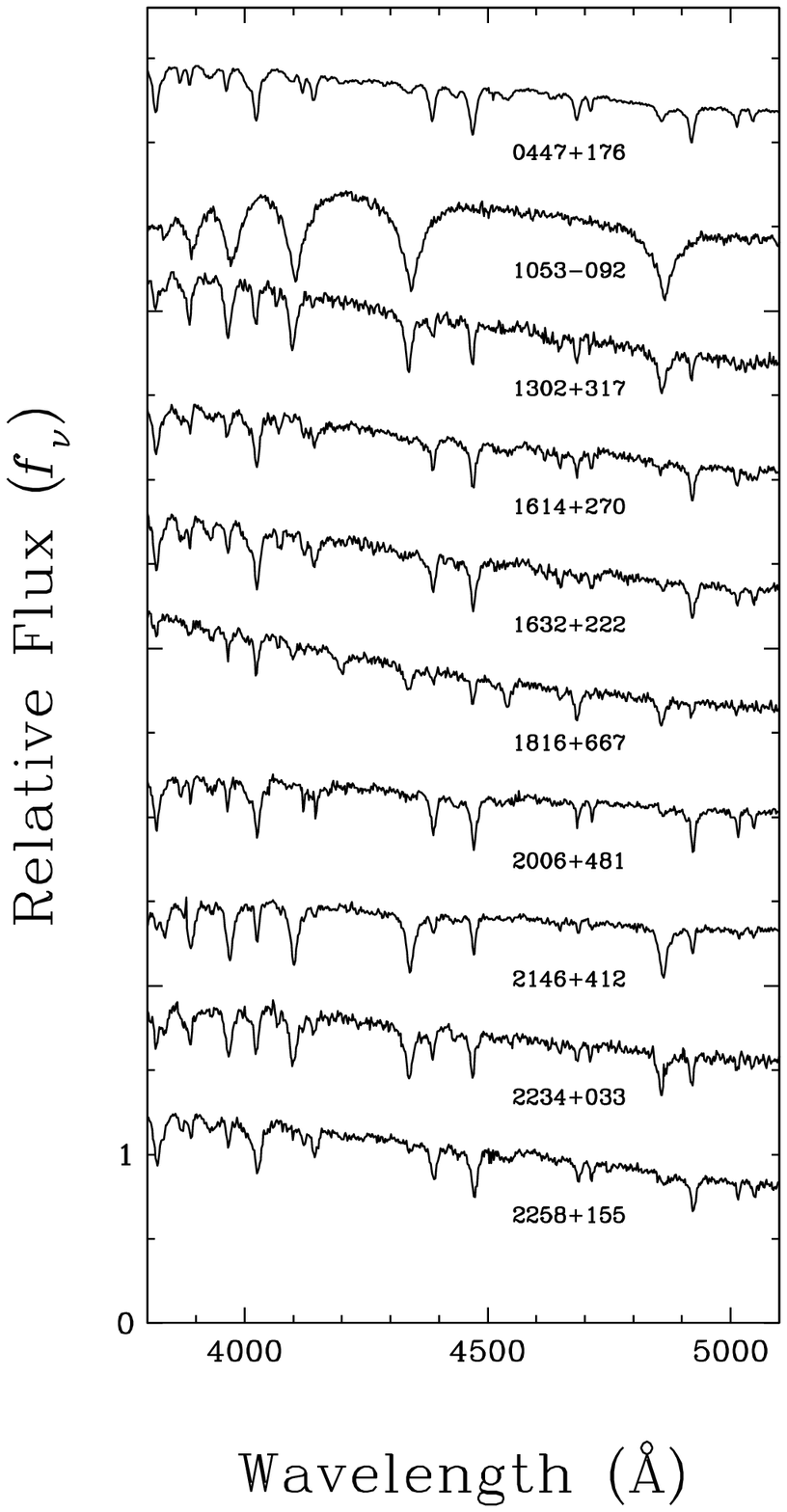] 
{Misclassified spectra in our original target list of DB white
dwarfs.\label{fg:f7}}

\figcaption[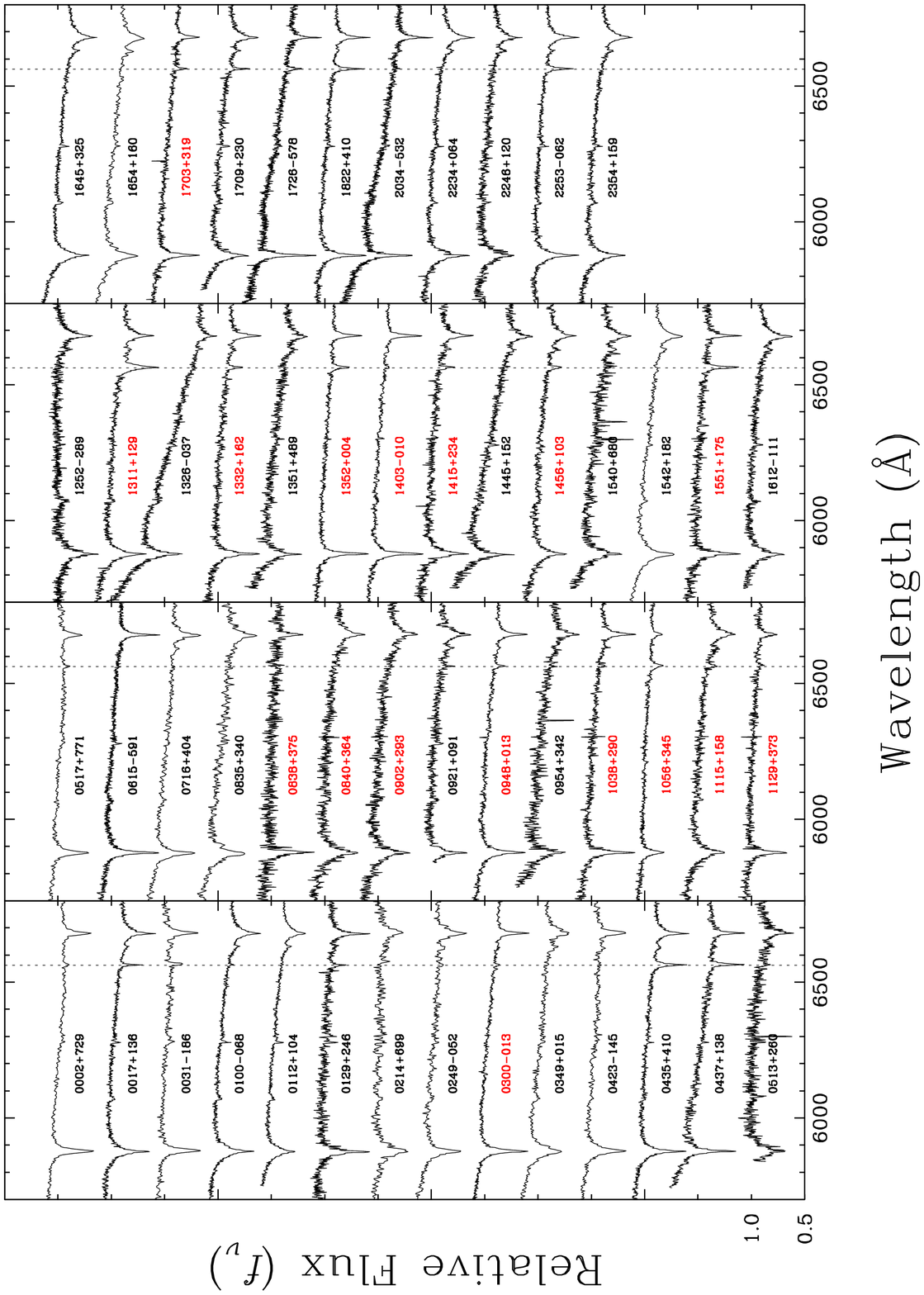] 
{Optical spectra in the red for the DB stars in our sample, when
available, ordered in right ascension. The spectra are normalized
at 6200 \AA\ and shifted vertically from each other by a factor of 0.5
for clarity. The dotted line indicates the location of
H$\alpha$. Spectra labeled in red are taken from the SDSS. H$\alpha$
spectra from the SPY survey, also used in our analysis of DBA stars,
are displayed in \citet{voss07}.\label{fg:f8}}

\figcaption[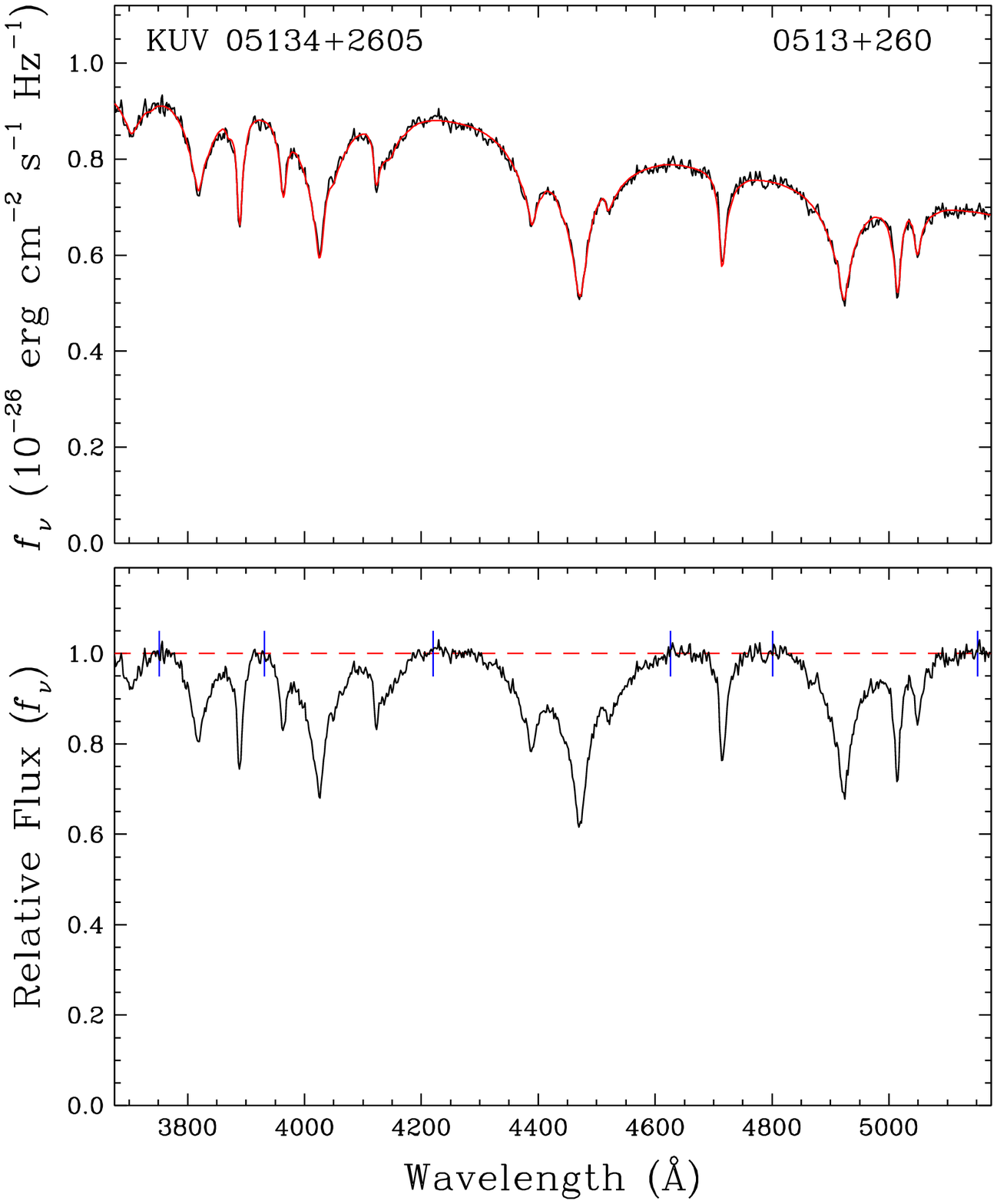] 
{Example of the procedure used to define the continuum. In the top
panel, the observed spectrum is fitted with a model spectrum multiplied
by a high order polynomial (up to $\lambda^5$) to achieve the best
possible match (shown in red); in this case the resulting atmospheric
parameters are meaningless since too many free parameters are used in
the fitting procedure. The continuum flux is then defined by this
fitting function at some predefined wavelength points (shown as blue
tick marks in the bottom panel), and these are then used to normalize each
segment of the spectrum to a continuum set to unity, as shown in the
bottom panel.\label{fg:f9}}

\figcaption[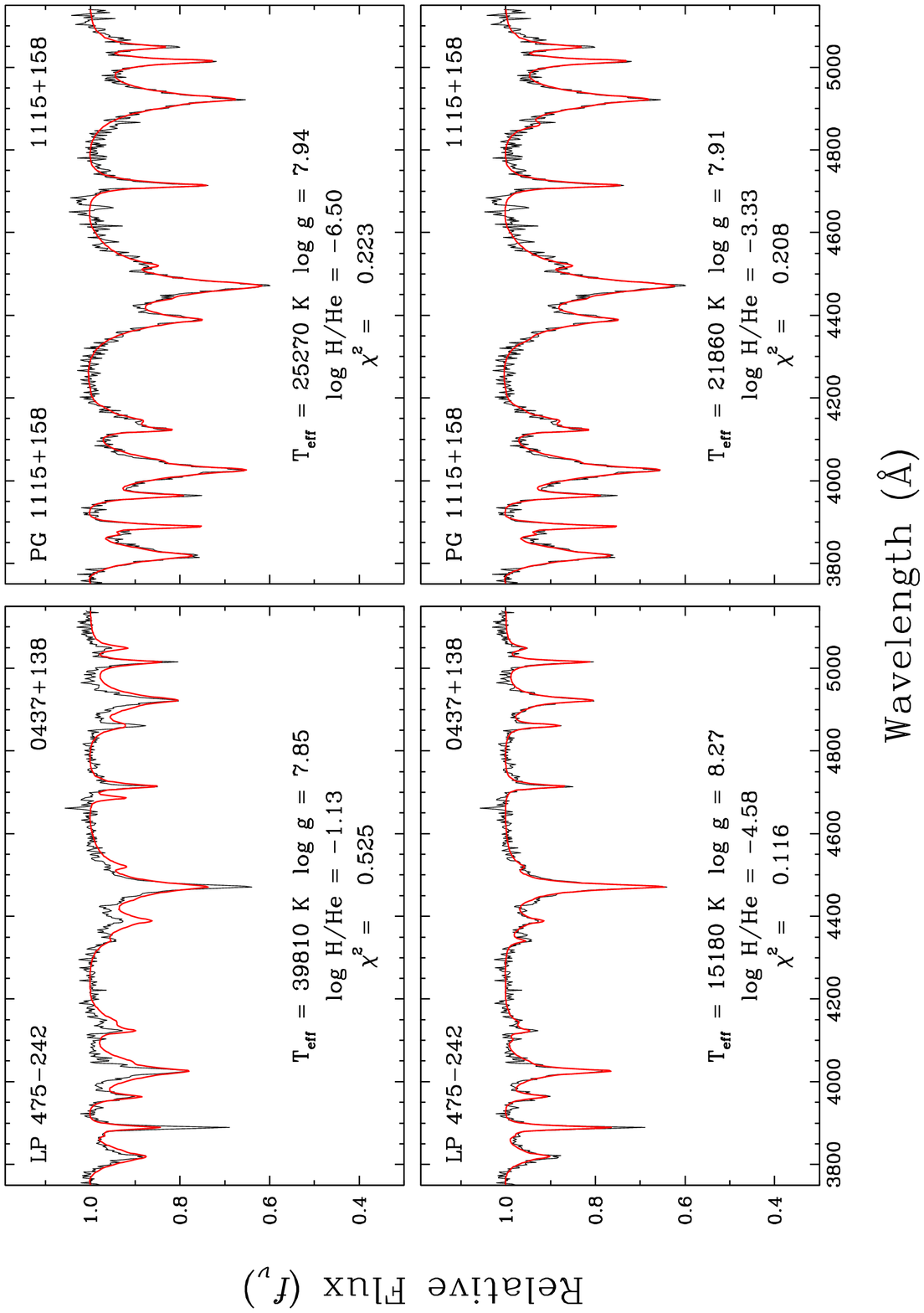] 
{Examples of spectroscopic fits for two DB white dwarfs using a cool (bottom
panels) and a hot (top panels) initial estimate of the effective
temperature in our fitting procedure.\label{fg:f10}}

\figcaption[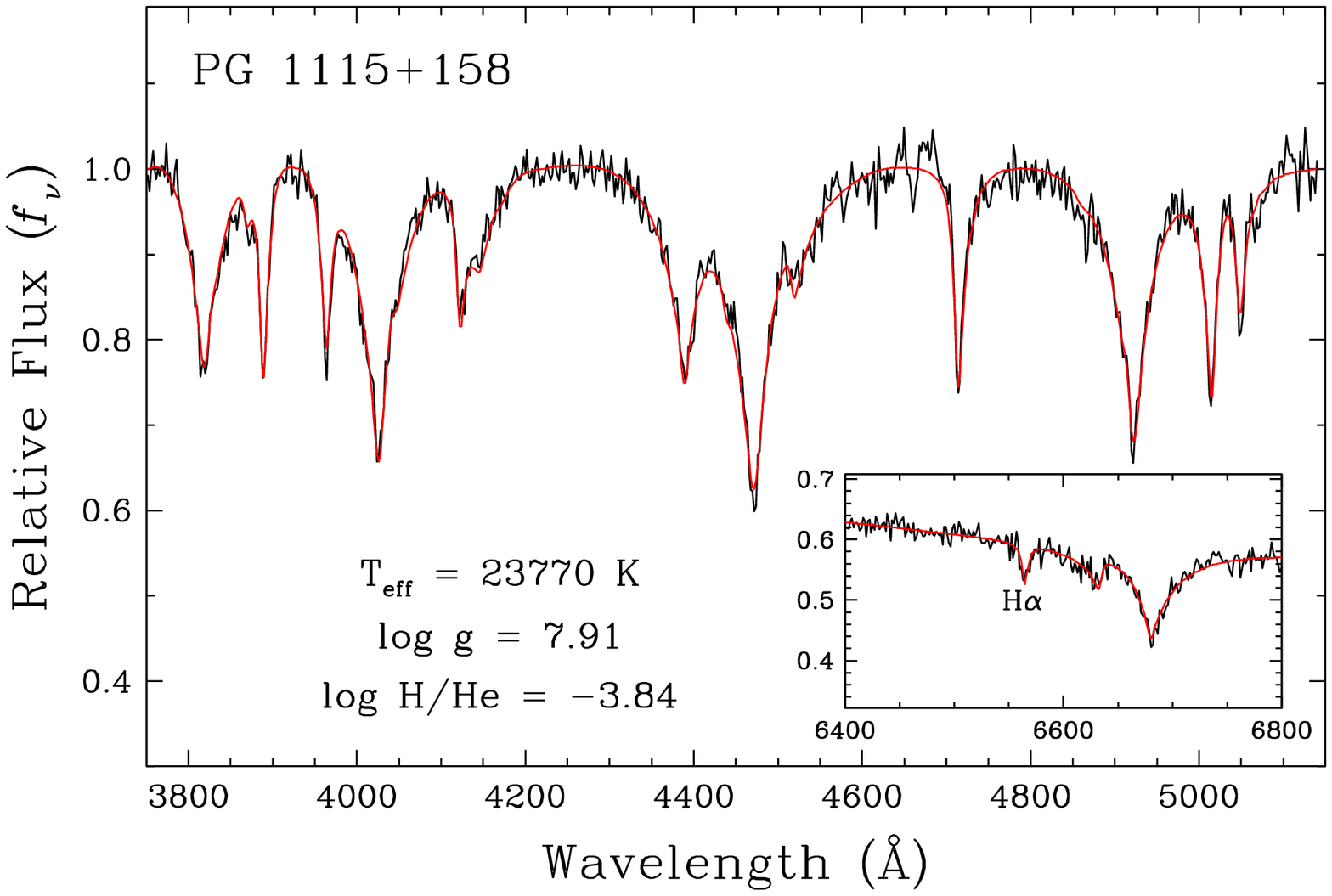] 
{Example of a full spectroscopic fit where the H$\alpha$ line profile,
shown in the insert, is used to measure, or constrain, the hydrogen
abundance of the overall solution.\label{fg:f11}}

\figcaption[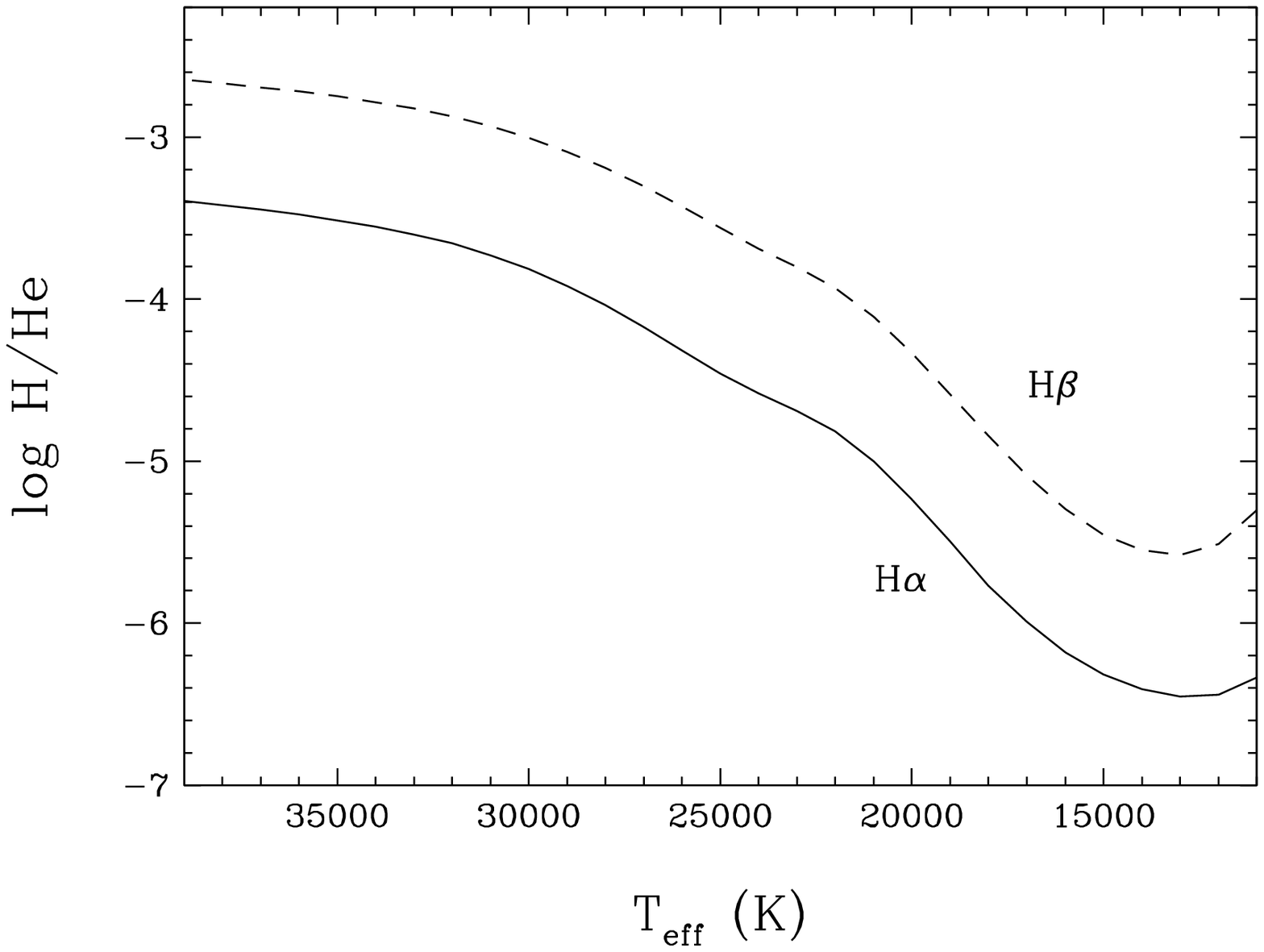] 
{Limits on the hydrogen abundance set by our spectroscopic
observations at H$\alpha$ and H$\beta$. We estimate the limit of
detectability at an equivalent width of 200 m\AA\ and 300 m\AA,
respectively.\label{fg:f12}}

\figcaption[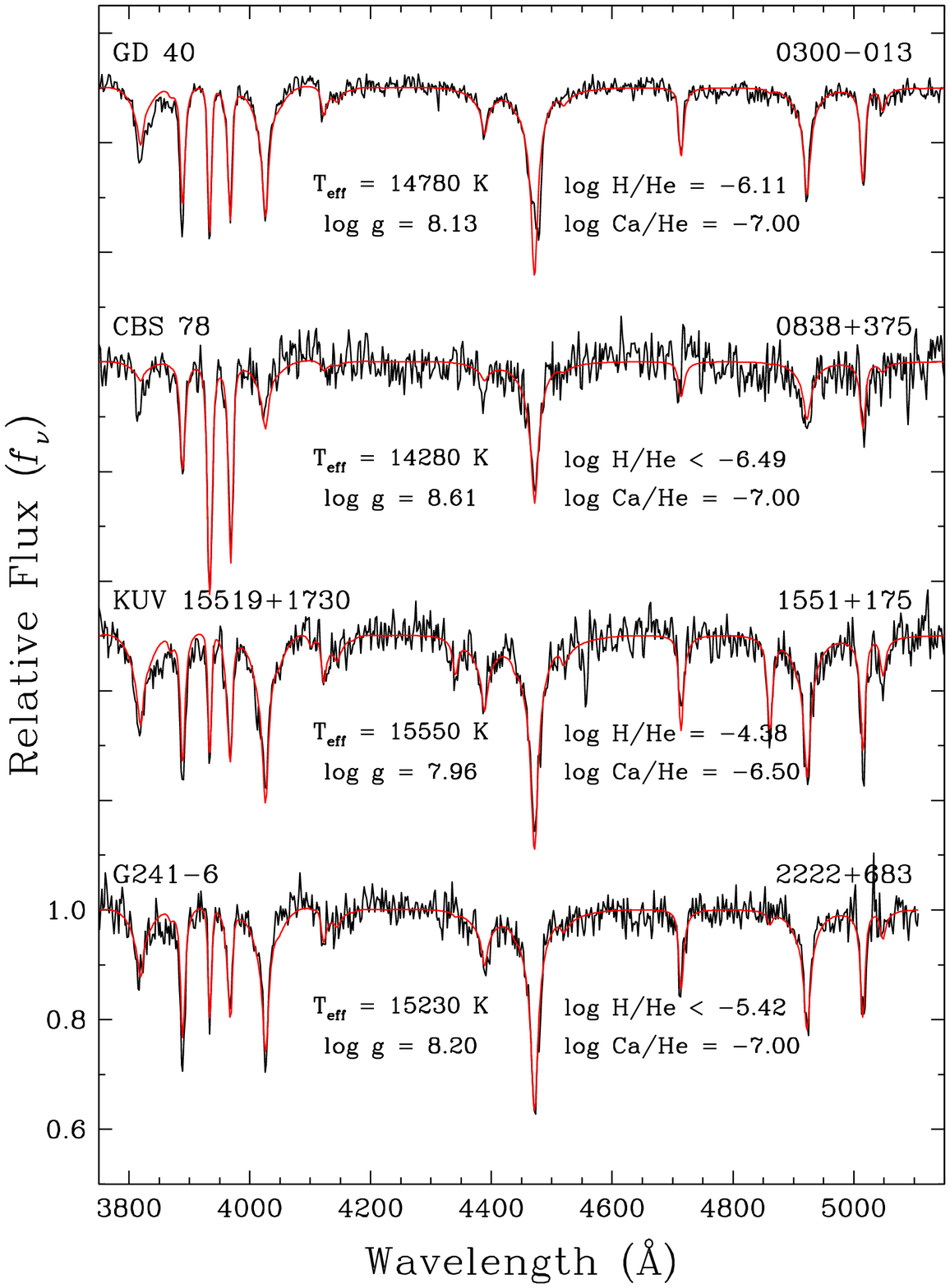] 
{Our best fits to the blue spectra of the strongest DBZ stars
in our sample. The region near H$\alpha$ is not displayed here.\label{fg:f13}}

\figcaption[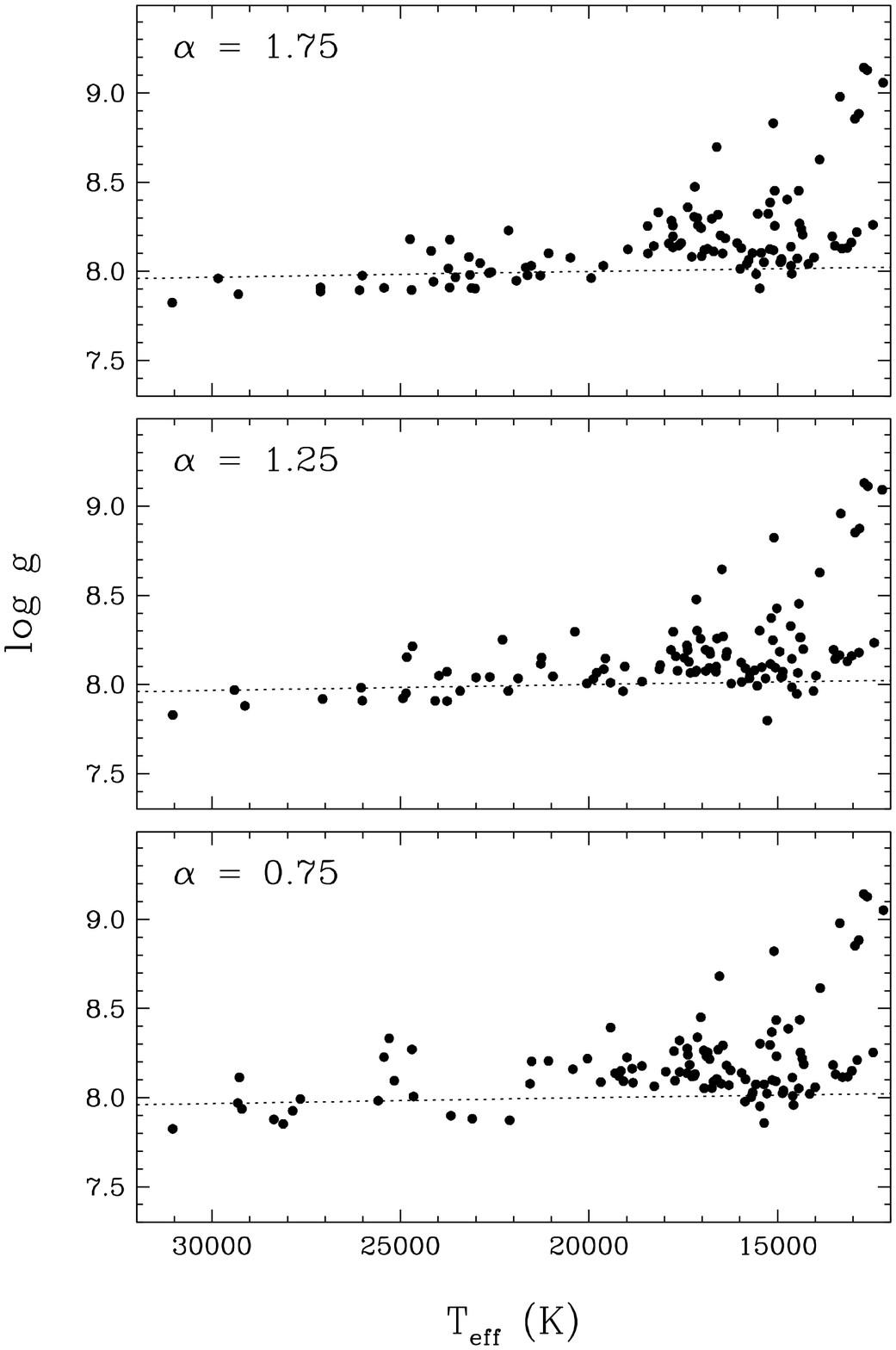] 
{Distribution of surface gravity as a function of effective
temperature for all DB white dwarfs in our sample for various
parameterization of the convective efficiency. Also shown
in each panel is a 0.6 \msun\ evolutionary sequence.\label{fg:f14}}

\figcaption[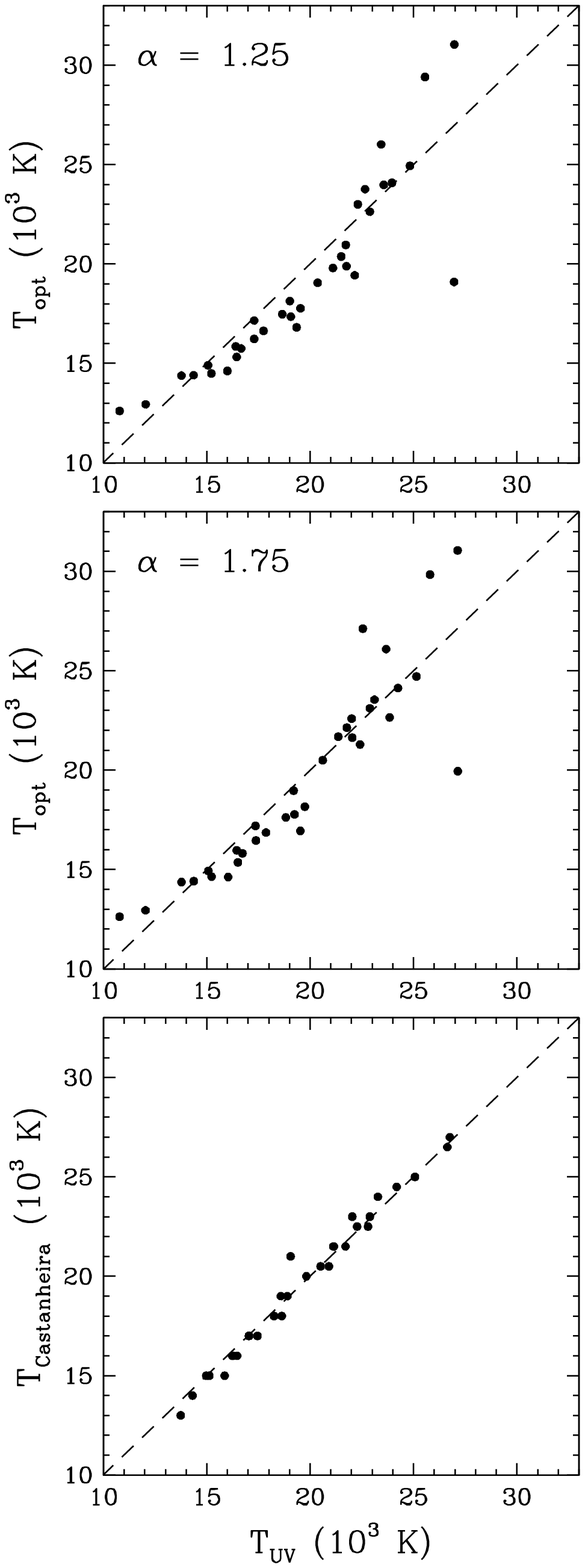] 
{Top two panels: Comparison of optical and UV temperatures for 34 DB
stars in our sample with available {\it IUE} spectra, using two
different parameterization of the convective efficiency. Bottom panel:
UV temperatures obtained by \citet{castan06} based on ML2/$\alpha=0.6$
models compared with our own temperature estimates based on
similar models.\label{fg:f15}}

\figcaption[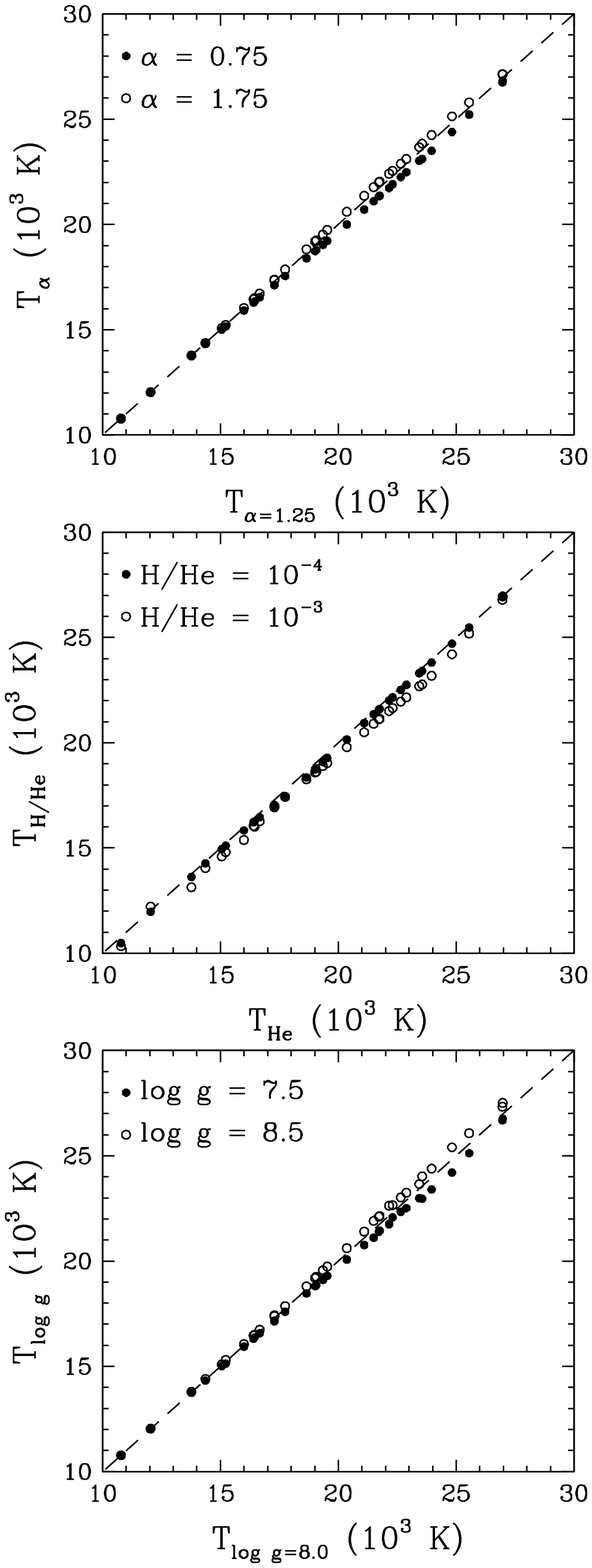] 
{Comparison of UV temperatures obtained from model spectra calculated
with different convective efficiencies (top panel), different hydrogen
abundances (middle panel), and different surface gravities (bottom panel). In
each panel, the UV temperatures on the y-axis, obtained from models
with parameters given in the legend, are compared against UV
temperatures obtained from pure helium model atmospheres at $\logg=8$
and ML2/$\alpha=1.25$.\label{fg:f16}}

\figcaption[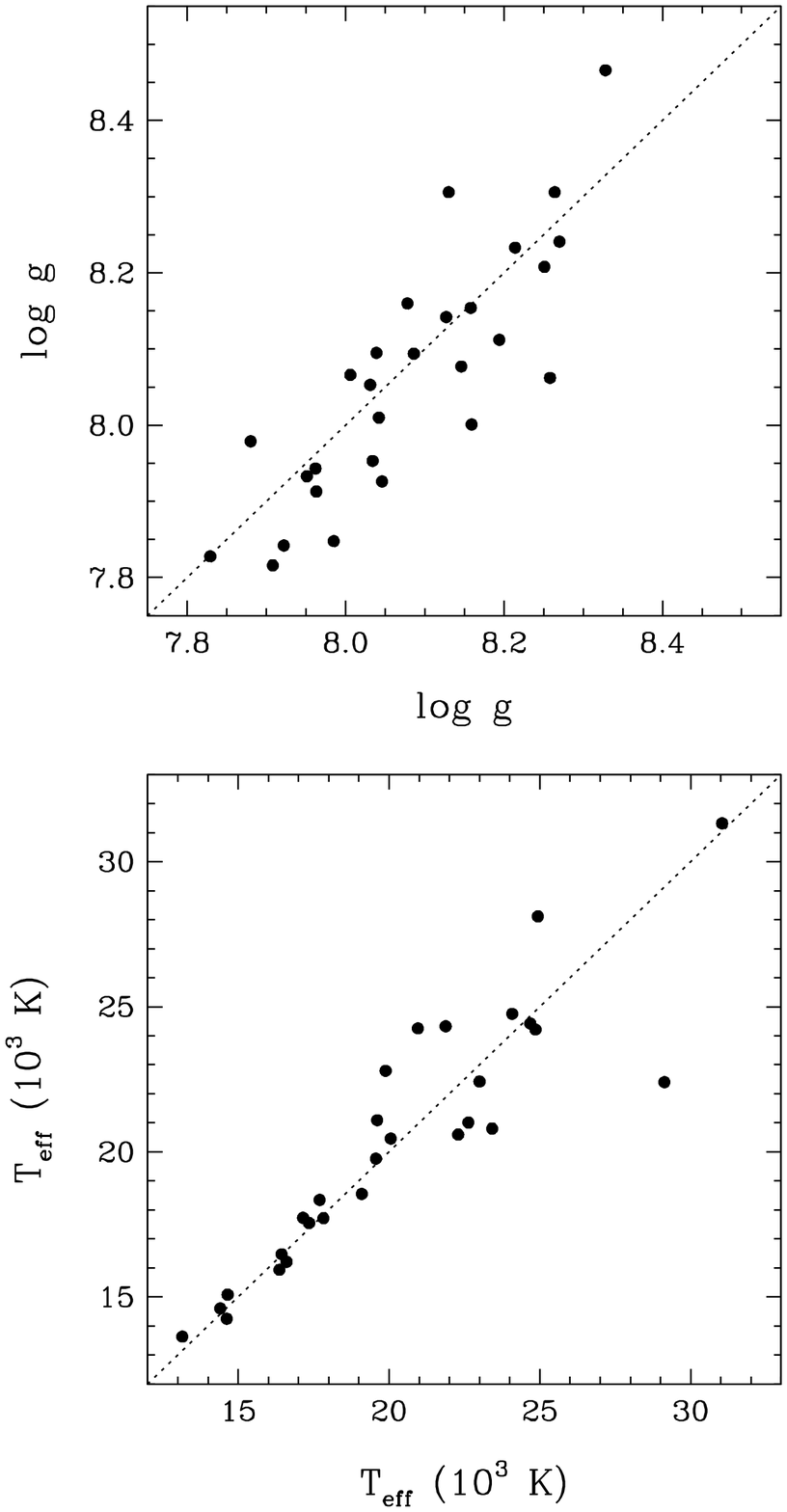] 
{Comparison of $\Te$ and $\logg$ determinations for 28 DB stars in our
sample with mutliple spectroscopic observations.\label{fg:f17}}

\figcaption[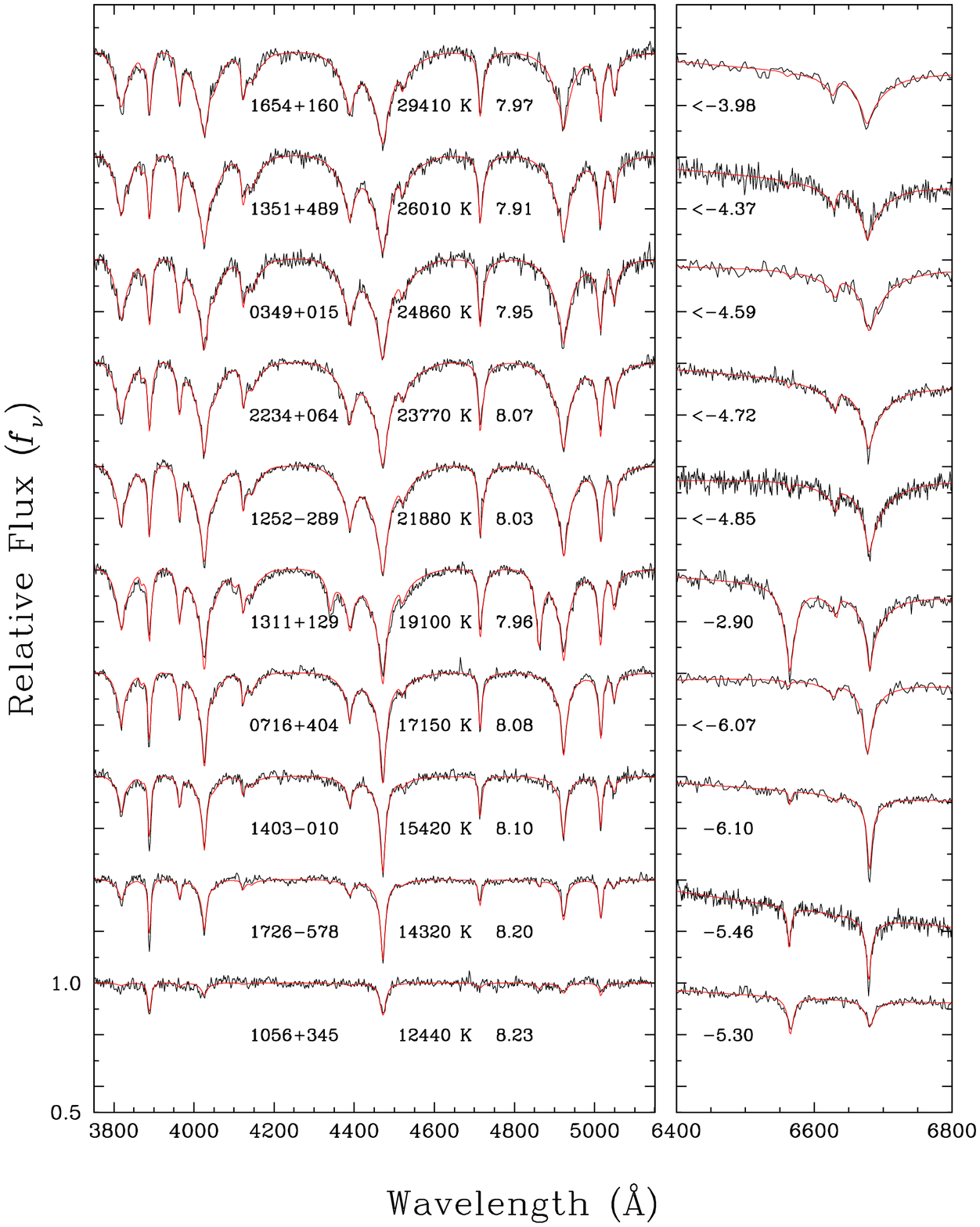] 
{Sample fits for several DB and DBA stars in our sample; the
atmospheric parameters ($\Te$, $\logg$, and log H/He) of each object
are given in the figure. The region near H$\alpha$ (right panel) is
used to measure, or constrain, the hydrogen abundance. In the case of
DB stars, these high signal-to-noise spectra provide {\it upper
limits} on the hydrogen-to-helium abundance
ratio.\label{fg:f18}}

\figcaption[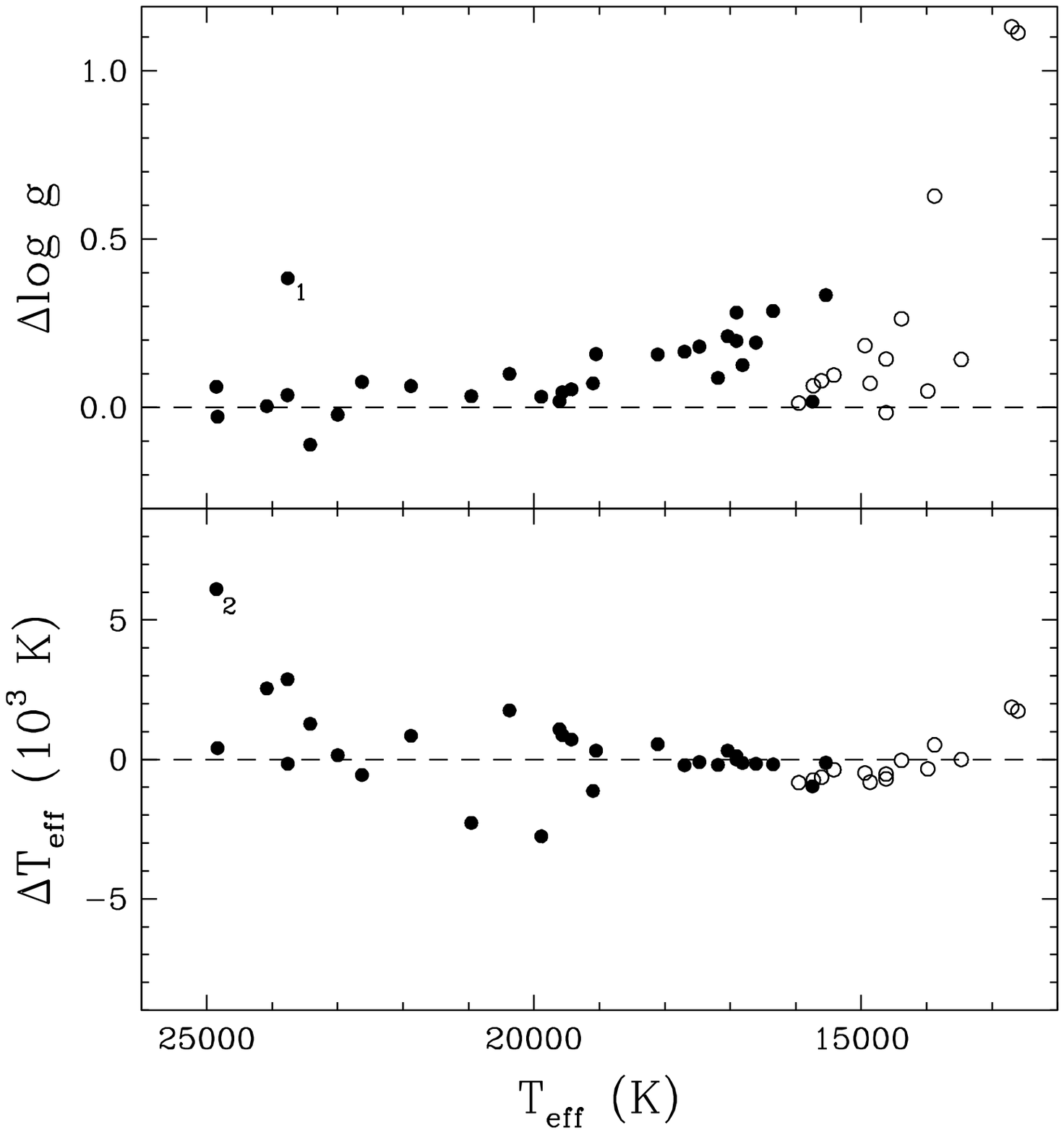] 
{Comparison of $\Te$ and $\logg$ measurements between our analysis and
that of \citet{voss07} as a function of our $\Te$ values (this work $-$ 
Voss in each case). The dashed lines represent a perfect
agreement. Objects shown as open circles correspond to cool DB white
dwarfs for which Voss et al.~assumed a value of $\logg=8$. The two
objects labeled in the figure correspond to (1) PG 1115+128 and (2)
KUV 03493+0131 (0349+015).\label{fg:f19}}

\figcaption[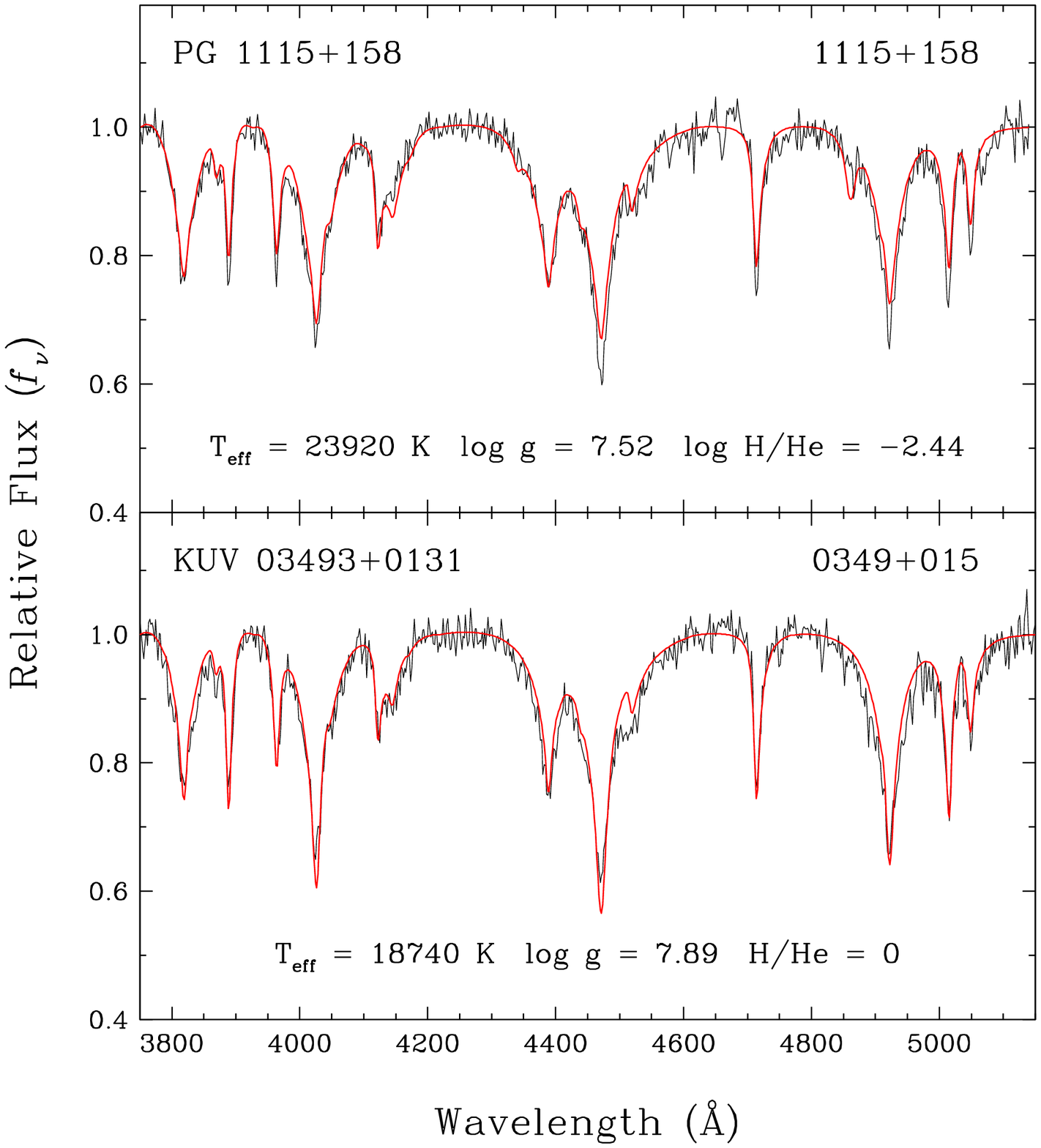] 
{Spectroscopic solutions obtained by \citet{voss07} for PG 1115+158
and KUV 03493+0131. These should be contrasted with our solutions
displayed in Figures \ref{fg:f11} and
\ref{fg:f18}, respectively.\label{fg:f20}}

\figcaption[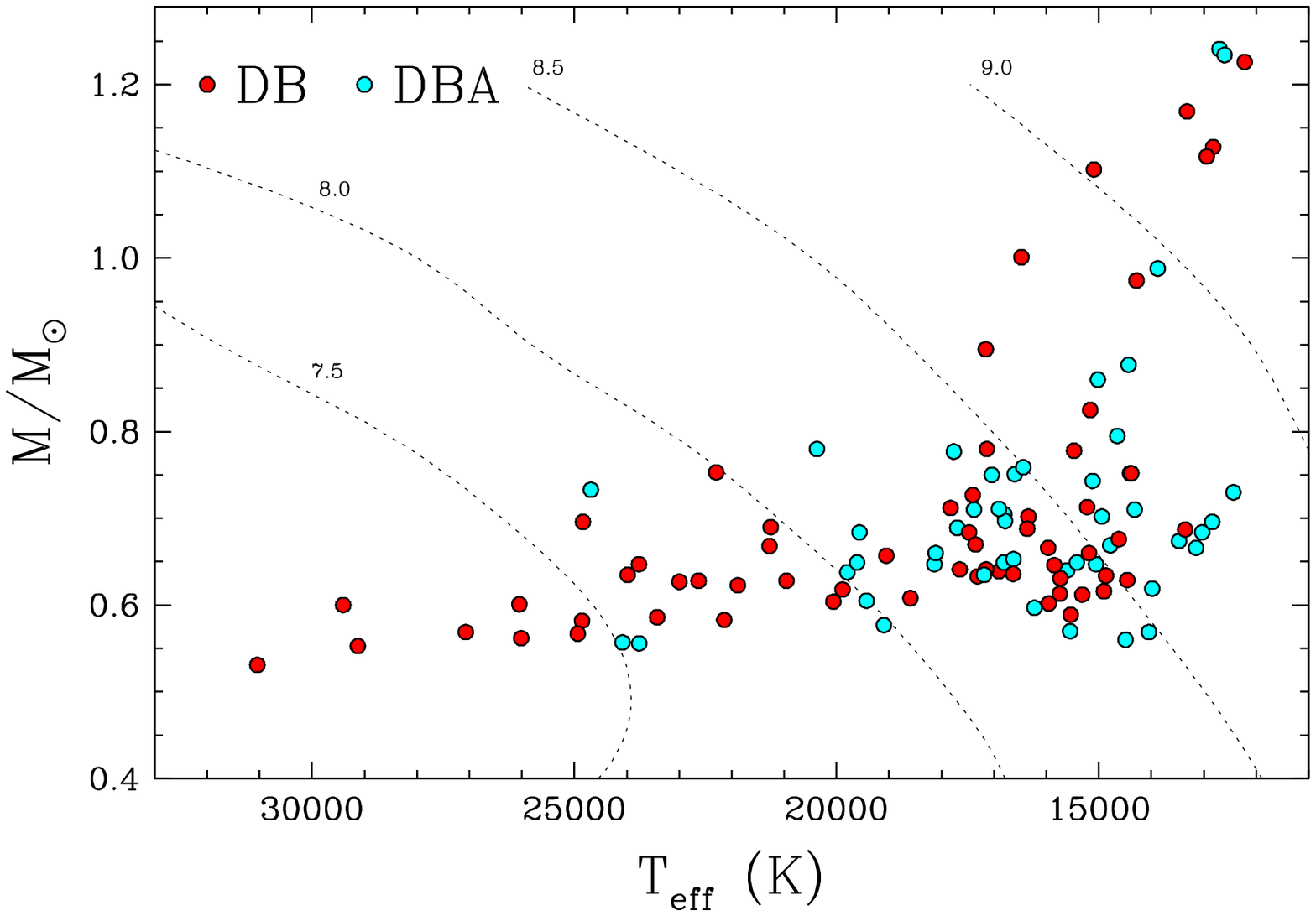] 
{Distribution of mass as a function of effective temperature for
the 61 DB (red symbols) and 47 DBA (blue symbols) stars in our
sample. Also shown as dotted lines are the theoretical isochrones from our
evolutionary models, labeled as $\log\tau$ where $\tau$ is the white
dwarf cooling age in years.\label{fg:f21}}

\figcaption[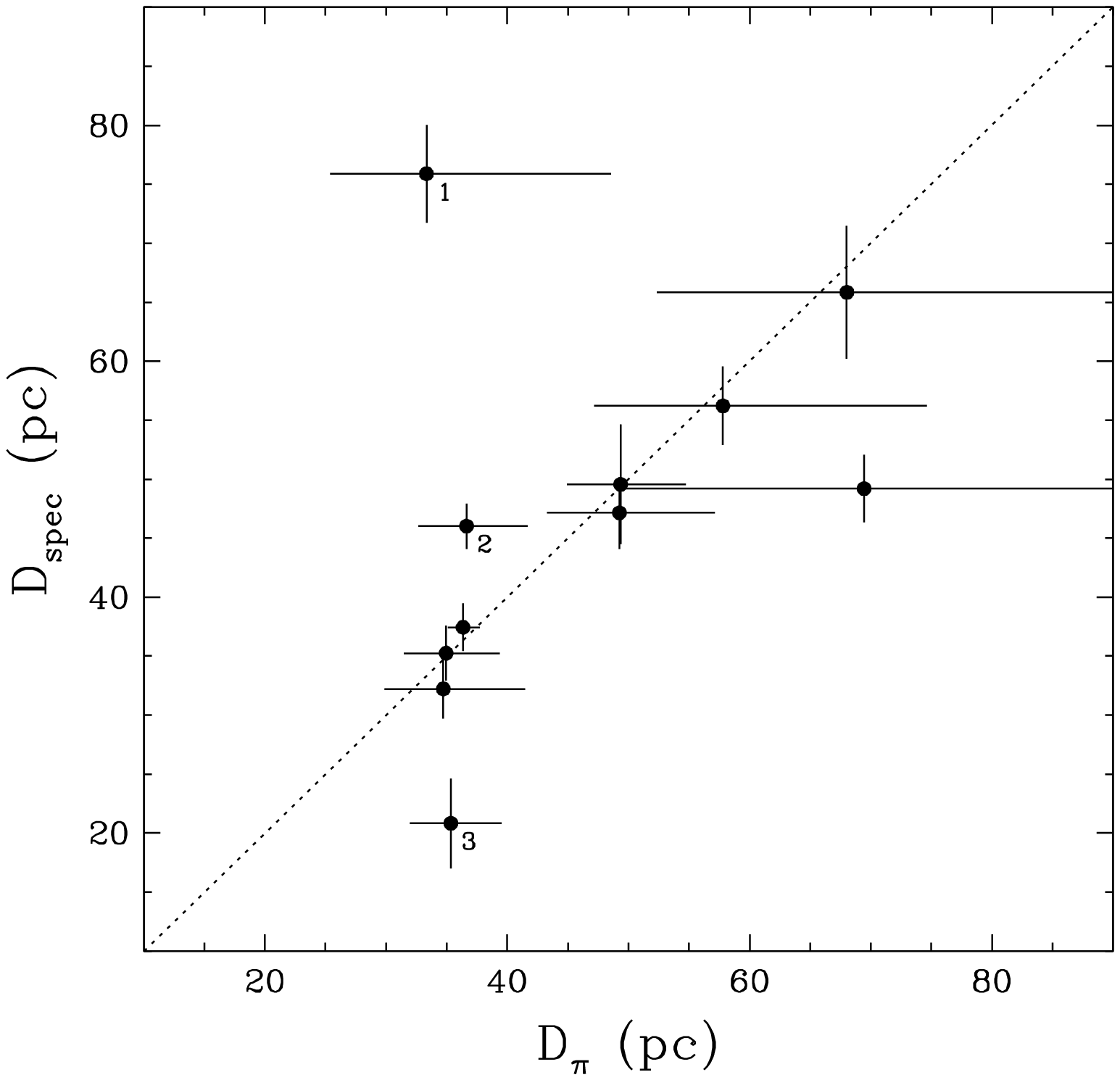] 
{Comparison of distances inferred from trigonometric parallax
measurements ($D_\pi$) with those derived from spectroscopy ($D_{\rm
spec}$) taken from Table 2.  The objects labeled in the figure correspond to
(1) Feige 4 (0017+136), (2) GD 358 (1645+325), and (3) G188-27
(2147+280).\label{fg:f22}}

\figcaption[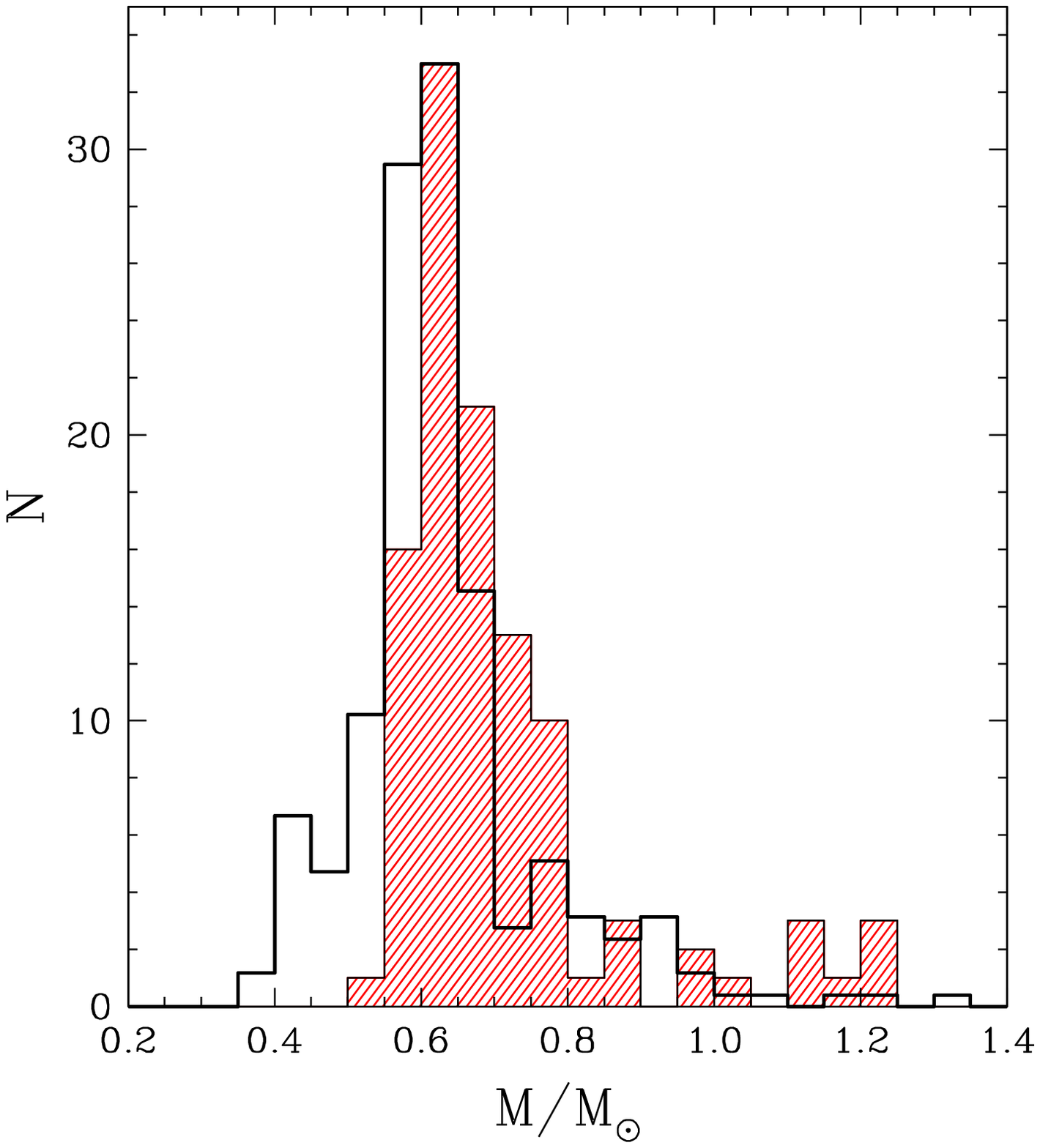] 
{Mass distribution of all DB and DBA white dwarfs in our sample
(hatched histogram). If the most massive DB stars
in this distribution are excluded (see text), the mean mass becomes
of $\langle M\rangle=0.671$ \msun\ with a standard
deviation of $\sigma_M=0.091$ \msun. Also shown as a
thick solid line is the mass distribution of the DA stars in the PG
sample.\label{fg:f23}}

\figcaption[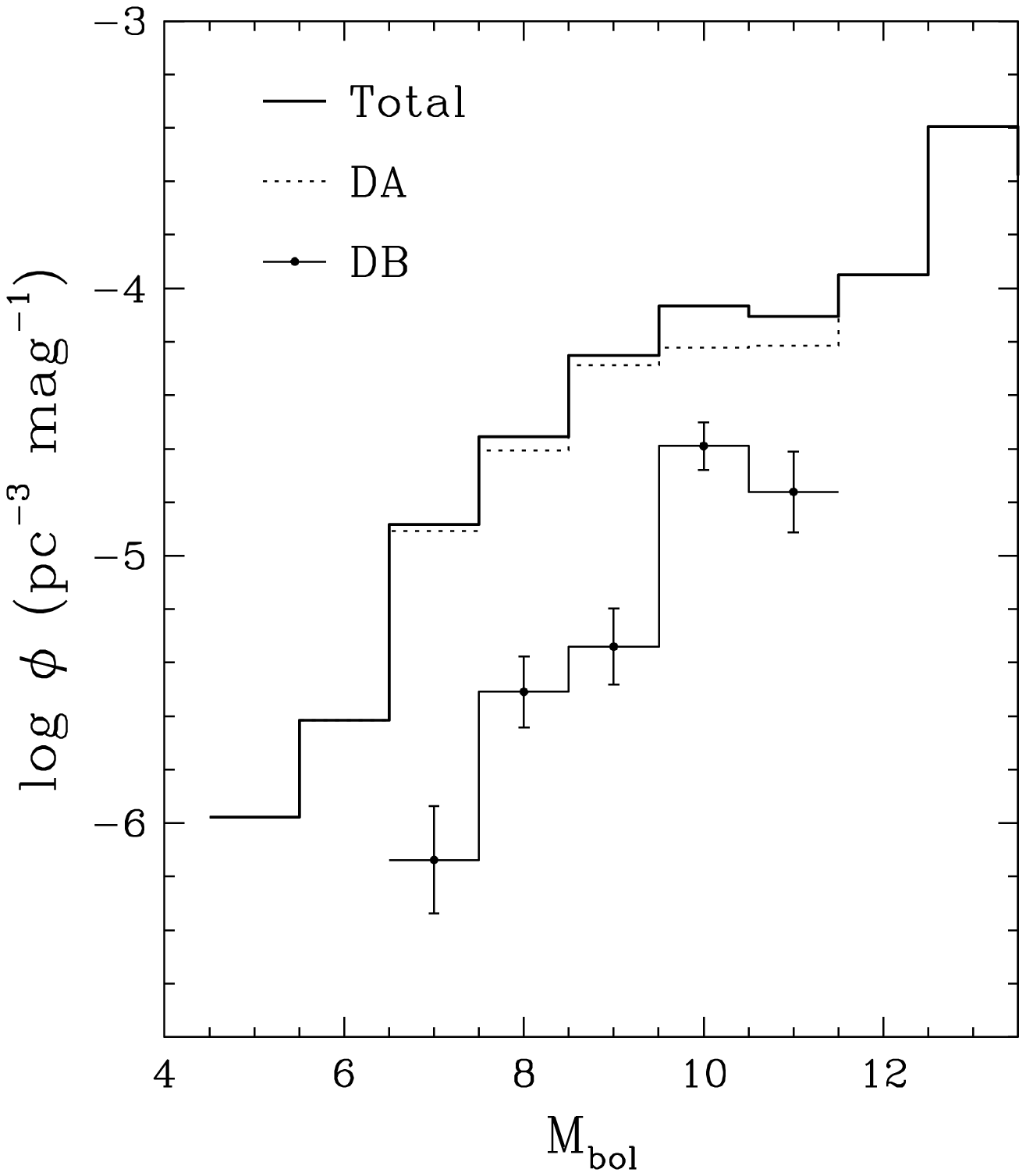] 
{Luminosity function derived for all DB and DA stars from the
complete PG sample presented in bolometric magnitude
bins, assuming a scale height for the Galaxy of $z_0=250$
pc (see \citealt{LBH05} for details).\label{fg:f24}}

\figcaption[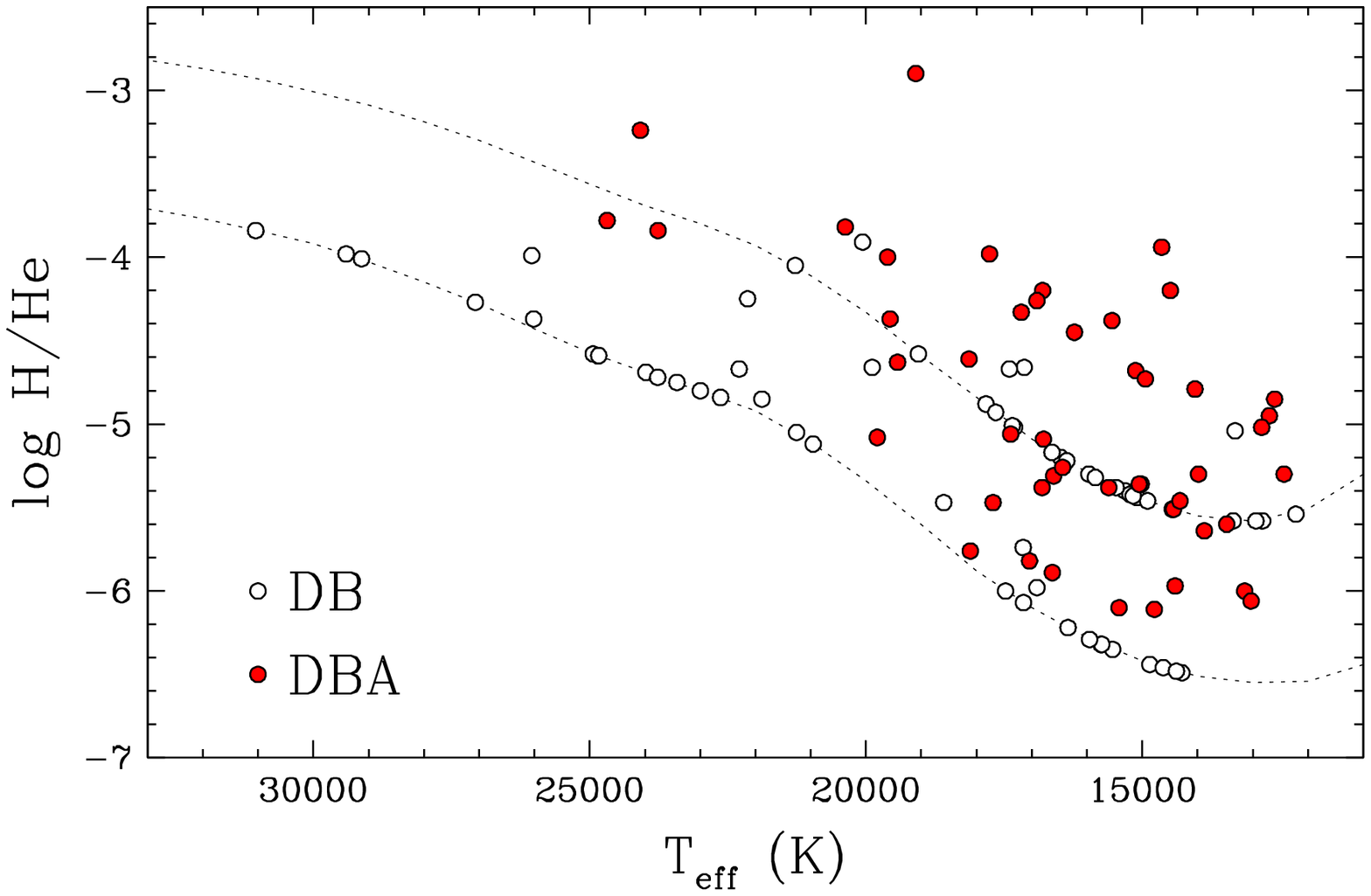] 
{Hydrogen-to-helium abundance ratio as a function of effective
temperature for all DB (white symbols) and DBA (red symbols) white
dwarfs in our sample. Limits on the hydrogen abundance set by our
spectroscopic observations at H$\alpha$ (lower dotted line) and
H$\beta$ (upper dotted line) are reproduced from Figure
\ref{fg:f12}. The hydrogen abundances for DB stars represent only
upper limits.\label{fg:f25}}

\figcaption[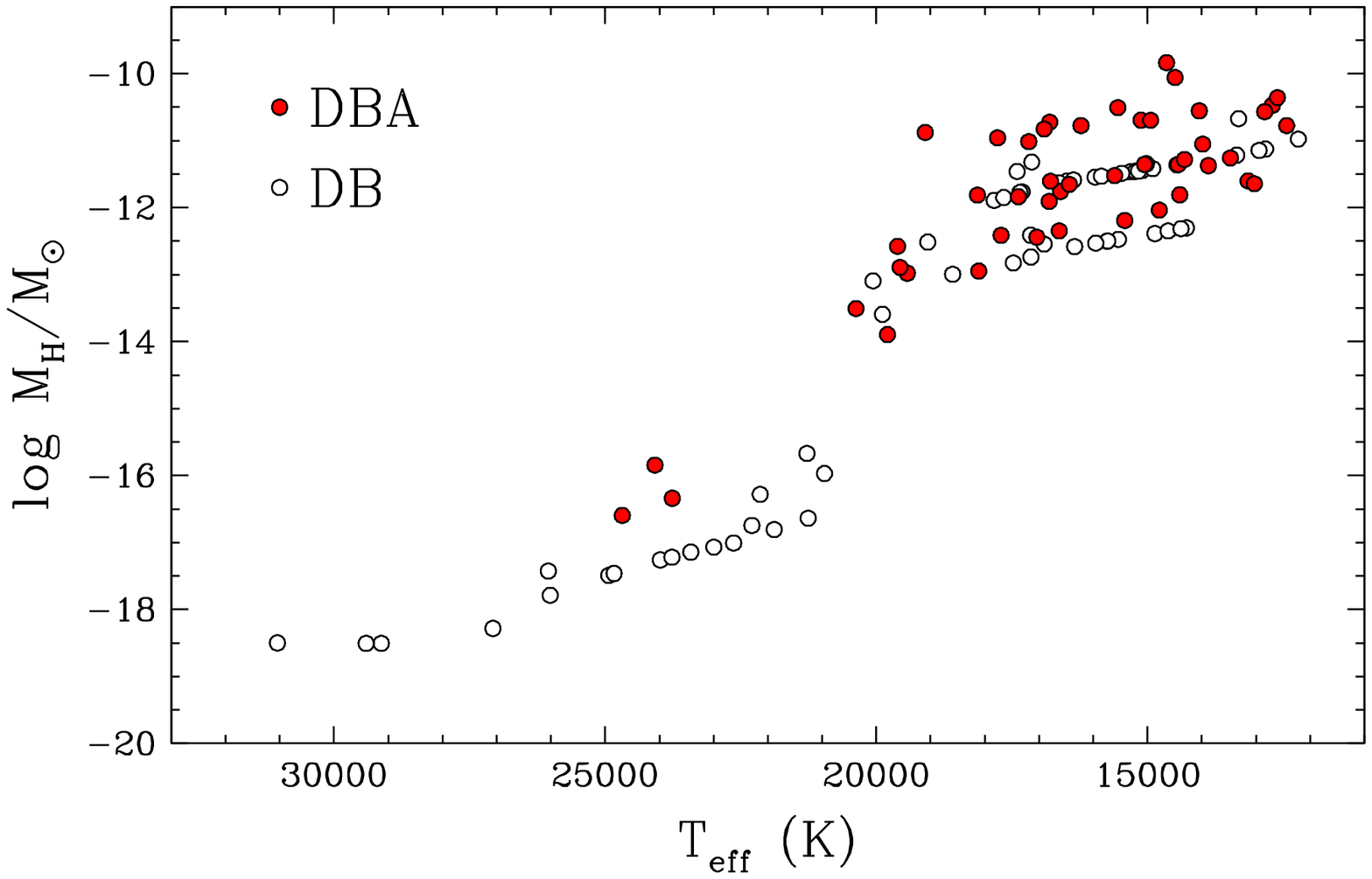] 
{Total hydrogen mass as a function of effective temperature for all DB
(white symbols) and DBA (red symbols) in our sample. The values for DB
stars represent only upper limits.\label{fg:f26}}

\figcaption[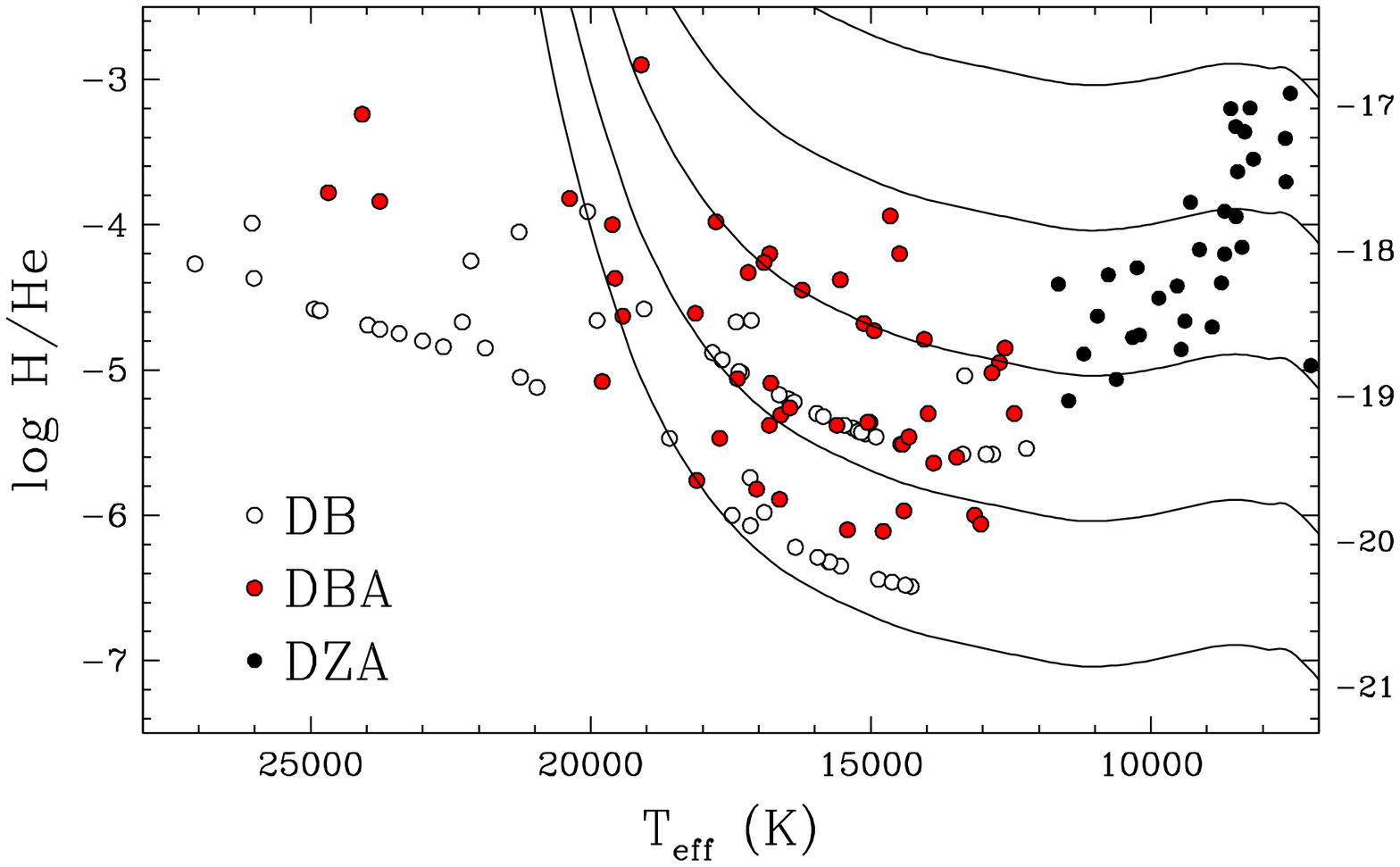] 
{Same as Figure \ref{fg:f25}, but including also the results for
DZA stars from \citet{dufour07a}. Solid lines represent the expected
abundances for continuous accretion of material from the interstellar
medium with accretion rates of $10^{-21}$ to $10^{-17}$ \msun\
yr$^{-1}$.\label{fg:f27}}

\figcaption[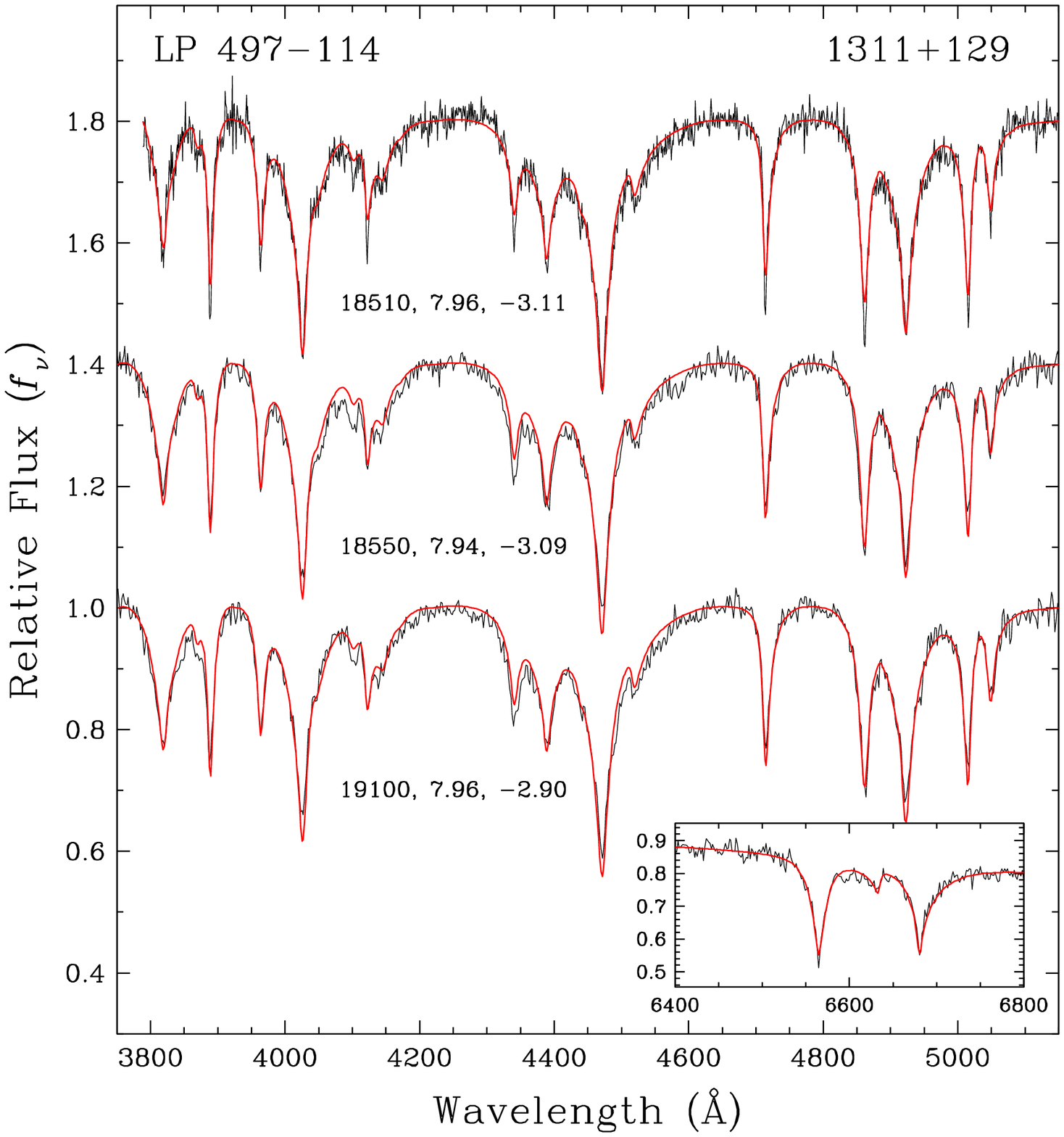] 
{Our best fit to three independent spectroscopic observations of LP
497-114 (1311+129), the DBA star with the largest hydrogen abundance
in our sample. Our adopted solution is derived from the bottom
spectrum, obtained in 2010. The other spectra have been secured in 1995
(middle) and 2004 (top; from SDSS).\label{fg:f28}}

\figcaption[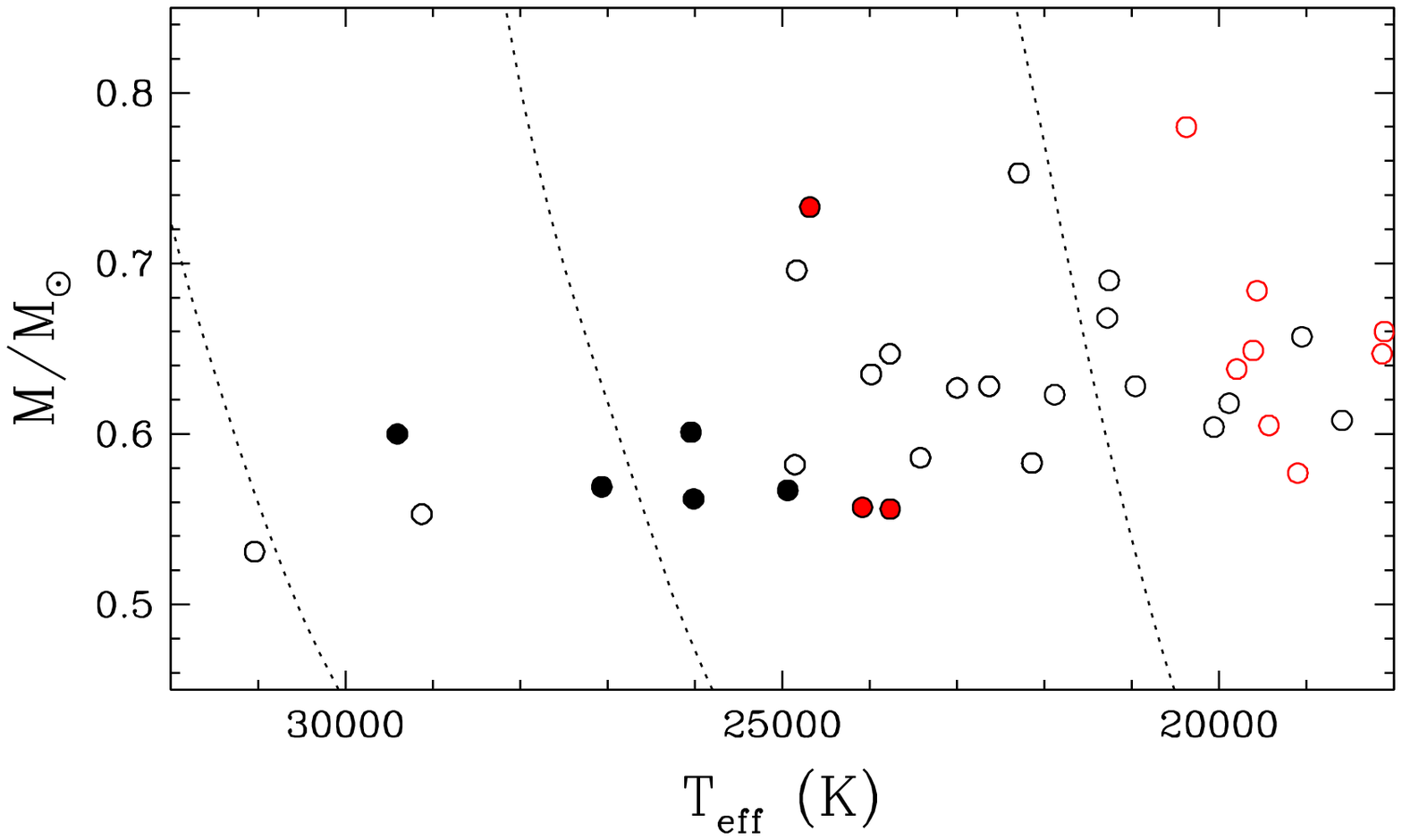] 
{Location of the V777 Her stars in a mass versus effective temperature
diagram. Variable DB white dwarfs are represented by filled symbols
while open symbols correspond to photometrically constant, or unknown,
objects. DB and DBA stars are shown in black and red, respectively.
The dashed lines correspond to theoretical blue edges for pure helium
envelope models with convective efficiencies given by, from left to right, ML3,
ML2, and ML1.\label{fg:f29}}

\figcaption[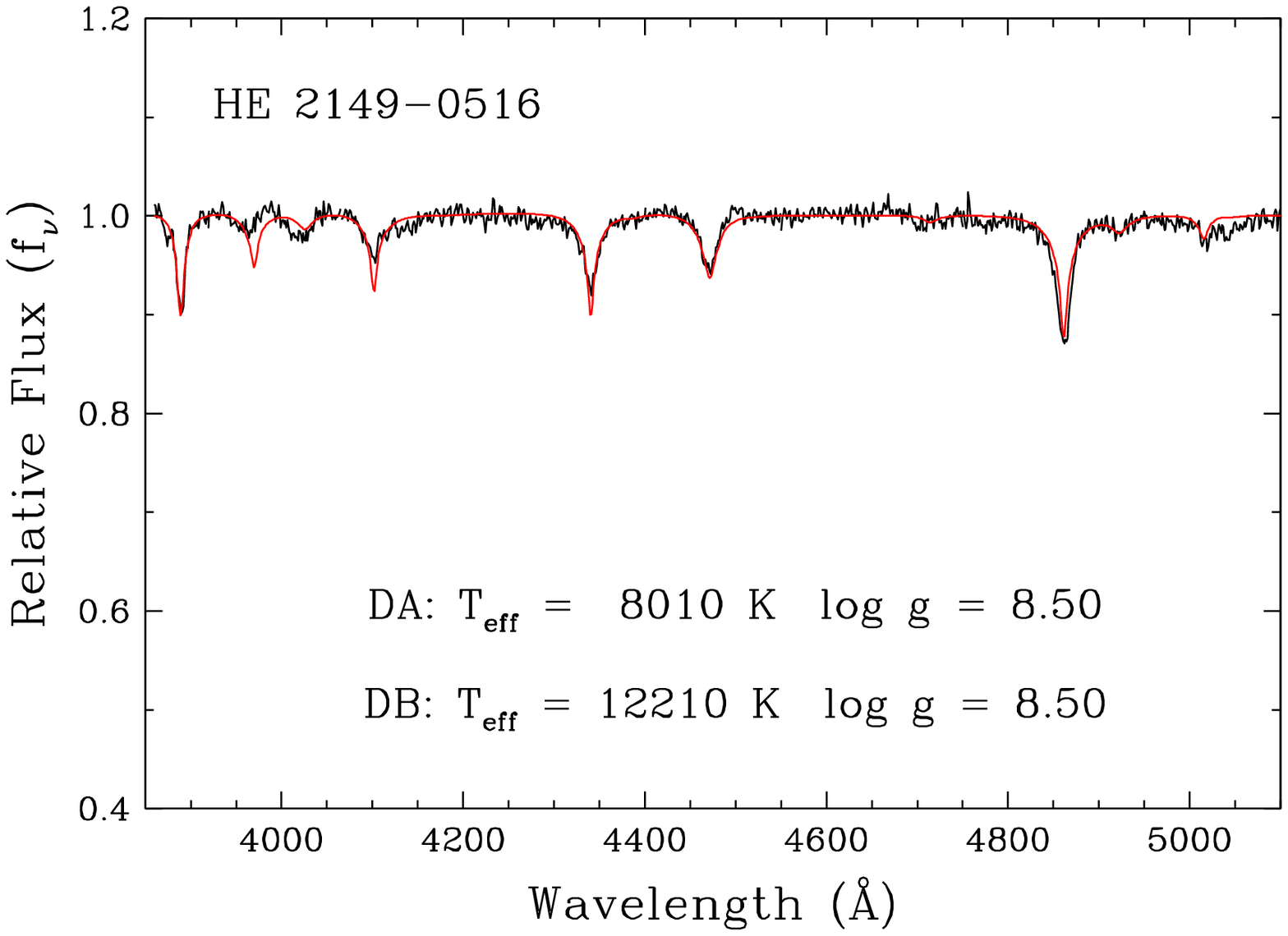] 
{Our best fit to HE 2149$-$0516 assuming composite DA + DB models. The
atmospheric parameters of each component are given in the figure; the
$\logg$ values have been kept constant in the fitting procedure, and a
hydrogen abundance of H/He = $10^{-6}$ was also assumed for the DB
component. Note that the relative flux scale starts at 0.4 so the
absorption lines are particularly weak as a result of the dilution by
each component of the system.\label{fg:f30}}

\clearpage
\begin{figure}[p]
\plotone{f1.eps}
\begin{flushright}
Figure \ref{fg:f1}
\end{flushright}
\end{figure}

\clearpage
\begin{figure}[p]
\plotone{f2.eps}
\begin{flushright}
Figure \ref{fg:f2}
\end{flushright}
\end{figure}

\clearpage
\begin{figure}[p]
\plotone{f3.eps}
\begin{flushright}
Figure \ref{fg:f3}
\end{flushright}
\end{figure}

\clearpage
\begin{figure}[p]
\plotone{f4.eps}
\begin{flushright}
Figure \ref{fg:f4}
\end{flushright}
\end{figure}

\clearpage
\begin{figure}[p]
\plotone{f5a.eps}
\begin{flushright}
Figure \ref{fg:f5}a
\end{flushright}
\end{figure}

\clearpage
\begin{figure}[p]
\plotone{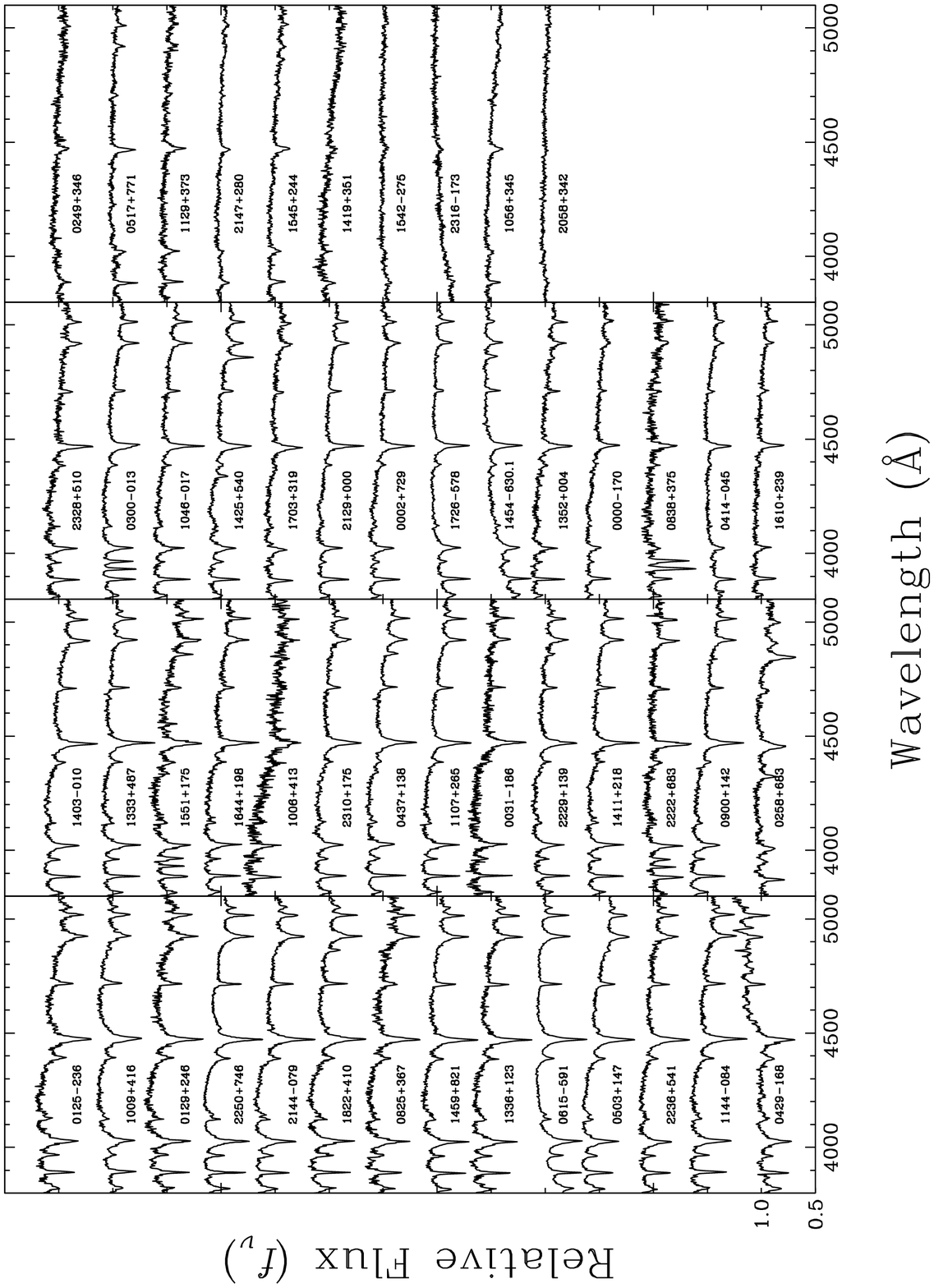}
\begin{flushright}
Figure \ref{fg:f5}b
\end{flushright}
\end{figure}

\clearpage
\begin{figure}[p]
\plotone{f6.eps}
\begin{flushright}
Figure \ref{fg:f6}
\end{flushright}
\end{figure}

\clearpage
\begin{figure}[p]
\plotone{f7.eps}
\begin{flushright}
Figure \ref{fg:f7}
\end{flushright}
\end{figure}

\clearpage
\begin{figure}[p]
\plotone{f8.eps}
\begin{flushright}
Figure \ref{fg:f8}
\end{flushright}
\end{figure}

\clearpage
\begin{figure}[p]
\plotone{f9.eps}
\begin{flushright}
Figure \ref{fg:f9}
\end{flushright}
\end{figure}

\clearpage
\begin{figure}[p]
\plotone{f10.eps}
\begin{flushright}
Figure \ref{fg:f10}
\end{flushright}
\end{figure}

\clearpage
\begin{figure}[p]
\plotone{f11.eps}
\begin{flushright}
Figure \ref{fg:f11}
\end{flushright}
\end{figure}

\clearpage
\begin{figure}[p]
\plotone{f12.eps}
\begin{flushright}
Figure \ref{fg:f12}
\end{flushright}
\end{figure}

\clearpage
\begin{figure}[p]
\plotone{f13.eps}
\begin{flushright}
Figure \ref{fg:f13}
\end{flushright}
\end{figure}

\clearpage
\begin{figure}[p]
\plotone{f14.eps}
\begin{flushright}
Figure \ref{fg:f14}
\end{flushright}
\end{figure}

\clearpage
\begin{figure}[p]
\plotone{f15.eps}
\begin{flushright}
Figure \ref{fg:f15}
\end{flushright}
\end{figure}

\clearpage
\begin{figure}[p]
\plotone{f16.eps}
\begin{flushright}
Figure \ref{fg:f16}
\end{flushright}
\end{figure}

\clearpage
\begin{figure}[p]
\plotone{f17.eps}
\begin{flushright}
Figure \ref{fg:f17}
\end{flushright}
\end{figure}

\clearpage
\begin{figure}[p]
\plotone{f18.eps}
\begin{flushright}
Figure \ref{fg:f18}
\end{flushright}
\end{figure}

\clearpage
\begin{figure}[p]
\plotone{f19.eps}
\begin{flushright}
Figure \ref{fg:f19}
\end{flushright}
\end{figure}

\clearpage
\begin{figure}[p]
\plotone{f20.eps}
\begin{flushright}
Figure \ref{fg:f20}
\end{flushright}
\end{figure}

\clearpage
\begin{figure}[p]
\plotone{f21.eps}
\begin{flushright}
Figure \ref{fg:f21}
\end{flushright}
\end{figure}

\clearpage
\begin{figure}[p]
\plotone{f22.eps}
\begin{flushright}
Figure \ref{fg:f22}
\end{flushright}
\end{figure}

\clearpage
\begin{figure}[p]
\plotone{f23.eps}
\begin{flushright}
Figure \ref{fg:f23}
\end{flushright}
\end{figure}

\clearpage
\begin{figure}[p]
\plotone{f24.eps}
\begin{flushright}
Figure \ref{fg:f24}
\end{flushright}
\end{figure}

\clearpage
\begin{figure}[p]
\plotone{f25.eps}
\begin{flushright}
Figure \ref{fg:f25}
\end{flushright}
\end{figure}

\clearpage
\begin{figure}[p]
\plotone{f26.eps}
\begin{flushright}
Figure \ref{fg:f26}
\end{flushright}
\end{figure}

\clearpage
\begin{figure}[p]
\plotone{f27.eps}
\begin{flushright}
Figure \ref{fg:f27}
\end{flushright}
\end{figure}

\clearpage
\begin{figure}[p]
\plotone{f28.eps}
\begin{flushright}
Figure \ref{fg:f28}
\end{flushright}
\end{figure}

\clearpage
\begin{figure}[p]
\plotone{f29.eps}
\begin{flushright}
Figure \ref{fg:f29}
\end{flushright}
\end{figure}

\clearpage
\begin{figure}[p]
\plotone{f30.eps}
\begin{flushright}
Figure \ref{fg:f30}
\end{flushright}
\end{figure}


\clearpage
\begin{thebibliography}{}

\bibitem[Achilleos et al.(1992)]{achilleos92} Achilleos, N., Wickramasinghe, D.~T., Liebert, J., Saffer, R.~A., \& Grauer, A.~D. 1992, \apj, 396, 273

\bibitem[Beauchamp(1995)]{B95} Beauchamp, A. 1995, Ph.~D.~thesis, Universit\'e de Montr\'eal

\bibitem[Beauchamp(1998)]{beauchamp98} Beauchamp, A. 1998, \jrasc, 92, 126

\bibitem[Beauchamp \& Wesemael(1998)]{BW98} Beauchamp, A., \& Wesemael, F. 1998, \apj, 496, 395

\bibitem[Beauchamp et al.(1997)]{BWB97} Beauchamp, A., Wesemael, F., \& Bergeron, P. 1997, \apj, 108, 559

\bibitem[Beauchamp et al.(1999)]{beauchamp99} Beauchamp, A., Wesemael, F., Bergeron, P., Fontaine, G., Saffer, R.~A., Liebert, J., \& Brassard, P. 1999, \apj, 516, 887

\bibitem[Beauchamp et al.(1995)]{beauchamp95} Beauchamp, A., Wesemael, F., Bergeron, P., \& Liebert, J. 1995, \apj, 441, L85

\bibitem[Beauchamp et al.(1996)]{beauchamp96} Beauchamp, A., Wesemael, F., Bergeron, P., Liebert, J., \& Saffer, R.~A. 1996, in {\it Hydrogen-Deficient Stars}, edited by C.S. Jeffery \& U. Heber, ASP Conference Series 96, San Francisco, 295

\bibitem[Bergeron, Leggett, \& Ruiz(2001)]{BLR01} Bergeron, P., Leggett, S.~K., \& Ruiz, M.~T. 2001, \apjs, 133, 413

\bibitem[Bergeron \& Liebert(2002)]{BL02} Bergeron, P., \& Liebert, J. 2002, \apj, 566, 1091

\bibitem[Bergeron et al.(1992)]{BSL92} Bergeron, P., Saffer, R.~A., \& Liebert, J.  1992, \apj, 394, 228

\bibitem[Bergeron et al.(1995)]{bergeron95} Bergeron, P., Wesemael, F., Lamontagne, R., Fontaine, G., Saffer, R.~A., \& Allard, N.~F. 1995, \apj, 449, 258 

\bibitem[B\"ohm \& Cassinelli(1971)]{ML2} B\"ohm, K.-H., \& Cassinelli, J.~P. 1971, \aap, 12, 21

\bibitem[Brassard \& Fontaine(1997)]{brassard97} Brassard, P., \& Fontaine, G. 1997, in Proc.~3rd Conf.~on Faint Blue Stars, eds A.~G. Davis Philip, J.~Liebert, \& R.~A. Saffer (Schenectady NY: L. Davis Press), 485 

\bibitem[Castanheira et al.(2006)]{castan06} Castanheira, B.~G., Kepler, S.~O., Handler, G., \& Koester, D. 2006, A\&A, 450, 331

\bibitem[Dahn et al.(1988)]{dahn88} Dahn, C.~C., et al.~1988, \aj, 95, 237

\bibitem[Deridder \& Van Rensbergen(1976)]{deridder76} Deridder, G., \& Van Rensbergen, W.~1976, \aaps, 23, 147

\bibitem[Desharnais et al.(2008)]{desharnais08} Desharnais, S., Wesemael, F., Chayer, P., Kruk, J.~W., \& Saffer, R.~A. 2008, \apj, 672, 540

\bibitem[Dufour et al.(2007a)]{dufour07a} Dufour, P., Bergeron, P., Liebert, J., Harris, H.~C., Knapp, G.~R., Anderson, S.~F., Hall, P.~B., Strauss, M.~A., Collinge, M.~J., \& Edwards, M.~C. 2007a, \apj, 663, 1291

\bibitem[Dufour et al.(2010a)]{dufour10a} Dufour, P., Desharnais, S., Wesemael, F., Chayer, P., Lanz, T., Bergeron, P., Fontaine, G., Beauchamp, A., Saffer, R.~A., Kruk, J.~W., \& Limoges, M.-M. 2010a, \apj, 718, 647

\bibitem[Dufour et al.(2008)]{dufour08} Dufour, P., Fontaine, G, Liebert, J., Schmidt, G.~D., \& Behara, N. 2008, \apj, 683, 978

\bibitem[Dufour et al.(2010b)]{dufour10b} Dufour, P., Kilic, M., Fontaine, G., Bergeron, P., Lachapelle, F.-R., Kleinman, S.~J., \& Leggett, S.~K. 2010b, \apj, 719, 803 

\bibitem[Dufour et al.(2007b)]{dufour07b} Dufour, P., Liebert, J., Fontaine, G, \& Behara, N. 2007b, Nature, 450, 522

\bibitem[Dupuis et al.(1993)]{dupuis93} Dupuis, J., Fontaine, G., Pelletier, C., \& Wesemael, F.\ 1993, \apjs, 84, 73

\bibitem[Eisenstein et al.(2006)]{eisen06} Eisenstein D.~J. et al. 2006, \aj, 132, 676

\bibitem[Farihi et al.(2010)]{farihi10} Farihi, J., Barstow, M.~A., Redfield, S., Dufour, P., \& Hambly, N.~C. 2010, \mnras, 404, 2123

\bibitem[Fontaine \& Brassard(1997)]{fontaine97} Fontaine, G., \& Brassard, P. 1997, in White Dwarfs: Proc. 10th European Workshop on White Dwarfs, ed. J. Isern, M. Hernanz, \& E. Garcia-Berro (Dordretch: Kluwer), 451

\bibitem[Fontaine \& Brassard(2008)]{fontaine08} Fontaine, G., \& Brassard, P. 2008, \pasp, 120, 1043

\bibitem[Fontaine et al.(2001)]{fon01} Fontaine, G., Brassard, P., \& Bergeron, P. 2001, \pasp, 113, 409

\bibitem[Fontaine \& Wesemael(1987)]{fontaine87} Fontaine, G., \& Wesemael, F. 1987, in IAU Colloquium 95, The Second Conference on Faint Blue Stars, eds. A. G. Davis Philip, D. S. Hayes, \& J. Liebert (Schenectady: L. Davis), 319 

\bibitem[Friedrich et al.(2000)]{friedrich00} Friedrich, S., Koester, D., Christlieb, N., Reimers, D., \& Wisotzki L. 2000, \aap, 363, 1040

\bibitem[Gould \& Chanam\'e(2004)]{gould04} Gould, A., \& Chanam\'e, J. 2004, \apjs, 150, 455

\bibitem[Green et al.(1986)]{PG} Green, R. F., Schmidt, M., \& Liebert, J. 1986, \apjs, 61, 305 

\bibitem[Handler(2001)]{handler01} Handler, G. 2001, \mnras, 323, L43

\bibitem[Holberg et al.(2003)]{holberg03} Holberg, J.~B., Barstow, M.~A., \& Burleigh, M.~R. 2003, \apjs, 147, 145

\bibitem[Holberg \& Bergeron(2006)]{holberg06} Holberg, J.~B., \& Bergeron, P. 2006, \apj, 132, 1221

\bibitem[Holberg et al.(2008)]{holberg08} Holberg, J.~B., Sion, E.~M., Oswalt, T., McCook, G.~P., Foran, S., \& Subasavage, J.~P. 2008, AJ, 135, 1225

\bibitem[Hummer \& Mihalas(1988)]{HM88} Hummer, D.~G., \& Mihalas, D. 1988, \apj, 331, 794

\bibitem[Hunter et al.(2001)]{hunter01} Hunter, C., Wesemael, F., Saffer, R.~A., Bergeron, P., \& Beauchamp, A. 2001, in {\it 12$^{th}$ European Conference on White Dwarfs}, edited by J.~L. Provencal, H.~L. Shipman, J.~MacDonald, \& S.~Goodchild, ASP Conference Series 226, San Francisco, 153

\bibitem[John(1994)]{john94} John, T.~L.\ 1994, \mnras, 269, 871

\bibitem[Kepler et al.(2007)]{kepler07} Kepler, S.~O., Kleinman, S.~J., Nitta, A., Koester, D., Castanheira, B.~G., Giovannini, O., Costa, A.~F.~M., \& Althaus, L. 2007, \mnras, 375, 1315

\bibitem[Koester et al.(2009)]{koester09} Koester, D., Kepler, S.~O., Kleinman, S.~J., \& Nitta, A.\ 2009, Journal of Physics Conference Series, 172, 012006

\bibitem[L\'epine et al.(2011)]{lepine11} L\'epine, S., Bergeron, P., \& Lanning, H.~H. 2011, \aj, 141, 96

\bibitem[Liebert et al.(2005)]{LBH05} Liebert, J., Bergeron, P., \& Holberg, J.~B. 2005, \apjs, 156, 47

\bibitem[Limoges \& Bergeron(2010)]{limoges10} Limoges, M.-M., \& Bergeron, P. 2010, \apj, 714, 1037

\bibitem[Limoges et al.(2009)]{limoges09} Limoges, M.-M., Bergeron, P., \& Dufour, P. 2009, \apj, 696, 1461

\bibitem[MacDonald \& Vennes(1991)]{MV91} MacDonald, J., \& Vennes, S. 1991, \apj, 371, 719

\bibitem[Massa \& Fitzpatrick(2000)]{massa} Massa, D., \& Fitzpatrick, E.~L. 2000, \apjs, 126, 517

\bibitem[Petitclerc et al.(2005)]{petitclerc05} Petitclerc, N., Wesemael, F., Kruk, F., Chayer, P., \& Bill\`eres, M. 2005, \apj, 624, 317

\bibitem[Press et al.(1986)]{press86} Press, W. H., Flannery, B. P.,
Teukolsky, S. A., \& Vetterling, W. T. 1986, Numerical Recipes
(Cambridge: Cambridge University Press)

\bibitem[Robinson \& Winget(1983)]{rw83} Robinson, E.~L. \& Winget, D.~E. 1983, \pasp, 95, 386

\bibitem[Tassoul et al.(1990)]{tassoul90} Tassoul, M., Fontaine, G., \& Winget, D.E. 1990, ApJS, 72, 335  

\bibitem[Tremblay \& Bergeron(2009)]{TB09} Tremblay, P.-E., \& Bergeron, P. 2009, \apj, 696, 1755

\bibitem[Tremblay \& Bergeron(2011)]{TB11} Tremblay, P.-E., Bergeron, P., \& Gianninas, A. 2011, \apj, 730, 128

\bibitem[Tremblay et al.(2010)]{TB10} Tremblay, P.-E., Bergeron, P., Kalirai, J.~S., \& Gianninas, A.\ 2010, \apj, 712, 1345 

\bibitem[van Altena et al.(1994)]{ypc} van Altena, W.~F., Lee, J.~T., \& Hoffleit, E.~D. 1994, The General Catalogue of Trigonometric Parallaxes (New Haven: Yale University Observatory) (YPC)

\bibitem[Voss et al.(2007)]{voss07} Voss, B., Koester, D., Napiwotzki, R., Christlieb, N., \& Reimers, D. 2007, \aap, 470, 1079

\bibitem[Werner \& Herwig(2006)]{WH06} Werner, K., \& Herwig, F. 2006, \pasp, 118, 183

\bibitem[Wesemael et al.(2001)]{wesemael01} Wesemael, F., Liebert, J., Schmidt, G.~D., Beauchamp, A., Bergeron, P., \& Fontaine, G. 2001, \apj, 554, 1118

\bibitem[Wickramasinghe(1979)]{W79} Wickramasinghe, D.~T. 1979, in {\it White Dwarfs and Variable Degenerate Stars}, edited by H.~M. Van Horn \& V.~Weidemann, University of Rochester, Rochester, 35

\end{thebibliography}
\end{document}